\documentclass[12pt]{iopart}
\usepackage{iopams} 
\usepackage{graphicx}
\usepackage[T1]{fontenc}

\newcommand{\al}{\alpha}
\newcommand{\bt}{\beta}
\newcommand{\gm}{\gamma}
\newcommand{\lb}{\lambda}
\newcommand{\beq}{\begin{equation}}
\newcommand{\eeq}{\end{equation}}
\newcommand{\ba}{\begin{array}}
\newcommand{\ea}{\end{array}}
\newcommand{\bn}{\begin{eqnarray}}
\newcommand{\en}{\end{eqnarray}}

\DeclareMathSymbol{\varGamma}{\mathord}{letters}{"00}
\DeclareMathSymbol{\varDelta}{\mathord}{letters}{"01}
\DeclareMathSymbol{\varTheta}{\mathord}{letters}{"02}
\DeclareMathSymbol{\varLambda}{\mathord}{letters}{"03}
\DeclareMathSymbol{\varXi}{\mathord}{letters}{"04}
\DeclareMathSymbol{\varPi}{\mathord}{letters}{"05}
\DeclareMathSymbol{\varSigma}{\mathord}{letters}{"06}
\DeclareMathSymbol{\varUpsilon}{\mathord}{letters}{"07}
\DeclareMathSymbol{\varPhi}{\mathord}{letters}{"08}
\DeclareMathSymbol{\varPsi}{\mathord}{letters}{"09}
\DeclareMathSymbol{\varOmega}{\mathord}{letters}{"0A}

\newcommand{\wec}[1]{\boldsymbol #1}
\newcommand{\bpmax}{\left(\begin{array}{cc}}
\newcommand{\epmax}{\end{array}\right)}
\newcommand{\nuc}[2]{$^{ #1}$#2}
\newcommand{\nucs}{\nuc{178}{Hf} --- \nuc{200}{Hg}}
\newcommand{\cR}{{\cal R}}
\newcommand{\cU}{{\cal U}}
\newcommand{\cF}{{\cal F}}
\newcommand{\cZ}{{\cal Z}}
\newcommand{\cS}{{\cal S}}
\newcommand{\cW}{{\cal W}}
\newcommand{\cP}{{\cal P}}
\newcommand{\hb}{\hbar}
\newcommand{\Rel}{{\rm Re}}

\newcommand{\warp}{{\wec{\alpha''}=\wec{\alpha'}=\wec{\alpha}}}
\newcommand{\wara}{|_{\warp}}

\eqnobysec

\begin{document}

\topical[Bohr collective Hamiltonian]{Quadrupole collective states within the Bohr collective Hamiltonian}

\author{L Pr\'ochniak$^1$ and S G Rohozi\'nski$^2$}

\address{$^1$ Institute of Physics, Maria Curie-Sk{\l}odowska University, \\ pl. M. Curie-Sk{\l}odowskiej 1, 20-031 Lublin, Poland}
\address{$^2$ Institute of Theoretical Physics, University of Warsaw, \\ Ho\.za 69, 00-681
Warsaw,  Poland}
\ead{leszek.prochniak@umcs.lublin.pl,Stanislaw-G.Rohozinski@fuw.edu.pl}

\begin{abstract}
The article reviews the general version of the Bohr collective model for the
description of quadrupole collective states, including a detailed discussion
the model's kinematics.  The quadrupole coordinates, momenta and angular
momenta are defined and the structure of the isotropic tensor fields as
functions of the tensor variables is investigated.
After the comprehensive discussion of the quadrupole kinematics, the general
form of the classical and quantum Bohr Hamiltonian is presented.  The
electric and magnetic multipole moments operators acting in the collective
space are constructed and the collective sum rules are given.  A discussion
of the tensor structure of the collective wave functions and a review of
various methods of solving the Bohr Hamiltonian eigenvalue equation are also
presented.  Next, the methods of derivation of the classical and quantum
Bohr Hamiltonian from the microscopic many-body theory are recalled.  
Finally, the microscopic approach to the Bohr Hamiltonian is applied to interpret
collective properties of twelve heavy even-even nuclei in the Hf-Hg region.  
Calculated energy levels and  E2 transition probabilities are compared with
experimental data.  

\end{abstract}

\submitto{\jpg}
\maketitle

\section{Introduction}\label{intro} The story of the Bohr collective
Hamiltonian seems to go back to 1879, long before the discovery of atomic
nuclei, when Lord Rayleigh \cite{Ray879} showed that the time dependent
coefficients, $\al_{\lb\mu}(t)$, in equation \beq\label{ldsurf} R(\theta
,\phi ;t)=R_0\left(1+ \sum_{\lb ,\mu}\al_{\lb\mu}(t)Y^{\ast}_{\lb\mu}(\theta
,\phi )\right) \eeq of the surface of an incompressible liquid drop in the
spherical coordinates $R, \theta,\phi$ play the role of normal modes of
small oscillations of the surface around a spherical shape.  The next
milestone was the liquid drop model of an atomic nucleus \cite{Wei35} --- an
object known already at that time.  It allowed Fl\"ugge \cite{Flu41} to
apply the Rayleigh normal modes to a classical description of low-energy
excitations of spherical nuclei.  The model of Rayleigh and Fl\"ugge was
quantized by Aage Bohr \cite{Boh52} who formulated in this way the quantum
model of surface vibrations of spherical nuclei.  It was Aage Bohr also, who
introduced the concept of the intrinsic frame of reference for the
quadrupole nuclear surface and replaced the variables $\al_{2\mu}$ by the
Euler angles and parameters $\bt$ and $\gm$ (nowadays often called the Bohr
deformation parameters) which describe the surface in the intrinsic system. 
Afterwards, Bohr and Mottelson \cite{Boh53} generalized the model to
vibrations and rotations of deformed nuclei.  A generalization of the Bohr
Hamiltonian to describe large-amplitude collective quadrupole excitations of
any even-even nuclei was proposed by Belyaev \cite{Bel65} and Kumar and
Baranger \cite{Kum67}.  This general form of the Bohr collective Hamiltonian is
sketched in Volume 2 of the Bohr-Mottelson book \cite{BM75}.  In the
meantime several specific forms of the collective Hamiltonian
intended to describe the collective excitations in nuclei of different types
were proposed \cite{Wil56,Dav58,Dav60,Fa62a,Fa62b,Fae65}.  All that belongs
to the past.  During the last decades of the twentieth century and in the
beginning of the twenty first century, a great progress in the development
of microscopic many-body theories of nuclear systems took place.  However,
it is still not possible for the self-consistent Hartree-Fock-Bogolyubov
approach, employing single-particle degrees of freedom, to directly describe
the collective phenomena in nuclei.  One of the methods which can be used in
order obtain such a description is the Adiabatic Time Dependent
Hartree-Fock-Bogolyubov theory (ATDHFB) \cite{Bel65,x68ba01,x78ba01}, which
in the case of the quadrupole coordinates leads to the Bohr Hamiltonian.  On
the other hand, the Random Phase Approximation (RPA) (cf \cite {Bar61,Saw61}
for its application to nuclear physics) is not able to explain a large
amplitude collective motion because it assumes harmonicity of the
excitations.  The other method which can extend a microscopic approach to
the collective phenomena is the Generator Coordinate Method (GCM)
\cite{Hill53}.  In its Gaussian Overlap Approximation (GOA)
\cite{Bri68,Oni75} the GCM also yields the Bohr Hamiltonian
\cite{Une76,Goz85}.  The integral Hill-Wheeler equation of the GCM without
this approximation for all
five quadrupole generator coordinates is considerably more difficult to
solve \cite{2008BE29}.

It is not an intention of the present article to review all applications of
the Bohr Hamiltonian used to describe the nuclear quadrupole collective
excitations which have been done hitherto.  Neither is the full list of
publications on this subject compiled here.  Rather the review focuses
 on the Bohr Hamiltonian itself.  \Sref{quadrcoor} contains a detailed
discussion of the kinematics of the quadrupole degrees of freedom.  Tensor
properties of the coordinates and momenta themselves are a subject, which
goes far beyond the Bohr Hamiltonian model.   
 The discussion is based on the
concept of the isotropic tensor fields presented in \ref{tensor}. 
The
tensor algebra and analysis allow us to investigate in \sref{collham} the
possible most general form of the Hamiltonian and other observables.
Moreover, the tensor structure of the collective wave function with a
definite spin is presented.  In \sref{collham} we also recall some
analytically solvable examples of the Bohr Hamiltonian and we briefly review
different methods of solving the eigenvalue equation of the general
Hamiltonian.  A specific version of the collective model is determined by
several scalar functions which define the Hamiltonian and all other
observables.  Within the phenomenological approach to the collective model
the parameters of all these scalar functions are fixed using experimental
data on properties of the collective excitations.  Obviously, there is
another possibility, that all the necessary tensor fields, such as the
collective potential, the inertial functions, the moments of charge
distribution and the gyromagnetic tensors, are derived from an underlying
more fundamental theory.  Among others, the classical or semi-classical theories or
models, such as different versions of the liquid drop model (e.g. 
\cite{Flu41,MS66,MS69}) or the Thomas-Fermi model \cite{Tho27,Fer28} can be
used and  because of such a possible classical background the Bohr Hamiltonian 
is usually qualified  as the
geometrical model.  However, we focus here on the quantum methods of
derivation of the Bohr Hamiltonian from a microscopic many-body theory.  In
\sref{sec:micro} the 
ATDHFB theory, which  leads to the classical 
(at  the intermediate stage)
collective Hamiltonian, and
the Generator Coordinate Method, which gives the quantum Hamiltonian
directly, are recalled and their application to the collective quadrupole
Bohr Hamiltonian is presented.  The section ends with an example of
calculations which illustrates the entire reviewed approach that leads from
a microscopic theory through the Bohr Hamiltonian to the description of the
collective quadrupole excitations in several even-even nuclei (\nucs).  Some
important properties of the ATDHFB inertial functions are discussed in
\ref{app:iner}.

\section{Quadrupole collective coordinates}\label{quadrcoor}

The fundamental assumption of the Bohr collective model applied to the
description of the quadrupole collective states is that the dynamical
variables (coordinates) form a real quadrupole electric, i.e.  of the
positive parity (cf \ref{sphten}), tensor $\al_2^{(\mathrm{E})}$ (the
superscript (E) will be omitted below).  It is not necessary to give the
tensor $\al_2$ a geometrical interpretation.  In particular, the
interpretation of $\al_2$ coming from \eref{ldsurf}, even if often used, is
not needed.  Thus, referring to the model as to the geometrical model is, in
general, unjustified.  Indeed, at the present time the Bohr model with some
special forms of the collective Hamiltonian is formulated and treated simply
as an algebraic collective model (cf e.g.  \cite{Row04,RoTu05,Row09}).

\subsection{The laboratory coordinates}\label{labcoor} 
Let the covariant components $\al_{2\mu}$ ($\mu =-2,\,\dots ,\, 2$) of a tensor
$\al_2$ in the laboratory system U$_{\mathrm{lab}}$ form a set of  five
independent dynamical variables.  Although in general, it is convenient to
use  complex coordinates, there are, in fact,  five real
coordinates $a_{2\mu}\ (\mu =0,\, 1,\, 2)$ and $b_{2\mu}\ (\mu =1,\, 2)$
which are respectively  the real and imaginary parts of $\al_{2\mu}$ (see
\eref{reim} for definition).  The volume element in the space of
these coordinates is (cf \eref{vol})
\beq\label{vol2}
\rmd\Omega =\prod_{\mu =0}^2\rmd a_{2\mu}\prod_{\mu =1}^2\rmd b_{2\mu} .
\eeq 
The covariant momentum tensor and the 
contravariant momentum Hermitian adjoint to it (cf \eref{momcov} and \eref{momcontra}) are
$\hat{\pi}_{2\mu}=-\rmi\hbar\partial /\partial\al^{\ast}_{2\mu}$ and
$\hat{\pi}_{2\mu}^{\dag} =-\rmi\hbar\partial /\partial\al_{2\mu}$,
respectively.  The angular momentum vector connected with rotations in the
physical three-dimensional space, expressed by the five components of the
quadrupole tensor, reads (cf \eref{angmom})
\beq\label{angmom2}
\hat{L}_{1\nu}\equiv \hat{L}^{(2)}_{1\nu}=\rmi\sqrt{10}\left[\al_2\times\hat{\pi}_2\right]_{1\nu} .
\eeq

\subsection{The intrinsic coordinates}\label{intrcoor}

\begin{figure}
\begin{center}
\resizebox{!}{7.5cm}
{  \includegraphics{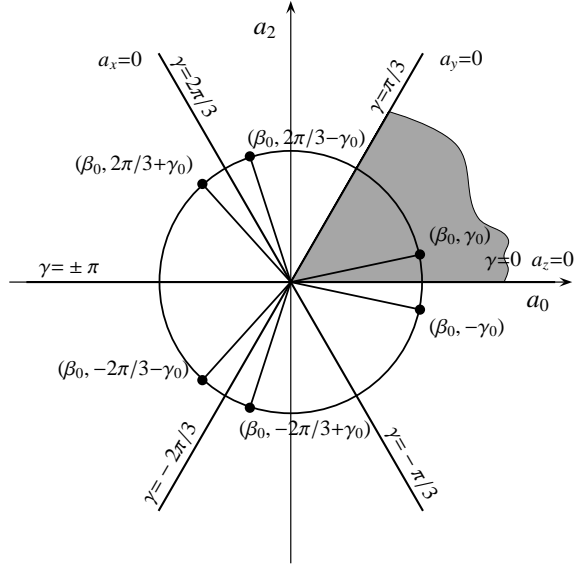}}
\end{center}
\caption{ Three straight lines: $a_x=0$, $a_y=0$, $a_z=0$ (or six rays: $\gm
=0$, $\gm =\pm\pi /3$, $\gm =\pm 2\pi /3$, $\gm =\pm\pi$) going through the
origin of the coordinate system and dividing the $(a_0a_2)$ plane into the six
sectors.  Each point  of the polar coordinates $(\bt_0,\gm_0)$ located  in the sector
$0\leq\gm\leq\pi /3$ has  five twin points in the remaining five sectors. 
All these six points of polar coordinates $(\bt_0,\pm\gm_0)$, $(\bt_0,\pm
2\pi /3\pm\gm_0)$, marked with  full circles, are equivalent to each
other. This  means
that any  intrinsic component of a tensor field at one of these points
is a linear combination with constant coefficients of the intrinsic
components of the same field at any other of the remaining five points. 
It is therefore sufficient to consider tensor fields in  one sector only. 
Usually, the shadowed sector, $0\leq\gm\leq\pi /3$, is considered.}
\label{figsec}
\end{figure}

Apart from, perhaps, the case of the five-dimensional harmonic oscillator 
Hamiltonian
\cite{Boh52}, the laboratory coordinates are not used in practice.  Usually,
one introduces the system of principal axes of the tensor $\al_2$,
U$_{\mathrm{in}}$, called the intrinsic system.  The orientation of
U$_{\mathrm{in}}$ with respect to U$_{\mathrm{lab}}$ is given by the three
Euler angles, $\varphi$, $\vartheta$, $\psi$.  For given laboratory
components $\al_{2\mu}$ the Euler angles are defined through the three
implicit relations:
\numparts
\bn
\sum_{\mu}D^{2\ast}_{\mu\pm 1}(\varphi ,\vartheta ,\psi )\al_{2\mu} &=& 0, \label{ali1}\\
\sum_{\mu}D^{2\ast}_{\mu 2}(\varphi ,\vartheta ,\psi )\al_{2\mu} & = & \sum_{\mu}D^{2\ast}_{\mu -2}(\varphi ,\vartheta ,\psi )\al_{2\mu} \label{ali2}
\en
\endnumparts
which imply vanishing of 
the following real and imaginary parts 
of the intrinsic components $\al^{(\mathrm{in} )}_{2\mu '}$ of $\al_2$:
\beq\label{intr}
a^{(\mathrm{in} )}_{21}=b^{(\mathrm{in} )}_{21}=b^{(\mathrm{in} )}_{22}=0.
\eeq
The two remaining real parts of the intrinsic components are
\numparts
\bn
a_{20}^{(\mathrm{in} )}\equiv a_0 &=&\sum_{\mu}D^{2\ast}_{\mu 0}(\varphi ,\vartheta ,\psi )\al_{2\mu}, \label{a0}\\
a_{22}^{(\mathrm{in} )}\equiv a_2&=&\frac{1}{\sqrt{2}}\sum_{\mu}\left(D^{2\ast}_{\mu 2}(\varphi ,\vartheta ,\psi ) +D^{2\ast}_{\mu -2}(\varphi ,\vartheta ,\psi )\right)\al_{2\mu}. \label{a2}
\en
\endnumparts
These coordinates are usually parametrized by two parameters, $\bt$ and $\gm$ ($0\leq\bt<\infty$, $-\pi\leq\gm\leq\pi$) defined by the relations
\beq\label{bg}
a_0=\bt\cos{\gm} \qquad a_2= \bt\sin{\gm}.
\eeq
The parameters $\bt$ and $\gm$, commonly known as the Bohr deformation parameters, play
the role of polar coordinates in plane $(a_0a_2)$ (see \fref{figsec}).  In
equations \eref{ali1}, \eref{ali2}, \eref{a0} and \eref{a2} the Wigner
functions with $\nu \neq 0$ appear in the following combinations denoted as
\beq
D^{2(\pm)}_{\mu\nu}(\varphi ,\vartheta ,\psi )=\frac{1}{\sqrt{2}}\left( D^{2}_{\mu\nu}(\varphi ,\vartheta ,\psi )
\pm (-1)^{\nu}D^{2}_{\mu -\nu}(\varphi ,\vartheta ,\psi )\right)\ .
\eeq 

Relations \eref{ali1}, \eref{ali2}, \eref{a0} and \eref{a2} define the transformation
\beq\label{trans}
 \al_{2\mu}(a_0,a_2,\varphi ,\vartheta ,\psi ) =D^{2}_{\mu 0}(\varphi ,\vartheta ,\psi )a_0 +D^{2(+)}_{\mu 2}(\varphi ,\vartheta ,\psi )a_2
 \eeq
from the intrinsic coordinates $a_0,\, a_2 ,\, \varphi ,\, \vartheta ,\, \psi$ to the laboratory ones --- $\al_{2\mu}\; (\mu =-2,\, \dots ,\, 2)$.
The Jacobian of this transformation is, up to a constant coefficient, equal to (cf \cite{Roh82})
\numparts
\bn
&& D(a_0,a_2 ,\varphi ,\vartheta ,\psi)=\left|\frac{\mathrm{D}(\al_{2-2},\dots ,\al_{22})}{\mathrm{D}(a_0 ,a_2 ,\varphi ,\vartheta ,\psi )}\right|
=|X(a_0,a_2)|\sin{\vartheta} \label{jacob1} \\
&& X(a_0,a_2)=a_2(3a_0^2-a_2^2)a_2(\sqrt{3}a_0 +a_2)(\sqrt{3}a_0 -a_2)=\bt^3\sin{3\gm}. \label{jacob2}
\en
\endnumparts
The transformation \eref{bg} to the polar coordinates $\bt$ and $\gm$ adds the factor $\bt$ to
the Jacobian \eref{jacob2} since
$\rmd a_0\rmd a_2= \bt\rmd \bt\rmd\gm$.
For values of the intrinsic coordinates for which the Jacobian is not equal
to zero the transformation is invertible, meaning that the intrinsic frame
can be defined.  The inverse transformation is given by implicit
functions of the Euler angles.  The Jacobian vanishes on the three straight
lines in the $(a_0a_2)$ plane specified by the conditions
\bn\label{zerojac}
\fl a_x=0\ \mathrm{or}\ \gm =0, \pm\pi \qquad a_y=0\ \mathrm{or}\ \gm =\pi /3, -2\pi /3\qquad a_z=0\ \mathrm{or}\ \gm =2\pi /3, -\pi /3
\en
where
\bn
\fl a_x=-\frac{1}{2}\left(\sqrt{3}a_0+a_2\right)=\bt\sin{\gm_x} \qquad a_y=\frac{1}{2}\left(\sqrt{3}a_0-a_2\right)=\bt\sin{\gm_y} \nonumber \\
\fl a_z=a_2=\bt\sin{\gm_z}\qquad \mathrm{and}\qquad
\gm_x=\gm -2\pi /3 \qquad  \gm_y=\gm +2\pi /3\qquad  \gm_z=\gm .\label{axyz}
\en
The lines intersect  at the point $a_0=a_2=0\ (\bt =0)$ and divide the plane
into six sectors as is shown in \fref{figsec}.  At the boundaries of these
sectors the intrinsic frame cannot be uniquely defined.  The reason is that
the tensor $\al_2$ is then axially symmetric (cf \eref{rotz}).

As will become clear, the appearance of the three lines, instead of only one,
on which the Jacobian vanishes, reflects the fact that equations
\eref{ali1} and \eref{ali2} define, in general, three mutually perpendicular
axes in the three dimensional space but not their ordering (say $x,y,z$) nor their
arrows.  This means that the axes are defined up to a discrete group
of 48 transformations O$_k$, $k= 0,\,\dots ,\, 47$, changing the names and
arrows of the axes.  The group is the symmetry group of a cube
and is denoted by O$_{\mathrm{h}}$ (cf \cite{Ham64}).

Traditionally, one generates all the 48 transformations ${\mathrm O}_k \in
{\mathrm O}_{\mathrm h}$ by three so called Bohr's rotations, R$_1$,
R$_2$, R$_3$, and the inversion P (other sets of generators can also be
chosen).  Definitions of the Bohr's rotations are given in table
\ref{tabgen}.  The twenty four unimodular transformations (rotations) O$_k$
for $k= 0,\, \dots ,\, 23$, which are the products of the Bohr's rotations are
listed in \cite{Cor76} (but the numbering in \cite{Cor76} is different from
that used in this paper).  The remaining twenty four transformations can be
formed by composing the unimodular ones with the inversion.  

\begin{table}
\caption{\label{tabgen} Generators of the O$_{\mathrm{h}}$ group of
transformations of the coordinate system.  The index $k$ numbers the
transformations.  The angles $\varpi_k$, $\varrho_k$, $\varsigma_k$ are the
Euler angles defining the orientation of the coordinate system after the
transformation O$_k$.  The coordinates $x_k$, $y_k$, $z_k$ define the position 
with respect to the transformed system.}
\begin{indented}
\lineup
\item[]\begin{tabular}{rclrrrrrr}\br
     \centre{3}{Transformation}&\centre{3}{The Euler angles}&\centre{3}{Coordinates}\\
\crule{3}&\crule{3}&\crule{3}\\
$k$ & Description &O$_k$ & $\varpi_k$ & $\varrho_k$ & $\varsigma_k$ & $x_k$ & $y_k$ & $z_k$ \\
\ns
\mr
0 & Identity &E & 0 & 0 & 0 & $x$ & $y$ & $z$ \\
1 & Bohr's &R$_1$ & $\pi$ & $\pi$ & 0 & $x$ & $-y$ & $-z$ \\
2 & rotat- &R$_2$ & 0 & 0 & $\pi /2$ & $y$ & $-x$ & $z$ \\
3 & ions & R$_3$ & 0 & $\pi /2$ & $\pi /2$ & $y$ & $z$ & $x$ \\
24 & Inversion &P & 0 & 0 & 0 & $-x$ & $-y$ & $-z$ \\
\br
\end{tabular}
\end {indented}
\end{table}

The transformation rules of the tensor $\al_2$ under the O$_\mathrm{h}$
transformations are presented in table \ref{tabtr}.  Since the tensor
$\al_{2}$ has the positive parity, it is insensitive to the inversion of the
coordinate system.  For $\al_{2}$ it is therefore sufficient to consider
only the SO(3) group of rotations.  The five-dimensional irreducible
representation $D^{(2)}$ of SO(3) can be decomposed into  two irreducible
representations,  $E$ and $F_2$, of the octahedral group O.  These
irreducible representations are identical with representations of
O$_\mathrm{h}$ \cite{Ham64}.  This is clear from table \ref{tabtr}, where
the transformation rules of the real and imaginary parts of $\al_{2\mu}$
under the Bohr's rotations are shown.  The bases of the irreducible
representations $E$ and $F_2$ are easily identified.  Clearly, $a_{20}$ and
$a_{22}$ form a basis of the two-dimensional irreducible representation $E$
of O, whereas the components $-a_{21}$, $-b_{21}$, $b_{22}$ form a basis of
the three-dimensional irreducible representation $F_2$.  This makes obvious
that equalities \eref{intr} are invariant under the O$_\mathrm{h}$
transformations.

\begin{table}
\caption{\label{tabtr}Transformation rules of $a_{2\mu}$ and $b_{2\mu}$
under R$_1$, R$_2$ and R$_3$.  The real and imaginary parts of the $\al_2$
components transformed by R$_k$ are denoted by $a_{2\mu}^{(k)}$ and
$b_{2\mu}^{(k)}$.  Decomposition into two irreducible representations of the
group O, $E$ and $F_2$, is also shown.}
\begin{indented}
\lineup
\item[]\begin{tabular}{clccccc}
\br
 & &\centre{2}{$E$}&\centre{3}{$F_2$}\\
& &\crule{2}&\crule{3}\\
$k$ & O$_k$ & $a^{(k)}_{20}$ & $a^{(k)}_{22}$ & $a^{(k)}_{21}$ & $b^{(k)}_{21}$ & $b^{(k)}_{22}$ \\
\mr
0 &E & $a_{20}$ & \m$a_{22}$ & \m$a_{21}$ & \m$b_{21}$ & \m$b_{22}$ \\
1 &R$_1$ &$a_{20}$ & \m$a_{22}$ & $-a_{21}$ & \m$b_{21}$ & $-b_{22}$ \\
2 &R$_2$ &$a_{20}$ & $-a_{22}$ & \m$b_{21}$ & $-a_{21}$ & $-b_{22}$ \\
3 &R$_3$ &$-\frac{1}{2}a_{20}+\frac{\sqrt{3}}{2}a_{22}$ & $-\frac{\sqrt{3}}{2}a_{20}-\frac{1}{2}a_{22}$ &
$-b_{21}$ & \m$b_{22}$ & $-a_{21}$ \\
24 &P & $a_{20}$ & \m$a_{22}$ & \m$a_{21}$ & \m$b_{21}$ & \m$b_{22}$ \\
\br
\end{tabular}
\end{indented}
\end{table}

When names and arrows of the intrinsic axes are changed,  intrinsic
coordinates $a_0$, $a_2$ (or $\bt$, $\gm$), $\varphi$, $\vartheta$, $\psi$
of \eref{trans} are transformed accordingly.    Changes of the deformation parameters and of the orientation
angles under the O transformations are given in table \ref{tabintr}.  Values of
the Euler angles $\varphi_3,\ \vartheta_3,\ \psi_3$ after the circular
permutation R$_3$ of the intrinsic axes can be found from the following
equations given in \cite{Var88}:
\bn
\fl \cos{\vartheta_3}= -\sin{\vartheta}\cos{\psi} \qquad \sin{\vartheta_3}\sin{\psi_3}= \cos{\vartheta} \qquad
\cos{(\varphi_3-\varphi)}=\cos{\psi}\sin{\psi_3}. \label{ang3}
\en
\begin{table}
\caption{\label{tabintr}Transformations of the Bohr deformation parameters 
and of the Euler angles defining the orientation of the intrinsic axes with respect to 
the laboratory axes under the Bohr's rotations R$_k$, $k=1,2,3$, of the intrinsic 
coordinate system.
The Euler angles $\varphi_3$, $\vartheta_3$,
$\psi_3$ giving the orientation of the intrinsic system with axes  permuted 
circularly can be found from equations \eref{ang3}.}
\begin{indented}
\lineup
\item[]\begin{tabular}{clcccccccc}
\br
\centre{5}{Bohr's rotation}&\centre{2}{Deformations}&\centre{3}{The Euler angles}\\
\crule{5}&\crule{2}&\crule{3}\\
$k$ &O$_k$ &$\varpi_k$ & $\varrho_k $ & $\varsigma_k$ & $\bt_k$ & $\gm_k$ & $\varphi_k$ & $\vartheta_k$ & $\psi_k$ \\
\mr
0 & E & 0 & 0 & 0 & $\bt$ & \m$\gm$ & $\varphi$ & $\vartheta$ &\m$\psi$ \\
1 & R$_1$ & $\pi$ & $\pi$ & 0 & $\bt$ & \m$\gm$ & $\varphi \pm\pi$ & $\pi -\vartheta$ & $-\psi$ \\
2 & R$_2$ & 0 & 0 & $\pi /2$ & $\bt$ & $-\gm$ & $\varphi$ & $\vartheta$ & $\psi +\pi /2$ \\
3 & R$_3$ & 0 & $\pi /2$ & $\pi /2$ & $\bt$ & $\gm -2\pi /3$ & $\varphi_3$ & $\vartheta_3$ & $\psi_3$ \\
\br
\end{tabular}
\end{indented}
\end{table}

The Wigner functions of the Euler angles $\varphi_k,\vartheta_k,\psi_k$
giving the 
orientation of the rotated intrinsic system are related to those of the
Euler angles $\varphi ,\vartheta ,\psi$ of the orientation of the initial
intrinsic system by the following superposition formula \cite{Var88}:
\beq\label{superpos}
D^{\lb}_{\mu\nu}(\varphi_k,\vartheta_k,\psi_k)=\sum_{\mu '}D^{\lb}_{\mu\mu '}(\varphi ,\vartheta ,\psi )
D^{\lb}_{\mu '\nu}(\varpi_k,\varrho_k,\varsigma_k)
\eeq
where $\varpi_k,\varrho_k,\varsigma_k$ with $k= 0,1,2,3$ 
are the Euler angles of the corresponding rotation.

In order to present the contravariant and covariant momentum tensors as
differential operators in the intrinsic variables, one should first
calculate the derivatives of the intrinsic coordinates with respect to the
components of the tensor $\al_2$ using  relations \eref{ali1},
\eref{ali2} and \eref{a0}, \eref{a2}.  The calculation is a bit tedious. 
Relations \eref{ali1} and \eref{ali2}, when differentiated with respect
to $\al_{2\mu}$, yield a system of three linear equations for the
derivatives of the Euler angles from which the derivatives themselves can be
calculated.  Useful formulae for the derivatives of the Wigner
functions with respect to the Euler angles can be found in \cite{Var88}.  By
differentiating relations \eref{a0} and \eref{a2} with respect to
$\al_{2\mu}$ one obtains the derivatives of $a_0$ and $a_2$.  The complete
calculation is performed in \cite{Roh82}.  The final result for the momentum
tensor reads:
\bn
\fl \hat{\pi}_{2\mu} = -\rmi\hbar\left(D^2_{\mu 0}(\varphi ,\vartheta ,\psi )\frac{\partial}{\partial a_0}
+D^{2(+)}_{\mu 2}(\varphi ,\vartheta ,\psi )\frac{\partial}{\partial a_2}\right)
 +\frac{\rmi D^{2(-)}_{\mu 2}(\varphi ,\vartheta ,\psi )}{2a_z}\hat{L}_z(\varphi ,\vartheta ,\psi ) \nonumber \\
-\frac{\rmi D^{2(-)}_{\mu 1}(\varphi ,\vartheta ,\psi )}{2a_x}\hat{L}_x(\varphi ,\vartheta ,\psi )
-\frac{D^{2(+)}_{\mu 1}(\varphi ,\vartheta ,\psi )}{2a_y}\hat{L}_y(\varphi ,\vartheta ,\psi ) \label{inmom}
\en
where $a_x$, $a_y$, $a_z$ are defined by \eref{axyz}, and
\bn
\hat{L}_x(\varphi ,\vartheta ,\psi ) &=&-\rmi\hbar\left(-\frac{\cos{\psi}}{\sin{\vartheta}}\frac{\partial}{\partial\varphi}
+\sin{\psi}\frac{\partial}{\partial\vartheta}+\cot{\vartheta}\cos{\psi}\frac{\partial}{\partial\psi}\right) \nonumber \\
\hat{L}_y(\varphi ,\vartheta ,\psi ) &=&-\rmi\hbar\left(\frac{\sin{\psi}}{\sin{\vartheta}}\frac{\partial}{\partial\varphi}
+\cos{\psi}\frac{\partial}{\partial\vartheta}-\cot{\vartheta}\sin{\psi}\frac{\partial}{\partial\psi}\right) \label{inangmom} \\
\hat{L}_z(\varphi ,\vartheta ,\psi ) &=&-\rmi\hbar\frac{\partial}{\partial\psi} \nonumber
\en
are the Cartesian intrinsic components of the angular momentum. Indeed, from
\eref{angmom}, \eref{trans} and \eref{inmom} one can derive the following
relations 
\beq
\hat{L}^{(\mathrm{in})}_{10} = \hat{L}_z \qquad
\hat{L}^{(\mathrm{in})}_{1\pm 1} = \mp\frac{1}{\sqrt{2}}\left(\hat{L}_x \pm \hat{L}_y\right) \label{sphinangmom}
\eeq
between the laboratory and the intrinsic spherical components
of the angular momentum vector:
\beq\label{trlabinam}
\hat{L}_{1\mu} = \sum_{\nu}D^1_{\mu\nu}\hat{L}^{(\mathrm{in})}_{1\nu}.
\eeq
It should be remembered that the spherical intrinsic components of angular
momentum fulfill the commutation relations with the sign opposite to that of
\eref{commll}, namely
\beq\label{commllin}
[\hat{L}_{1\mu}^{(\mathrm{in} )},\hat{L}_{1\nu}^{(\mathrm{in} )}]=+\hbar\sqrt{2}(1\mu 1\nu |1\kappa )\hat{L}_{1\kappa}^{(\mathrm{in} )}
\eeq
and that these components commute with the laboratory ones
\beq\label{commlinl}
[\hat{L}_{1\mu}^{(\mathrm{in} )},\hat{L}_{1\nu}]=0.
\eeq
Using the deformation parameters  instead of the intrinsic components one 
should replace in \eref{inmom} the derivatives with respect to $a_0$ and $a_2$ 
by the derivatives with respect to $\bt$ and $\gm$:
\bn\label{derbg}
\fl \frac{\partial}{\partial a_0}=\cos{\gm}\frac{\partial}{\partial\bt}-\sin{\gm}\frac{1}{\bt}\frac{\partial}{\partial{\gm}} \qquad
\frac{\partial}{\partial a_2}=\sin{\gm}\frac{\partial}{\partial\bt}+\cos{\gm}\frac{1}{\bt}\frac{\partial}{\partial{\gm}}.
\en
It should be noticed that both the laboratory and the intrinsic components
of angular momentum do not commute with the Wigner functions and that the
order of $D^{2(\pm)}_{\mu\nu}$ and $\hat{L}^{(\mathrm{in})}_{1\nu}$ in
\eref{inmom} is essential (this is not so in \eref{trlabinam} --- see
\cite{BM69}).  However, it is sometimes convenient to exchange their order
 using the commutation relations.  Doing so in the formulae for the
Hermitian adjoint momentum one obtains:
\bn\label{hcinmom}
\fl \hat{\pi}^{\dag}_{2\mu} = -\rmi\hbar\left(
\left(\frac{\partial}{\partial a_0}+\frac{\sqrt{3}}{2}\left(\frac{1}{a_y}-\frac{1}{a_x}\right)\right)
\left(D^2_{\mu 0}(\varphi ,\vartheta ,\psi )\right)^{\ast}\right. \nonumber \\
\fl +\left.\left(\frac{\partial}{\partial a_2}+\frac{1}{2}\left(\frac{2}{a_z}-\frac{1}{a_x}-\frac{1}{a_y}\right)\right)
\left( D^{2(+)}_{\mu 2}(\varphi ,\vartheta ,\psi )\right)^{\ast}\right)
 -\frac{\rmi}{2a_z}\hat{L}_z(\varphi ,\vartheta ,\psi )\left( D^{2(-)}_{\mu 2}(\varphi ,\vartheta ,\psi )\right)^{\ast}\nonumber \\
\fl +\frac{\rmi}{2a_x}\hat{L}_x(\varphi ,\vartheta ,\psi )\left( D^{2(-)}_{\mu 1}(\varphi ,\vartheta ,\psi )\right)^{\ast}
-\frac{1}{2a_y}\hat{L}_y(\varphi ,\vartheta ,\psi )\left(D^{2(+)}_{\mu 1}(\varphi ,\vartheta ,\psi )\right)^{\ast}.
\en
Using the deformation parameters $\bt$ and $\gm$ one should appropriately
express the derivatives and take the dependence of $a_0$ and $a_2$ on $\bt,\
\gm$ into account.  The following relations are useful to this end:
\bn\label{axyzbg}
\fl \frac{\sqrt{3}}{2}\left(\frac{1}{a_y}-\frac{1}{a_x}\right)=\frac{3\sin{2\gm}}{\bt\sin{3\gm}} \qquad
\frac{1}{2}\left(\frac{2}{a_z}-\frac{1}{a_x}-\frac{1}{a_y}\right)=\frac{3\cos{2\gm}}{\bt\sin{3\gm}}.
\en
At the end of the discussion of the intrinsic coordinates, let them be
functions of the parameter $t$ (time in the case of classical motion).  The
laboratory coordinates become then the functions of $t$ too.  The
derivatives of $\al_{2\mu}$ with respect to $t$ read:
\bn\label{vel}
\fl \dot{\al}_{2\mu}=\dot{a}_0D^2_{\mu 0}(\varphi ,\vartheta ,\psi )
+\dot{a}_2D^{2(+)}_{\mu 2}(\varphi ,\vartheta ,\psi ) \nonumber \\
\fl -2\rmi\omega_xa_xD^{2(-)}_{\mu 1}(\varphi ,\vartheta ,\psi )
-2\omega_ya_yD^{2(+)}_{\mu 1}(\varphi ,\vartheta ,\psi )
+2\rmi\omega_za_zD^{2(-)}_{\mu 2}(\varphi ,\vartheta ,\psi )
\en
where
\bn
\fl \omega_x =\dot{\vartheta}\sin{\psi}-\dot{\varphi}\sin{\vartheta}\cos{\psi} \qquad
\omega_y = \dot{\vartheta}\cos{\psi}+\dot{\varphi}\sin{\vartheta}\sin{\psi} \qquad
\omega_z = \dot{\psi}+\dot{\varphi}\cos{\vartheta} \label{omega}
\en
play the role of the intrinsic components of the angular velocity. The
derivatives $\dot{a}_0$ and $\dot{a}_2$ can be expressed by $\dot{\bt}$ and
$\dot{\gm}$:
\bn\label{dotbg}
\fl \dot{a}_0=\dot{\bt}\cos{\gm}-\bt\dot{\gm}\sin{\gm} \qquad \dot{a}_2=\dot{\bt}\sin{\gm}+\bt\dot{\gm}\cos{\gm}.
\en
 
\subsection{Functions of the coordinates}\label{funcoor}

The general rules presented in  \ref{tensor} for constructing isotropic
tensor fields of a given tensor are applied here to the case of the
quadrupole tensor $\al_2$.  In this case one has the following five
elementary tensors \cite{Olv99}:
\numparts
\begin{enumerate}
\item  two quadrupole tensors
\bn
\al_{2\mu} &=& \varepsilon^{(1)}_{2\mu}(\al_2) \label{e21} \\
\sigma_{2\mu}=-\sqrt{7/2}[\al_2\times\al_2]_{2\mu}&\propto &\varepsilon^{(2)}_{2\mu}(\al_2) \label{e22}
\en
\item  two scalars
\bn
\sigma_0=(\al_2\cdot\al_2) &\propto &\varepsilon^{(2)}_{00}(\al_2) \label{e02} \\
\chi_0= (\al_2\cdot\sigma_2) &\propto &\varepsilon^{(3)}_{00}(\al_2) \label{e03} 
\en
\item  one octupole tensor
\beq\label{e33}
\chi_{3\mu}=\sqrt{2}[\al_2\times\sigma_2]_{3\mu} \propto \varepsilon^{(3)}_{3\mu}(\al_2).
\eeq
\end{enumerate}
\endnumparts
The elementary tensors are related to each other by the single syzygy which
reads (cf \cite{Chac76}):
\bn\label{syz2}
\fl \left[\chi_3\times\chi_3\right]_{6M}=\frac{\sqrt{6}}{3}\left\{2\chi_0\left[\al_2\times\al_2\times\al_2\right]_{6M}\right. \nonumber \\ 
\left. -3\sigma_0\left[\al_2\times\al_2\times\sigma_2\right]_{6M}
+\left[\sigma_2\times\sigma_2\times\sigma_2\right]_{6M}\right\}.
\en
Since the intrinsic components $\al_{2\pm 1}^{(\mathrm{in})}=0$ (see
\eref{ali1}), the only  nonvanishing intrinsic components 
of the elementary tensors are
\numparts
\bn
\fl \al_{20}^{(\mathrm{in})}=a_0=\bt\cos{\gm} \qquad  \al_{2\pm 2}^{(\mathrm{in})}=\frac{1}{\sqrt{2}}a_2=\frac{1}{\sqrt{2}}\bt\sin{\gm} \label{e21in} \\
\fl \sigma_{20}^{(\mathrm{in})}=a_0^2-a_2^2=\bt^2\cos{2\gm} \qquad  
\sigma_{2\pm 2}^{(\mathrm{in})}=-\sqrt{2}a_0a_2=-\frac{1}{\sqrt{2}}\bt^2\sin{2\gm} \label{e22in} \\
\fl \chi_{32}^{(\mathrm{in})}=-\chi_{3-2}^{(\mathrm{in})}=\frac{1}{\sqrt{2}}a_2(3a_0^2-a_2^2)=\frac{1}{\sqrt{2}}\bt^3\sin{3\gm}.  \label{e33in}
\en
\endnumparts
The scalars $\sigma_0$ and $\chi_0$ are the following functions of $a_0$ and
$a_2$, or $\bt$ and $\gm$:
\bn\label{e00}
\fl \sigma_0=a_0^2+a_2^2=\bt^2 \qquad \chi_0=a_0(a_0^2-3a_2^2)=\bt^3\cos{3\gm}.
\en
Scalar functions depending on $\sigma_0$ and $\chi_0$ can be treated as
functions of $\bt$ and $\zeta =\cos{3\gm}$.  The fundamental tensors of a
given even rank $2L$ ($L=0,\,\dots $) are built up from elementary tensors
$\al_2$ and $\sigma_2$ as follows:
\numparts
\beq\label{ftev}
\tau_{k\, 2L}\equiv \tau_{2L}^{(n)} =\left[\underbrace{\al_2\times\cdots\times\al_2}_{2L-n}\times
\underbrace{\sigma_2\times\cdots\times\sigma_2}_{n-L}\right]_{2L}
\eeq
for $k=1,\,\dots ,L+1$ ($n=L,\,\dots ,2L$), where $k$ stands for the tensor
consecutive number and $n$ --- for its order in $\al_2$ ($n=L+k-1$).  To
construct the fundamental tensors of an odd rank $2L+3$ (the tensor with
$L=1$ does not exist) one aligns the even rank tensors \eref{ftev} with the
single elementary tensor $\chi_3$ (see \eref{syz2}):
\beq\label{ftod}
\tau_{k\, 2L+3}\equiv \tau_{2L+3}^{(n+3)} =\left[\underbrace{\al_2\times\cdots\times\al_2}_{2L-n}\times
\underbrace{\sigma_2\times\cdots\times\sigma_2}_{n-L}\times\chi_3\right]_{2L+3}.
\eeq
\endnumparts
The ranges of $l$ and $n$ do not change. Obviously, for the fundamental
tensors, only the intrinsic components with even magnetic numbers are
nonvanishing.  An analytic formula for the intrinsic components of the
fundamental tensors created by aligning  several $\al_2$'s alone is known
\cite{Tur73,Chac77}.  It reads:
\beq\label{alphas}
\tau_{1\, 2l\, 2m}^{(\mathrm{in})}=\left[\underbrace{\al_2\times\cdots\times\al_2}_{l}\right]_{2l\, 2m}^{(\mathrm{in})}
=\bt^lt_{2l\, 2m}(\gm )
\eeq
where
\bn
\fl t_{2l\, 2m}(\gm )=\sqrt{\frac{(2l+2m)!(2l-2m)!}{(4l)!}}{l \choose m}\left(\sqrt{6}\right)^{l-m}\left(\cos{\gm}\right)^{l-m}
\left(\frac{1}{\sqrt{2}}\sin{\gm}\right)^m \nonumber \\ 
\times {}_2F_1\left(-\frac{l-m}{2},-\frac{l-m-1}{2};m+1;\frac{1}{3}\tan^2{\gm}\right) \label{ChMo}
\en
with ${}_2F_1$ being the hypergeometric function \cite{Abr72}. Then,
comparing \eref{e22in} with \eref{e21in} one finds that
\beq\label{sigmas}
\tau_{l+1\, 2l\, 2m}^{(\mathrm{in})}=\left[\underbrace{\sigma_2\times\cdots\times\sigma_2}_{l}\right]_{2l\, 2m}^{(\mathrm{in})}
=\bt^{2l}t_{2l\,2m}(-2\gm ).
\eeq
The fundamental tensor with a given even rank $2L$ and the number $k=n+1\
(n=0,\ \dots ,L)$ is created by aligning  the tensors \eref{alphas} and
\eref{sigmas} with ranks $2l=2L-2n$ and $2l=2n$, respectively.  To form the
fundamental tensor of an odd rank, the third one, namely $\chi_3$ of
\eref{e33in} should be aligned.  For further use, it is convenient to
introduce the `dimensionless' intrinsic components of the fundamental
tensors which depend only on $\gm$:
\numparts
\bn
\fl t_{n+1\, 2L\, 2K}(\gm )=\frac{1}{\bt^{L+n}}\tau_{n+1\, 2L\, 2K}^{(\mathrm{in})}=\left[t_{2L-2n}(\gm)\times t_{2n}(-2\gm )\right]_{2L\, 2K} \label{tev}\\
\fl t_{n+1\, 2L+3\, 2K}(\gm )=\frac{1}{\bt^{L+n+3}}\tau_{n+1\, 2L+3\, 2K}^{(\mathrm{in})}=\left[t_{n+1\, 2L}(\gm)\times \frac{1}{\bt^3}\chi_{3}\right]_{2L+3\, 2K}.
\label{todd}
\en 
\endnumparts
An arbitrary isotropic tensor field of the quadrupole tensor $\al_2$ has the
following structure:
 \beq\label{field2}
T_{IM}(\al_2)=\sum_{k=1}^{k_ I}f_k(\bt ,\zeta )\tau_{kIM}(\al_2,\sigma_2,\chi_3)
\eeq
where 
\bn\label{noten}
k_I &= &\left\{\ba{ll}I/2+1& \qquad \mathrm{for}\;\mathrm{even}\; I \\(I-3)/2+1& \qquad \mathrm{for}\;\mathrm{odd}\; I \ea\right.
\en
and $f_k$ are arbitrary scalar functions.  There is no vector (tensor of
rank 1) field of $\al_2$.  In the special case of the quadrupole field $(I=2)$,
formula \eref{field2} can be deduced (see \cite{Eis87}) from the
Hamilton-Cayley theorem for symmetric matrices (cf e.g.  \cite{Birk65}).

A general form of a quadrupole symmetric bitensor field\footnote{A quadrupole 
nonsymmetric bitensor field is determined by specifying seven, and not six, 
arbitrary scalar functions. For such a field one more term, namely
$m^{(3)}_3(\bt ,\zeta )(2\mu 2\nu |3K)\chi_{3K}(\al_2)$, should be added to
the right-hand side of \eref{bi2decom}.}, 
$M_{2,2}$, can be found from \eref{field2} and \eref{bidecom}:
\beq\label{bi2decom}
M_{2\mu ,2\nu}(\al_2)=\sum_{\lb =0}^2\sum_{n=\lb}^{2\lb}m^{(2\lb )}_n(\bt ,\zeta )\sum_{K=-2\lb}^{2\lb}(2\mu 2\nu |2\lb K)\tau^{(n)}_{2\lb K}
\eeq
with the six arbitrary scalar functions $m^{(2\lb )}_n$.  Since the
fundamental tensors have the nonvanishing components only with even
projections in the intrinsic system, the intrinsic components
$M^{(\mathrm{in})}_{2\mu ,2\nu}$ with odd $|\mu
-\nu |$ are equal to zero.  Also, by virtue of \eref{e21in} and
\eref{e22in}, the components differing in signs of both projections are
equal to each other.  In consequence, an arbitrary symmetric quadrupole
bitensor has only six, and not fifteen, independent nonvanishing intrinsic
components.  It is convenient to work with their following combinations:
\bn\label{eq:mxm1}
 M_x=-M^{(\mathrm{in})}_{21, 21}-M^{(\mathrm{in})}_{21, 2-1} &\qquad & M_0=M^{(\mathrm{in})}_{20, 20} \nonumber \\
M_y=+M^{(\mathrm{in})}_{21, 21}-M^{(\mathrm{in})}_{21, 2-1} &\qquad & M_1=\sqrt{2}M^{(\mathrm{in})}_{20, 22}\label{bin} \\
M_z=-M^{(\mathrm{in})}_{22, 22}+M^{(\mathrm{in})}_{22, 2-2} & \qquad & M_2=M^{(\mathrm{in})}_{22, 22}+M^{(\mathrm{in})}_{22, 2-2}. \nonumber
\en
These combinations are expressed in terms of some scalar functions $m_0$,
$m_1$, $m_2$, $m_2^{\prime}$, $m_3$ and $m_4$ (cf \cite{Bel65}) in the
following way:
\numparts
\bn
M_u & = & m_0-m_1\bt\cos{\gm_u}-m_2\bt^2\cos{2\gm_u }\qquad \mathrm{for}\; u=x,y,z \label{brot}\\
M_0 & = & m_0 +m_2^{\prime}\bt^2+ m_3\bt^3\cos{3\gm}+(m_1+m_3\bt^2)\bt\cos{\gm} \nonumber \\
&&+(m_2+m_2^{\prime})\bt^2\cos{2\gm}+m_4\bt^4\cos^2{2\gm} \label{b0}\\
M_1 & = & -(m_1+m_3\bt^2)\bt\sin{\gm}+(m_2+m_2^{\prime})\bt^2\sin{2\gm} \nonumber \\
&&-m_4\bt^4\sin{2\gm}\cos{2\gm} \label{b1} \\
M_2 & = & m_0+m_2^{\prime}\bt^2+m_3\bt^3\cos{3\gm}-(m_1+m_3\bt^2)\bt\cos{\gm} \nonumber \\
&&-(m_2+m_2^{\prime})\bt^2\cos{2\gm}+m_4\bt^4\sin^2{2\gm} \label{b2} 
\en
\endnumparts
where the scalar functions appearing in \eref{brot}, \eref{b0}, \eref{b1}
and \eref{b2} are related to the original scalar functions $m^{(2\lb )}_n$
in \eref{bi2decom} by
\numparts
\bn
m_0 &=& \frac{1}{\sqrt{5}}\left(m_0^{(0)}-\frac{1}{\sqrt{5}}\left(m_2^{(4)}\bt^2+m_3^{(4)}\bt^3\cos{3\gm}+m_4^{(4)}\right)\right) \label{m0} \\
m_1 &=& -\sqrt{\frac{2}{7}}\left(m_1^{(2)}+\sqrt{\frac{2}{7}}\left(m_3^{(4)}\bt^2+2m_4^{(4)}\bt^3\cos{3\gm}\right)\right) \label{m1} \\
m_2 &=&  -\sqrt{\frac{2}{7}}\left(m_2^{(2)}+\sqrt{\frac{2}{7}}\left(m_2^{(4)} -m_4^{(4)}\bt^2\right)\right) \label{m2} \\
m_2^{\prime} &=& \frac{1}{2}m_2^{(4)}  \qquad m_3 = \frac{1}{2}m_3^{(4)} \qquad m_4 = m_4^{(4)}. \label{m4}
\en
\endnumparts
Notice that because of the explicit dependence on $\bt$ and $\gm$ given in
\eref{b0} to \eref{b2} the six combination of the intrinsic components of
$M$ defined in \eref{eq:mxm1} can be related to each other at some specific
values of $\bt$ and $\gm$.  For instance, at $\bt =0$ one has
$M_x=M_y=M_z=M_0=M_2$, $M_1=0$, and at $\gm =0$ and arbitrary $\bt$ it is
$M_x=M_y$, $M_z=M_2$ and $M_1=0$.

\section{The Bohr collective Hamiltonian}\label{collham}

The present-day notion of the Bohr Hamiltonian is not very precise.  It
encompasses a large class of Hamiltonians of which the original Bohr
Hamiltonian
\cite{Boh52} is only a very special case.  Here, the general Bohr Hamiltonian
means a generic second order differential Hermitian operator in the Hilbert space of
functions of quadrupole coordinates $\al_{2\mu}$, $\mu = -2,\ \dots ,\ 2$. 
Hamiltonians of a similar type but in the spaces of other collective
coordinates, such as, for instance, the octupole deformations \cite{Roh82}
or the pairing variables \cite{Bes70}, can be called  Bohr Hamiltonians
as well, but they are not considered in this review.

\subsection{General form of the Hamiltonian}\label{bohrham}

The collective model arose from a classical description of  nuclear
 collective phenomena (cf \cite{Flu41,Boh52}).  This is why a classical
Hamiltonian is often a starting point for its formulation.  Now we follow
this approach and construct the most general collective Hamiltonian 
using the  quadrupole coordinates and making some natural
assumptions. We assume the classical Hamiltonian to be:
\begin{itemize}
\item[(i)] a real function of coordinates $\al_{2\mu}$ and velocities $\dot{\al}_{2\mu}$
\item[(ii)] invariant under orthogonal transformations of the coordinate system (the O(3) scalar)
\item[(iii)] an isotropic field (does not contain material tensors)
\item[(iv)]  a positive-definite quadratic form in the real and imaginary part of velocities $\dot{a}_{2\mu}$ and $\dot{b}_{2\mu}$.
\end{itemize}
A general form of such a Hamiltonian reads
\bn\label{clham}
 H_{\mathrm{cl}}(\al_2,\dot{\al}_2)&=&
\frac{1}{2}\sum_{\mu ,\nu}\dot{\al}_{2\mu}B_{2\phantom{\mu},2\phantom{\nu}}^{\phantom{2}\mu\phantom{,2}\nu}(\al_2)\dot{\al}_{2\nu}
+V_{\mathrm{cl}}(\al_2)  \nonumber \\
&=&\frac{1}{2}\left(\dot{\al}_2\cdot B_{2,2}(\al_2 )\cdot\dot{\al}_2\right)+V_{\mathrm{cl}}(\al_2).
\en
The potential $V_{\mathrm{cl}}(\al_2)$ is a real and isotropic function of
scalars $\sigma_0$ and $\chi_0$.  The inertial or mass bitensor $B_{2,2}$ is an
isotropic symmetric field of the coordinates $\al_{2\mu}$.  The bitensor matrix
is such that the kinetic energy is positive.

The quantum Hamiltonian corresponding to the classical one of 
\eref{clham} is obtained by the Podolsky-Pauli prescription \cite{Pod28,Pau33}:
\bn\label{qham}
\fl \hat{H}_{\mathrm{coll}}(\al_2,\pi_2)
=\frac{1}{2\sqrt{G(\al_2)}}\sum_{\mu ,\nu}\hat{\pi}_2^{\phantom{2}\mu}\sqrt{G(\al_2)}G_{2\mu ,2\nu}(\al_2)\hat{\pi}_2^{\phantom{2}\nu}
+V_{\mathrm{coll}}(\al_2)
\en
where $G=|\det (B_{2,2})|$ and $G_{2,2}$ is the inverse matrix of the inertial matrix $B_{2,2}$:
\beq
\sum_{\nu}B_{2\phantom{\mu},2\phantom{\nu}}^{\phantom{2}\mu\phantom{,2}\nu}G_{2\nu,2\eta}=\delta^{\mu}_{\eta}.
\eeq
The Hamiltonian $\hat{H}_{\mathrm{coll}}$  \eref{qham} is
Hermitian with the weight $\sqrt{G}$.
Hofmann \cite{Hof72} proved that the differential part of the Hamiltonian
\eref{qham} is unique provided it has the properties which
will be specified below for the quantum Hamiltonian
$\hat{H}_{\mathrm{quant}}$, and fulfills the correspondence principle to the
classical kinetic energy.  However, the quantum kinetic energy is given 
only up to an additive scalar function (cf also \cite{Kap97}).  This means that
the quantum potential $V_{\mathrm{coll}}$ in \eref{qham} is, in general, not
equal to the classical potential $V_{\mathrm{cl}}$ of \eref{clham} and can
contain quantum corrections.  Usually, it is tacitly assumed that the
potentials before and after the quantization are related by:
$V_{\mathrm{coll}}(\al_2) =V_{\mathrm{cl}}(\al_2)+ \mathrm{const}$, but this does not
have any reasonable justification.  The inertial bitensor $B_{2,2}$ and its
inverse
 $G_{2,2}$  are examples of symmetric matrices $M_{2\mu ,2\nu}$ 
\eref{bi2decom}, whose properties have been already discussed in section
\ref{funcoor}.  Thus, they are defined by specifying the six scalar
functions.

There are two ways to express $\hat{H}_{\mathrm{coll}}$ in the intrinsic
coordinates.  One way is to convert the derivatives $\partial
/\partial\al_{2\mu}$ into derivatives with respect to the intrinsic
coordinates using \eref{inmom}, \eref{hcinmom} (and, if necessary, also
\eref{axyz}, \eref{derbg} \eref{axyzbg}).  The other way is to express the
classical Hamiltonian $H_{\mathrm{cl}}$ \eref{clham} in terms of the
intrinsic variables transforming the velocities using 
  \eref{vel} and \eref{dotbg}.  One then has
\beq\label{inclham}
H_{\mathrm{cl}}=H_{\mathrm{cl,rot}}+H_{\mathrm{cl,vib}}
\eeq
where
\bn\label{inclhrv}
\fl H_{\mathrm{cl,rot}}=\frac{1}{2}\left(I_x(\bt ,\gm )\omega_x^2+I_y(\bt ,\gm )\omega_y^2+I_z(\bt ,\gm )\omega_z^2\right) \\
\fl H_{\mathrm{cl,vib}}=\frac{1}{2}\left(B_0(a_0,a_2)\dot{a}_0^2 +2B_1(a_0,a_2)\dot{a}_0\dot{a}_2+B_2(a_0,a_2)\dot{a}_2^2\right)+
V_{\mathrm{cl}}(\sigma_0,\chi_0)
\nonumber \\
\fl =\frac{1}{2}\left(B_{\bt\bt}(\bt ,\gm )\dot{\bt}^2 +2B_{\bt\gm}(\bt ,\gm )\dot{\bt}\bt\dot{\gm}+B_{\gm\gm}(\bt ,\gm )\bt^2\dot{\gm}^2\right) 
+V_{\mathrm{cl}}(\bt ,\zeta ). \label{eq:hclvib}
\en
The vibrational inertial functions $B_{\bt\bt}$, $B_{\bt\gm}$, $B_{\gm\gm}$ and $B_0$, $B_1$, $B_2$  
are interrelated  in the following way:
\numparts
\bn
\fl B_{\bt\bt}(\bt ,\gm ) =  B_0(\bt ,\gm )\cos^2{\gm}+2B_1(\bt ,\gm )\sin{\gm}\cos{\gm}+B_2(\bt ,\gm )\sin^2{\gm} \label{bbb} \\
\fl B_{\bt\gm}(\bt ,\gm )  =  (B_2(\bt ,\gm )-B_0(\bt ,\gm ))\sin{\gm}\cos{\gm}+B_1(\bt ,\gm )(\cos^2{\gm}-\sin^2{\gm}) \label{bbg}\\
\fl B_{\gm\gm}(\bt ,\gm )  =  B_0(\bt ,\gm )\sin^2{\gm}-2B_1(\bt ,\gm )\sin{\gm}\cos{\gm}+B_2(\bt ,\gm )\cos^2{\gm} \label{bgg}
\en
\endnumparts
and the moments of inertia read:
\bn\label{moi}
\fl I_u(a_0,a_2) =4B_u(a_0,a_2)a_u^2=4B_u(\bt ,\gm )\bt^2\sin^2{\gm_u}\qquad \mathrm{for}\; u=x,y,z
\en
The functions $B_0$, $B_1$, $B_2$ and $B_x$, $B_y$, $B_z$ are
the corresponding combinations of the intrinsic components of inertial 
bitensor $B_{2,2}$ (cf \eref{bin}).
Expressed in this way, the classical Hamiltonian
\eref{inclham} can be quantized with the Podolsky-Pauli prescription (cf
\cite{Kum67,Eis87}).

Both ways lead to the collective Hamiltonian expressed in terms of the intrinsic
coordinates which consists of two parts:
\beq\label{inqham}
\hat{H}_{\mathrm{coll}}=\hat{H}_{\mathrm{rot}}+\hat{H}_{\mathrm{vib}}
\eeq  
The rotational Hamiltonian $\hat{H}_{\mathrm{rot}}$  is given by
\beq\label{hrot}
\hat{H}_{\mathrm{rot}}=\frac{1}{2}\sum_{u=x,y,z}\frac{\hat{L}^2_u(\varphi ,\vartheta ,\psi )}{I_u(\bt ,\gm )}
\eeq
and the vibrational Hamiltonian $\hat{H}_{\mathrm{vib}}$ has the form
\bn
\fl \hat{H}_{\mathrm{vib}} = -\frac{\hbar^2}{2\sqrt{G}}\frac{1}{X}\left(\frac{\partial}{\partial a_0}X\sqrt{G}
\frac{B_2}{G_{\mathrm{vib}}}\frac{\partial}{\partial a_0}+
\frac{\partial}{\partial a_2}X\sqrt{G}
\frac{B_0}{G_{\mathrm{vib}}}\frac{\partial}{\partial a_2}\right. \nonumber \\
-\left.\frac{\partial}{\partial a_0}X\sqrt{G}
\frac{B_1}{G_{\mathrm{vib}}}\frac{\partial}{\partial a_2}
-\frac{\partial}{\partial a_2}X\sqrt{G}
\frac{B_1}{G_{\mathrm{vib}}}\frac{\partial}{\partial a_0}\right) 
+V_{\mathrm{coll}}(\sigma_0,\chi_0 )\label{hvib}\\
\fl  = -\frac{\hbar^2}{2\sqrt{G}}\left(\frac{1}{\bt^4}\frac{\partial}{\partial\bt}\bt^4\sqrt{G}
\frac{B_{\gm\gm}}{G_{\mathrm{vib}}}\frac{\partial}{\partial\bt}+
\frac{1}{\bt^2\sin{3\gm}}\frac{\partial}{\partial\gm}\sin{3\gm}\sqrt{G}
\frac{B_{\bt\bt}}{G_{\mathrm{vib}}}\frac{\partial}{\partial\gm}\right. \nonumber \\
\fl -\left.\frac{1}{\bt^4}\frac{\partial}{\partial\bt}\bt^3\sqrt{G}
\frac{B_{\bt\gm}}{G_{\mathrm{vib}}}\frac{\partial}{\partial\gm}
-\frac{1}{\bt\sin{3\gm}}\frac{\partial}{\partial\gm}\sin{3\gm}\sqrt{G}
\frac{B_{\bt\gm}}{G_{\mathrm{vib}}}\frac{\partial}{\partial\bt}\right) 
+V_{\mathrm{coll}}(\bt ,\zeta ) \label{hvibbg}
\en
with $G_{\mathrm{vib}}=B_0B_2-B_1^2=B_{\bt\bt}B_{\gm\gm}-B_{\bt\gm}^2$, $G=B_xB_yB_zG_{\mathrm{vib}}$ and $X$ given by \eref{jacob2}.

If the correspondence principle is not assumed i.e. if the considered 
nuclear collective system
does not have a classical counterpart, one can construct 
a bit more general quantum Bohr Hamiltonian $\hat{H}_{\mathrm{quant}}$ 
possessing the following properties:
\begin{itemize}
\item[(i)] $\hat{H}_{\mathrm{quant}}$ is a second order differential operator in coordinates $\al_{2\mu}$ possessing the finite lowest eigenvalue
\item[(ii)] $\hat{H}_{\mathrm{quant}}$  is a real operator (invariant under the time
reversal) i.e.  $\hat{H}_{\mathrm{quant}}=\hat{H}_{\mathrm{quant}}^{\ast}$
\item[(iii)] $\hat{H}_{\mathrm{quant}}$ is a scalar operator with respect to the
rotation group O(3)
\item[(iv)] $\hat{H}_{\mathrm{quant}}$ is Hermitian with a weight $W(\sigma_0,\chi_0)\geq 0$.
\end{itemize}
Under above assumptions $\hat{H}_{\mathrm{quant}}$ can always be
presented in the form\footnote{When assumption (ii) is given up (reality of
the Hamiltonian is not demanded) one more term can be added on the
right-hand side of \eref{pqham}, namely $(1/W)(\hat{\pi}_2\cdot WT_2(\al_2))
+(T_2(\al_2)\cdot\hat{\pi}_2)$, where $T_2$ is an arbitrary quadrupole field
(cf \cite{Une76}).}:
\beq\label{pqham}
\hat{H}_{\mathrm{quant}}= -\frac{1}{2W}\sum_{\mu ,\nu}\hat{\pi}_2^{\phantom{2}\mu}WA_{2\mu ,2}^{\phantom{2\mu ,2}\nu}(\al_2)\hat{\pi}_{2\nu} 
+V_{\mathrm{quant}}(\sigma_0,\chi_0)
\eeq
where $A_{2\mu ,2\nu}$ is a symmetric positive-definite bitensor matrix
which does not need to be related to $W$.  Converted into the intrinsic
coordinates with the help of \eref{inmom} and \eref{hcinmom} and
\eref{axyz}, \eref{derbg} \eref{axyzbg}, the Hamiltonian $\hat{H}_{\mathrm{quant}}$ 
takes again the
form \eref{inqham}  with the rotational part \eref{hrot}. 
The moments of inertia are expressed by the intrinsic components $A_x$,
$A_y$ and $A_z$ of $A_{2,2}$ (see \eref{bin}) as follows:
\bn\label{moia}
\fl I_u(a_0,a_2) =\frac{4a_u^2}{A_u(a_0,a_2)}=\frac{4}{A_u(\bt ,\gm )}\bt^2\sin^2{\gm_u}\qquad \mathrm{for}\; u=x,y,z.
\en
The weight $W$ and the remaining intrinsic components of $A_{2,2}$, namely
$A_0$, $A_1$ and $A_2$ enter the vibrational Hamiltonian in the following
way:
 \bn\label{hviba}
\fl \hat{H}_{\mathrm{vib}} = -\frac{\hbar^2}{2WX}\left(\frac{\partial}{\partial a_0}XW
A_{0}\frac{\partial}{\partial a_0}+
\frac{\partial}{\partial a_2}XW
A_{2}\frac{\partial}{\partial a_2}\right. \nonumber \\
+\left.\frac{\partial}{\partial a_0}XW
A_{1}\frac{\partial}{\partial a_2}
+\frac{\partial}{\partial a_2}XW
A_{1}\frac{\partial}{\partial a_0}\right) \nonumber \\
+V_{\mathrm{quant}}(\sigma_0,\chi_0) \nonumber \\
\fl \phantom{\hat{H}_{\mathrm{vib}}} = -\frac{\hbar^2}{2W}\left(\frac{1}{\bt^4}\frac{\partial}{\partial\bt}\bt^4W
A_{\bt\bt}\frac{\partial}{\partial\bt}+
\frac{1}{\bt^2\sin{3\gm}}\frac{\partial}{\partial\gm}\sin{3\gm}W
A_{\gm\gm}\frac{\partial}{\partial\gm}\right. \nonumber \\
+\left.\frac{1}{\bt^4}\frac{\partial}{\partial\bt}\bt^3W
A_{\bt\gm}\frac{\partial}{\partial\gm}
+\frac{1}{\bt\sin{3\gm}}\frac{\partial}{\partial\gm}\sin{3\gm}W
A_{\bt\gm}\frac{\partial}{\partial\bt}\right) \nonumber \\
+V_{\mathrm{quant}}(\bt ,\zeta )
\en
where the relations between $A_{\bt\bt}$, $A_{\bt\gm}$, $A_{\gm\gm}$  and
$A_0$, $A_1$, $A_2$ are identical with those between the vibrational inertial
functions given in \eref{bbb} to \eref{bgg}.

The Hamiltonian \eref{qham} obtained from its classical counterpart
\eref{clham} by the Podolsky-Pauli quantization procedure is a special
version of \eref{pqham} in which $W=\sqrt{G}$ and $A_{2,2}=G_{2,2}$.  In
this case $W$ and $A_{2,2}$ are related to each other and both come from the
classical inertial bitensor $B_{2,2}$.  Below, to the end of \sref{collham},
the Hamiltonians $\hat{H}_{\mathrm{coll}}$ and $\hat{H}_{\mathrm{quant}}$
will not be distinguished from one another.  Both of them will be denoted
simply by $\hat{H}$ and their potential parts by $V$.

\subsection{Collective multipole operators}\label{colmom}

Nuclear collective states are most often investigated experimentally by means
of  gamma spectroscopy.  The collective model should therefore make it
possible to
calculate measured quantities such as the electromagnetic transition
probabilities as well as the spectroscopic and intrinsic moments.  To this
end the electromagnetic multipole operators in the collective space should
be constructed.  It is assumed that the electric multipole operators are
isotropic tensor fields dependent only on the collective coordinates
$\al_{2\mu}$.  Since the fields depending on $\al_2$ all have the positive
parity, only the electric multipole operators with even multipolarities can
be constructed.  The electric multipole operator with an even multipolarity
$\lb$ is proportional to the `dimensionless' $\lb$-pole moment of the charge
distribution, $Q^{\mathrm{(charge)}}_{\lb}$, and can be written in the
following form\footnote{The electric multipole operators are not the
differential operators.  They are denoted with the hat just to keep the
notation for all electromagnetic operators uniform.}:
\numparts
\bn
\hat{M}(\mathrm{E}0)&=&\frac{1}{\sqrt{4\pi}}ZeR_0^{2}q_{0}^{(0)}(\sigma_0,\chi_0) \label{E0} \\
\fl \nonumber \mbox{for $\lb=0$ and}\\
\hat{M}(\mathrm{E}\lb ;\mu )&=&\sqrt{\frac{2\lb +1}{16\pi}}ZeR_0^{\lb}Q^{\mathrm{(charge)}}_{\lb\mu}(\al_2) \nonumber \\
&=&\sqrt{\frac{2\lb +1}{16\pi}}ZeR_0^{\lb}\sum_{k=1}^{\lb /2+1}q^{(\lb )} _k(\sigma_0,\chi_0 )
\tau_{k\lb\mu}(\al_2,\sigma_2) \label{Elambda}
\en
\endnumparts
for $\lb =2,\  4,\; \dots$. In these formulae the functions $q^{(\lb )}_k$
are scalar functions and $R_0=r_0A^{1/3}$.  The fundamental tensors
$\tau_{k\lb}$ appearing in \eref{Elambda} are given by \eref{ftev}.  The
intrinsic components of the moments of charge distribution are given by
\beq
Q_{\lb\mu}^{\mathrm{(charge)(in)}}(\bt ,\gm )=\sum_{k=1}^{\lb /2+1}q^{(\lb )} _k(\bt ,\zeta  )
\tau^{\mathrm{(in)}}_{k\lb\mu}(\bt ,\gm ). \label{Qin}
\eeq
According to \eref{alphas}--\eref{sigmas} the non-zero intrinsic components of the quadrupole and hexadecapole moments read:
\numparts
\bn
Q_{20}^{\mathrm{(charge)(in)}}(\bt ,\gm ) &=& q^{(2)}_1\bt\cos{\gm} +q^{(2)}_2\bt^2\cos{2\gm} \label{Q20in} \\
Q_{22}^{\mathrm{(charge)(in)}}(\bt ,\gm ) &=&\frac{1}{\sqrt{2}}( q^{(2)}_1\bt\sin{\gm} -q^{(2)}_2\bt^2\sin{2\gm}). \label{Q22in} 
\en
\endnumparts
\numparts
\bn
\fl Q_{40}^{\mathrm{(charge)(in)}}(\bt ,\gm ) = \frac{1}{\sqrt{70}}(q^{(4)}_1\bt^2(5\cos^2{\gm}+1)+2q^{(4)}_2\bt^3\cos{\gm}(7\cos^2{\gm}-4) \nonumber \\
+q^{(4)}_3\bt^4(5\cos^2{2\gm}+1)) \label{Q40in} \\
\fl Q_{42}^{\mathrm{(charge)(in)}}(\bt ,\gm ) = \frac{1}{2}\sqrt{\frac{3}{7}}(2q^{(4)}_1\bt^2\sin{\gm}\cos{\gm}-q^{(4)}_2\bt^3\sin{\gm}
-2q^{(4)}_3\bt^4\sin{2\gm}\cos{2\gm}) \label{Q42in} \\
\fl Q_{44}^{\mathrm{(charge)(in)}}(\bt ,\gm ) = \frac{1}{2}(q^{(4)}_1\bt^2\sin^2{\gm}-2q^{(4)}_2\bt^3\sin^2{\gm}\cos{\gm}
+q^{(4)}_3\bt^4\sin^2{2\gm}). \label{Q44in} 
\en
\endnumparts
For the E2 moment it is often assumed that, in approximation, 
$q^{(2)}_1= \mathrm{const}$ and $q^{(2)}_2=0$.

Similarly, only the magnetic multipole operators with  the odd
multipolarities can be constructed.  It is assumed that they all are the
odd-rank tensor fields dependent linearly on the angular momentum operator
\eref{angmom2}.  They can be written in the form
\beq\label{Mlambda}
\hat{M}(\mathrm{M}\lb ;\mu )=\sqrt{\frac{2\lb+1}{4\pi}}\frac{Z}{A}\mu_{\mathrm{N}}R_0^{\lb -1}\sum_{\kappa =\lb-1}^{\lb +1}[\Gamma_{\kappa}(\al_2)
\times\frac{1}{\hbar}\hat{L}_1]_{\lb\mu}
\eeq
for $\lb =1,\  3,\;\dots$, where, according to \eref{field2}, the gyromagnetic tensor 
fields $\Gamma_{\kappa}$ for $\kappa \geq 2$ are given by
\beq\label{Gamma}
\Gamma_{\kappa\mu}(\al_2)=\sum_{k=1}^{k_ {\kappa}}g^{(\kappa )}_k(\sigma_0 ,\chi_0 )\tau_{k\kappa\mu}(\al_2,\sigma_2,\chi_3)
\eeq
and $\Gamma_0=g^{(0)}_0(\sigma_0,\chi_0)$.  The functions $g^{(\kappa)}_k$'s 
are called the scalar gyromagnetic functions.  It
follows from \eref{Mlambda} that the intrinsic Cartesian components of the
M1 operator read
\numparts
\bn
 \hat{M}_u(\mathrm{M}1)&=&\sqrt{\frac{3}{4\pi}}\frac{Z}{A}\frac{\mu_{\mathrm{N}}}{\hbar}g_u(\bt ,\gm )\hat{L}_u \label{M1} \\
g_u(\bt ,\gm )&=& g_0^{(0)}-\sqrt{\frac{2}{5}}g_1^{(2)}\bt\cos{\gm_u}
-\sqrt{\frac{2}{5}}g_2^{(2)}\bt^2\cos{2\gm_u} \label{M1g}
\en
\endnumparts
for $u=x,\ y,\ z$ (cf \cite{Kum67} for the $\bt$- and $\gm$-dependence of
$g_u$).  It is often assumed that $g_0^{(0)}=1$ and $g_1^{(2)}=g_2^{(2)}=0$
(cf e.g.  equation (1.37) in \cite{RiSch80}).  The coefficients in front of
the right-hand sides of \eref{E0}, \eref{Elambda} and \eref{Mlambda} are
chosen to ensure the proper physical dimensions but their specific form
is, to a large extent, a matter of convention.

\subsection{Collective wave functions}\label{colwf}

In the previous sections the kinematics and dynamics of the collective model
have been formulated.  The model is determined up to a number of scalar
functions defining the collective kinetic energy as well as the potential
and the electromagnetic multipole operators.  The question arises whether
the collective wave functions can also be
expressed through a number of scalar functions.  The answer is `yes'. 
However, the search for such scalar functions can turn out to be impractical.

The collective wave functions are common eigenfunctions of the three
operators, $\hat{H}$, $\hat{L}^2=(\hat{L}_1\cdot\hat{L}_1)$ and
$\hat{L}_{10}$, namely
\numparts
\bn
\hat{H}\Psi_{iIM}(\al_2) &=&E_{iI}\Psi_{iIM}(\al_2) \label{evH} \\
\hat{L}^2\Psi_{iIM}(\al_2) &=&\hbar^2I(I+1)\Psi_{iIM}(\al_2) \label{evL2} \\
\hat{L}_{10}\Psi_{iIM}(\al_2) &=&\hbar M\Psi_{iIM}(\al_2) \label{evL0} 
\en
\endnumparts
where $i$ simply numbers (numerical) solutions or stands for a
set of three additional quantum numbers (in the case of analytical solutions).
 Since  equations \eref{evL2} and \eref{evL0} are automatically
fulfilled by the wave functions of the form
\beq\label{labwf}
\Psi_{iIM}(\al_2)=\sum_k\upsilon_{kiI}(\sigma_0,\chi_0)\tau_{kIM}(\al_2)
\eeq
where $\tau_{kI}$ are fundamental tensors \eref{field2} and $\upsilon_{kiI}$
are arbitrary {\em scalar} functions, it only remains to determine
the functions $\upsilon_{kiI}$ using \eref{evH}.  In the intrinsic coordinates the wave
function
\eref{labwf} has the form
\bn\label{intrwf}
\fl \Psi_{iIM}(\bt ,\gm ,\varphi ,\vartheta ,\psi )= \sum_{K=-I}^I\Psi^{\mathrm{(in)}}_{iIK}(\bt ,\gm )D^I_{MK}(\varphi ,\vartheta ,\psi ) \nonumber \\
=\sum_K\sum_k\upsilon_{kiI}(\bt ,\zeta )\tau_{kIK}^{(\mathrm{in})}(\bt ,\gm )D^I_{MK}(\varphi ,\vartheta ,\psi ) \nonumber \\
=\sum_K\sum_k\phi_{kiI}(\bt ,\zeta )t_{kIK}(\gm )D^I_{MK}(\varphi ,\vartheta ,\psi ) .
\en

The exact analytical solutions of \eref{evH} are known for several special
forms of the Hamiltonian \eref{inqham}.  In all of these cases the
full inertial bitensor is replaced by only one constant mass parameter which
(in view of \eref{brot} to \eref{b2}) leads to
$B_x=B_y=B_z=B_{\bt\bt}=B_{\gm\gm}=B= \mathrm{const}$, $B_{\bt\gm}=0$
(kinetic energy used originally by Bohr \cite{Boh52}).  A comprehensive
review of the exact and approximate solutions for different potentials is
given in \cite{For05}.  Here we discuss only a few cases with analytical
solutions.  

We start with an obvious remark that if the potential has the
form of $V(\bt,\zeta )=V_{\bt}(\bt )+V_{\gm}(\zeta )/\bt^2$, the variable
$\bt$ can be separated from the remaining angular coordinates in the
eigenvalue equations \eref{evH}, \eref{evL2}, \eref{evL0}.  In consequence,
the scalar factors in the wave function \eref{intrwf}  factorize as
follows
\beq\label{fact}
\phi_{kn\Lambda\nu I}(\bt ,\zeta )= u^{\Lambda}_n(\bt)w_{kI}^{\Lambda\nu}(\zeta )
\ .
\eeq
Here the index $i$ is replaced with the set of quantum numbers $(n\Lambda\nu
)$ with $n$ labelling the energy levels $E_{n\Lambda}$, $\Lambda$ being here
the separation constant, and $\nu$ --- an additional quantum number
introduced by Arima (cf \cite{Kish71}) to distinguish different solutions
with a given $\Lambda$.  The variable $\bt$ plays the role of the radial
coordinate in the five-dimensional space and $u^{\Lambda}_n(\bt )$ stands
for the corresponding radial wave function.

The angular part of the
wave function \eref{intrwf}
\beq\label{so5y}
Y^{\Lambda\nu}_{IM}(\gm ,\varphi ,\vartheta ,\psi )=\sum_K\sum_kw^{\Lambda\nu}_{kI}(\zeta )
t_{kIK}(\gm )D^I_{MK}(\varphi ,\vartheta ,\psi ) 
\eeq
fulfills the equation
\bn\label{angulareq}
\fl \left(-\frac{1}{\sin{3\gm}}\frac{\partial}{\partial\gm}\sin{3\gm}\frac{\partial}{\partial\gm}
+\frac{1}{\hbar^2}\sum_{u=x,y,z}\frac{\hat{L}^2_u(\varphi ,\vartheta ,\psi )}{4\sin^2{\gm_u}}
+\frac{2B}{\hbar^2}V_{\gm}(\cos{3\gm})\right)Y^{\Lambda\nu}_{IM}(\gm ,\varphi ,\vartheta ,\psi ) \nonumber \\
= \Lambda Y^{\Lambda\nu}_{IM}(\gm ,\varphi ,\vartheta ,\psi ).
\en
The most important and most extensively studied is the case 
with $V_{\gm}(\zeta )=0$, in which the separation constant $\Lambda$ is
equal to $\lb (\lb +3)$,
where  $\lb$, called the seniority \cite{Rak57}, is a
natural number.  In this case a set of $k_I$ ordinary second order
differential equations for functions $w_{kI}^{\lb\nu}$ ($k=1,\
\dots ,k_I$) with a given $\lb$ and $I=2L+3s$ ($s=0,\ 1$ for even, odd $I$,
respectively) was derived in  \cite{Bes59,Chac77,Eis87}:
\bn
\fl 9\frac{\rmd}{\rmd\zeta}(1-\zeta^2)\frac{\rmd w_{k\, 2L+3s}^{\lb\nu}(\zeta )}{\rmd\zeta}-6(L+k+3s)\zeta\frac{\rmd w_{k\, 2L+3s}^{\lb\nu}(\zeta )}{\rmd\zeta}
\nonumber \\
\fl +(\lb-3s-L-k)(\lb+3s +L+k+3)w^{\lb\nu}_{k\, 2L+3s}(\zeta ) +4(k+1)(k+2)w^{\lb\nu}_{k+2\, 2L+3s}(\zeta ) \nonumber \\
\fl +6(L-k+1)\frac{\rmd w_{k-1\, 2L+3s}^{\lb\nu}(\zeta )}{\rmd\zeta}+12(k+1)\frac{\rmd w_{k+1\, 2L+3s}^{\lb\nu}(\zeta )}{\rmd\zeta}=0 \label{polw}
\en
($\nu$ labels the possible different solutions).  The system of equations
\eref{polw} can be solved by means of the method of polynomials.  B\`es
solved this system for $I\leq 6$, $\lb\leq 9$ in \cite{Bes59}.  Polynomials
$w_{kI}^{\lb\nu}$ for all values of the quantum numbers were found in
\cite{Chac77,Ghe78} using the theory of harmonic homogeneous polynomials. 
The functions $Y^{\lb\nu}_{IM}(\gm ,\varphi ,\vartheta ,\psi )$ form a set
of spherical harmonics in the five-dimensional coordinate space (cf
\cite{El05}).  Another set of the five-dimensional spherical harmonics as
functions of the biharmonic coordinates in the laboratory frame was
constructed in \cite{Roh80}.  Moreover, note that the operator in the l.h.s. 
of \eref{angulareq} is proportional to the Casimir operator of the SO(5)
group and algebraic methods can be applied to calculate the SO(5)
Clebsch-Gordan coefficients and reduced matrix elements of tensor operators,
see \cite{Row04,RoTu05,Row05,Baer07}.  Let us add that the solutions of
\eref{angulareq} for some simple nonzero $V_{\gm}(\zeta )$ potentials are
discussed e.g.  in \cite{For06,Bon07,Row09}.

The radial equation reads
\beq\label{radeq}
\left(-\frac{1}{\bt^4}\frac{\partial}{\partial\bt}\bt^4\frac{\partial}{\partial\bt}+\frac{\Lambda}{\bt^2}
+\frac{2B}{\hbar^2}(V_{\bt}(\bt )-E_{n\Lambda})\right)u^{\Lambda}_n(\bt )=0.
\eeq
The elementary quantum mechanics (cf e.g.  \cite{Lan65}) provides us with a
few relevant solvable potentials $V_{\bt}(\bt )$ and we discuss some of them
below.

The Davidson modified oscillator potential
$V_{\bt}(\bt )=\frac{1}{2}C(\bt^2+\bt^4_{\mathrm{eq}}/\bt^2)$ has been
applied  for
the first time in \cite{Roh74} to solve analytically the Wilets-Jean model \cite{Wil56}.  
Later on the Davidson potential has been used by several authors
 \cite{El86a,El86b,Row98}. The solution of the radial equation for this
potential then reads:
\numparts
\bn
\fl u^{\lb}_n(\bt ) 
 =\sqrt{\frac{2(n!)}{\Gamma (n+\kappa +\frac{5}{2})}}
\left(\sqrt{\frac{BC}{\hbar^2}}\right)^{\kappa +\frac{5}{2}}
\bt^{\kappa}\rme^{-\frac{1}{2}\sqrt{\frac{BC}{\hbar^2}}\bt^2}
L^{(\kappa +\frac{3}{2})}_n\left(\sqrt{\frac{BC}{\hbar^2}}\bt^2\right) \label{oscrad}\\
E_{n\lb}=\sqrt{\frac{C}{B}}\hbar (2n+\kappa +\frac{5}{2}) \label{osclev}
\en
\endnumparts
where $L^{(\al)}_n(x)$ are the Laguerre polynomials \cite{Abr72} and  
\beq
\kappa (\kappa +3)=\frac{BC}{\hbar^2}\bt_{\mathrm{eq}}^4+\lb (\lb +3).\label{sen}
\eeq
The modified oscillator potential was also treated 
using algebraic methods in \cite{Row09}.  

The Kratzer potential
$V_{\bt}(\bt)=\frac{1}{2}C\bt_{\mathrm{eq}}^2((\bt_{\mathrm{eq}}/{\bt})^2-2\bt_{\mathrm{eq}}/{\bt})$
was introduced to the Bohr Hamiltonian in \cite{For04}.  The Hamiltonian
with this potential has a discrete spectrum only for negative
energies.  The radial functions and the energy levels of the bound states
read:
\numparts
\bn
u^{\lb}_n(\bt ) &=& N_{n\lb}\bt^{\kappa}\rme^{-k_{n\lb}\bt}L^{(2\kappa +3)}_n(\frac{\bt}{2k_{n\lb}}) \label{kratzrad} \\
E_{n\lb} &=& -\frac{1}{2}C\bt_{\mathrm{eq}}^2\left(\frac{BC}{\hbar^2}\bt_{\mathrm{eq}}^4\right)\frac{1}{(n+\kappa +2)^2} \label{kratzlev}
\en
\endnumparts
where $k_{n\lb} = \sqrt{2B|E_{n\lb}|/\hbar^2}$, $N_{n\lb}$ is a
normalization constant, and $\kappa$ is given formally by \eref{sen}.

The Bohr model with the potential $V_{\bt}(\bt )$ in the form of an infinite
square well has been discussed in \cite{Wil56,Ia00}.  The corresponding
radial wave functions and the energy levels are given by:
\numparts
\bn
u^{\lb}_n(\bt )&=&N_{n\lb}\frac{1}{\bt^{3/2}}J_{\lb +3/2}(x_{n,\lb}\bt/\bt_{\mathrm{well}}) \label{wellrad} \\
E_{n\lb} &=&\frac{1}{2B}\frac{\hbar^2x^2_{n,\lb}}{\bt^2_{\mathrm{well}}} \label{welllev}
\en
\endnumparts
where $N_{n\lb}$ is a normalization factor, $\bt_{\mathrm{well}}$ is the well
radius, $J_{\lb +3/2}(x) $ is the Bessel function \cite{Abr72} and 
$x_{n,\lb}$ is its $n$-th zero.

The modified oscillator potential for $\bt_{\mathrm{eq}}{=}0$ turns into the
five-dimensional harmonic oscillator (the case of original Bohr Hamiltonian
\cite{Boh52}), whose properties were studied most extensively.  In this case
the radial wave functions and the energy levels are still given by
\eref{oscrad} and \eref{osclev}, respectively, with $\kappa =\lb$ by virtue
of \eref{sen}. For the angular part of the solutions one can take the functions 
$Y^{\Lambda\nu}_{IM}(\gm ,\varphi ,\vartheta ,\psi )$ discussed in the
paragraph following \eref{polw}.
 The wave functions of the five-dimensional harmonic
oscillator, but in a form which does not make their tensor structure
explicit, were found also in \cite{Cor76} and \cite{Goz80}.  
Obviously, the eigenvalue problem for the harmonic oscillator Hamiltonian  can be solved immediately
in the laboratory coordinates.  The most convenient way to present the
solution is to introduce the notion of the quadrupole phonon.  The phonon
creation and annihilation operators, $\hat{\bt}^{\dag}_{2\mu}$ and
$\hat{\bt}_2^{\phantom{2}\mu}$, respectively, are defined as follows:
\numparts
\bn
 \hat{\bt}^{\dag}_{2\mu}&=&\frac{1}{\sqrt{2}}\left(\left(\frac{BC}{\hbar^2}\right)^{1/4}\al_{2\mu}-\rmi\left(\frac{BC}{\hbar^2}\right)^{-1/4}\hat{\pi}_{2\mu}\right)
\label{phcr}\\ 
\hat{\bt}^{\phantom{2}\mu}_{2}&=&\frac{1}{\sqrt{2}}\left(\left(\frac{BC}{\hbar^2}\right)^{1/4}\al_{2}^{\phantom{2}\mu}
-\rmi\left(\frac{BC}{\hbar^2}\right)^{-1/4}\hat{\pi}_{2}^{\phantom{2}\mu}\right). \label{phan}
\en
\endnumparts 
The Hamiltonian expressed in terms of the phonon operators has the well known form:
\beq\label{pham}
\hat{H}=\hbar\sqrt{\frac{C}{B}}\left((\hat{\bt}^{\dag}_{2}\cdot\hat{\bt}_{2})+\frac{5}{2}\right)
\eeq
and its eigenstates have the structure
\bn
 |N\mathrm{[c]}IM\rangle =\left[\underbrace{\hat{\bt}^{\dag}_2\times \dots \times\hat{\bt}^{\dag}_2}_{N}\right]_{IM}^{\mathrm{[c]}}|0\rangle 
\label{phst} 
\en
where the symbol [c] stands for a given coupling scheme and $|0\rangle$ is
the ground state annihilated by $\hat{\bt}_2^{\phantom{2}\mu}$:
$\hat{\bt}_2^{\phantom{2}\mu}|0\rangle =0$.  In section \ref{funcoor} it was
shown how to establish independent coupling schemes.  However, to have the
states with a definite seniority $\lb$ one should introduce the so called
traceless creation operators \cite{Chac76,Eis87}.  The methods of group
theory have been applied to calculate matrix elements of the multipole
moment operators between the basis states of the group chain
U(5)$\supset$SO(5)$\supset$SO(3) \cite {He65}.  Besides the exact solutions
of the harmonic oscillator presented above, the approximate harmonic
solutions of the Bohr Hamiltonian with the collective potentials possessing
well pronounced minimum for $\bt =\bt_{\mathrm{eq}}\neq 0$ are known from
the very beginning of the Bohr collective model (see e.g.  \cite{Eis87} for
a survey of them).

Special cases of the Bohr Hamiltonian for which equation \eref{evH}
possesses exact solutions are important for various reasons, but it is clear
that in order to solve \eref{evH} for Hamiltonians outside this class one needs to
use numerical methods.  \Eref{evH} can be solved by transformation into a
matrix eigenvalue problem using  an appropriate truncated basis in the
collective Hilbert space or by direct numerical methods suitable for the
second order partial differential equations.  In both approaches one must
take into account the specific properties of the functions belonging to the
domain of the Bohr Hamiltonian.  First of all, they must be invariant
against the octahedral group transformations.  Moreover, they must behave
appropriately (it will be explained below) at the boundaries (including
infinity) of the six sectors of the $(\bt,\gm)$ plane discussed in
\sref{intrcoor}.

The basis used to calculate the matrix of the Bohr Hamiltonian can be chosen as
the set of eigenfunctions of one of the exactly solvable cases, e.g. of the
harmonic oscillator which was employed by Dussel and B\`es
\cite{Dus70}, Gneuss and Greiner \cite{Gne71}, and Hess {\it et al}
\cite{He80,He81}.  Obviously, such bases automatically fulfill conditions
mentioned in the previous paragraph.  However, this  does not need to be true
for  `artificial' bases (used e.g. in \cite{Kum74,Lib82,Del89,Pro99,1999LI38})
which can be constructed from an arbitrary complete set of functions which are square integrable with
respect to the measure \eref{vol2}. 
We briefly discuss two examples of such a construction. The first one was
proposed by  Kumar \cite{Kum74} who, 
in order to ensure that the solutions have
the correct tensor structure 
and satisfy the mentioned conditions, 
expands  the scalar functions
$\upsilon_{kiI}(\bt ,\zeta )$ of \eref{intrwf} with given $k$ and $I$ in the
following basis functions:
\bn
\fl \upsilon^{(nm)}_{kI}(\bt ,\zeta ) 
=\rme^{-\sigma_{\bt} (\bt^2 -\bt^2_0)^l/2}\rme^{-\sigma_{\zeta}(\bt^3\zeta-\bt^3_0\zeta_0)/2}
(\bt^2-\bt^2_0)^n(\bt^3\zeta-\bt^3_0\zeta_0)^m  \label{kumb}
\en
which depend on the four variational parameters: $\bt_0$, $\zeta_0$,
$\sigma_{\bt}$ and $\sigma_{\zeta}$.  These parameters obey the following
auxiliary conditions: $l=1$ and $\sigma_{\zeta}=0$ for $\bt_0=0$, and $l=2$
for $\bt_0\neq 0$.
In the second example another complete set of functions is  adopted, this time in the space of functions of 
all five quadrupole variables $\bt ,\gm ,\varphi ,\vartheta ,\psi$ 

\bn \label{polb}
\Psi_{KIM}^{(nm)}(\bt ,\gm ,\varphi ,\vartheta ,\psi )=\rme^{-\sigma\bt^2/2}\bt^n
\left\{\ba{c}\cos{m\gm}\\ \sin{m\gm}\ea\right\}D^I_{MK}(\varphi ,\vartheta ,\psi)
\en
with $n=0,\ 1,\dots$, $m=n,\ n-2,\dots ,0$ or 1 and $-I\leq K\leq I$.
The modified, still non-orthogonal, basis after the appropriate symmetrization 
with respect to the octahedral group O of the transformations of the intrinsic frame 
(cf \eref{k1}, \eref{k2} and \eref{k3}) has the following form
\bn\label{modb}
\fl\tilde{\Psi}_{KIM}^{(nm)}(\bt ,\gm ,\varphi ,\vartheta ,\psi ) \nonumber \\
\fl =\rme^{-\sigma\bt^2/2}\bt^n\sum_{l=0}^{l^I_{\mathrm{max}}(m)}
\tilde{f}_{KI\, 2l}^{(m)}(\gm )\left(D^{I}_{M\, 2l}(\varphi ,\vartheta ,\psi )+(-1)^ID^{I}_{M\, -2l}(\varphi ,\vartheta ,\psi )\right)
\en
where functions  $\tilde{f}^{(m)}_{KI\, 2l}(\gm )$  are some specific linear
combinations of the trigonometric functions $\cos{p\gm}$ or $\sin{p\gm}$
with the possible values of $p= m,\ m-2,\ m-4,\ m-6$ and
$l^I_{\mathrm{max}}(m)$ is the maximal value of $l$ for given $I$ and $m$
(see \cite{Pro99} and \cite{1999LI38} for details). The functions  \eref{modb} are then
numerically orthonormalized and the parameter $\sigma$ is chosen so as to make
the basis optimal.

Before we write down a system of partial differential equations for
\eref{intrwf} derived from \eref{evH} and which can be solved by direct
numerical methods \cite{Kum67,Roh77,Tro91,Tro92}, we discuss the consequences of
the invariance of $\Psi_{iIM}$ with respect to the
octahedral group transformations of the intrinsic system.
This invariance leads to
\beq\label{ohinv}
\Psi_{iIM}(\bt ,\gm ,\varphi ,\vartheta ,\psi )= \sum_{K=-I}^I\Psi^{\mathrm{(in)}}_{iIK}(\bt_k ,\gm_k )D^I_{MK}(\varphi_k ,\vartheta_k ,\psi_k ) 
\eeq
for $k=0,\ 1,\ 2,\ 3$, where the intrinsic coordinates with the index $k$ are
given in \tref{tabintr}.  It follows from \eref{superpos} that
\beq\label{tranfi}
\Psi^{\mathrm{(in)}}_{iIK}(\bt ,\gm )= \sum_{K'=-I}^I\Psi^{\mathrm{(in)}}_{iIK'}(\bt_k ,\gm_k )D^I_{KK'}(\varpi_k ,\varrho_k ,\varsigma_k )
\eeq 
For $k=0$, this is, of course, the identity. 
Formula \eref{tranfi}  implies for $k=1,2,3$ the following
 properties of $\Psi^{\mathrm{(in)}}_{iIK}$ 
\numparts
\bn
\Psi^{\mathrm{(in)}}_{iIK}(\bt ,\gm )&=&(-1)^I\Psi^{\mathrm{(in)}}_{iI-K}(\bt ,\gm ) \label{k1}\\
\Psi^{\mathrm{(in)}}_{iIK}(\bt ,\gm )&=&(-1)^{K/2}\Psi^{\mathrm{(in)}}_{iIK}(\bt ,-\gm ) \label{k2} \\
\Psi^{\mathrm{(in)}}_{iIK}(\bt ,\gm )&=& \sum_{K'=-I}^I\Psi^{\mathrm{(in)}}_{iIK'}(\bt ,\gm -2\pi /3 )D^I_{KK'}(0 ,\pi /2 ,\pi /2 ) \label{k3}
\ .
\en
\endnumparts
From \eref{k1}, \eref{k2} it follows that
$\Psi^{\mathrm{(in)}}_{iI0}=0$ for odd $I$'s and
that  the components with odd $K$ all vanish.  Moreover, it is seen from \eref{k1}, \eref{k2}
and \eref{k3} that the value of the function $\Psi^{\mathrm{(in)}}_{iI\pm K}(\bt ,\gm)$ for an
arbitrary $\gm\ (-\pi\leq\gm\leq\pi)$ can be determined by the values of
$\Psi^{\mathrm{(in)}}_{iIK'}(\bt ,\gm )$ with different even, nonnegative
values of $K'$ and the corresponding $\gm$
from the sector $0\leq\gm\leq\pi /3$ shown in \fref{figsec}.  Thus, it is
sufficient to find $\Psi^{\mathrm{(in)}}_{iIK}(\bt ,\gm )$ with even $K\geq 0$ in
the sector
$0\leq\gm\leq\pi /3$ to determine the wave function for all values of
$\gm$.  Finally, the wave
function  \eref{intrwf} can be presented in the form
\bn\label{intrwf0}
\fl \Psi_{iIM}(\bt ,\gm ,\varphi ,\vartheta ,\psi )= \sum_{k=0}^{[I/2]}\frac{\Psi^{\mathrm{(in)}}_{iI2k}(\bt ,\gm )}{1+\delta_{k0}}\left(D^I_{M2k}(\varphi ,\vartheta ,\psi )
+(-1)^I D^I_{M-2k}(\varphi ,\vartheta ,\psi )\right)
\en
where $[I/2]$ stands for the integer part of $I/2$. 
If the wave functions $\Psi_{iIm}(\al_2)$ are normalized to unity
\beq\label{normpsi}
\langle i'I'M'|iIM\rangle =\int\Psi^{\ast}_{i'I'M'}(\al_2)\Psi_{iIM}(\al_2)W\rmd\Omega = \delta_{i'i}\delta_{I'I}\delta_{M'M}
\eeq
the normalization of the functions $\Psi^{\mathrm{(in)}}_{iI2k}(\bt ,\gm )$ is
\bn\label{normin}
\fl 6\int_0^{\infty}\int_0^{\pi /3}\sum_{k=0}^{[I/2]}\frac{2}{1+\delta_{k0}}\Psi^{\mathrm{(in)}\ast}_{i'I2k}(\bt ,\gm )\Psi^{\mathrm{(in)}}_{iI2k}(\bt ,\gm )W(\bt^2,\bt^3\cos{3\gm} )
\bt^4\sin{3\gm}\rmd\bt\rmd\gm \nonumber \\
=\frac{2I+1}{8\pi^2}\delta_{i'i}.
\en
It is also convenient to introduce  the intrinsic wave functions
\beq\label{Fi}
\Phi_{iI2k}(\bt ,\gm )=\sqrt{\frac{16\pi^2}{(2I+1)(1+\delta_{k0})}}\Psi^{\mathrm{(in)}}_{iI2k}(\bt ,\gm )
\eeq
which can be referred to as the normalized intrinsic wave functions because
by virtue of \eref{normin} and \eref{Fi} they obey:
\bn\label{normfi}
 \sum_{k=0}^{[I/2]}(\Phi_{i'I2k}|\Phi_{iI2k})=\delta_{i'i}
\en
where the scalar product $(\Phi_{i'I2k}|\Phi_{iI2k})$ is meant as
\bn
\fl (\Phi_{i'I2k}|\Phi_{iI2k})=
6\int_0^{\infty}\int_0^{\pi /3}\Phi^{\ast}_{i'I2k}(\bt ,\gm )\Phi_{iI2k}(\bt ,\gm )W(\bt
,\cos{3\gm})\bt^4\sin{3\gm}\rmd\bt\rmd\gm \ .
\en
The  wave function \eref{intrwf0} is expressed in terms of the normalized intrinsic
wave functions as
\bn\label{intrwf1}
\fl \Psi_{iIM}(\bt ,\gm ,\varphi ,\vartheta ,\psi )= \nonumber \\
\fl \sqrt{\frac{2I+1}{16\pi^2}}\sum_{k=0}^{[I/2]}\Phi_{iI2k}(\bt ,\gm )\sqrt{1+\delta_{k0}}\left(D^I_{M2k}(\varphi ,\vartheta ,\psi )
+(-1)^I D^I_{M-2k}(\varphi ,\vartheta ,\psi )\right)
\en

A set of the differential equations for $\Phi_{iI2k}$ obtained from \eref{evH} is the following:
\bn\label{setfi}
\fl \left(\hat{H}_{\mathrm{vib}}(\bt ,\gm )-E_{iI}+\frac{1}{4}\left(I(I+1)-(2k)^2\right)\left(\frac{1}{I_x(\bt ,\gm )}+\frac{1}{I_y(\bt ,\gm )}\right)+
\frac{(2k)^2}{2I_z(\bt ,\gm )}\right)\Phi_{iI2k}(\bt ,\gm ) \nonumber \\
\fl +\frac{1}{8}\left(\frac{1}{I_x(\bt ,\gm )}-\frac{1}{I_y(\bt ,\gm )}\right)\left(C_{Ik+1}\Phi_{iI2k+2}(\bt ,\gm )+C_{Ik}\Phi_{iI2k-2}(\bt ,\gm )\right)=0
\en
where $k=0(1),\ \dots ,I/2(I/2-1/2)$ for even(odd) $I$ and 
\beq
C_{Ik}=\sqrt{(1+\delta_{k1})(I+2k)(I+2k-1)(I-2k+1)(I-2k+2)}.
\eeq
By virtue of the symmetry conditions \eref{k2} and \eref{k3} and of  equations
\eref{setfi} the normalized intrinsic wave functions $\Phi_{iI2k}(\bt ,\gm )$ fulfill
the following boundary conditions 
\begin{itemize}
\item[(i)] At $\gm =0$:
\beq\label{bc0}
\left.\frac{\partial\Phi_{iI0}(\bt ,\gm )}{\partial\gm}\right|_{\gm =0}=0 \qquad \Phi_{iI2k}(\bt ,0)=0 \quad \mathrm{for}\  k\neq 0.
\eeq
\item[(ii)] At $\gm =\pi /3$:
\numparts
\bn
 \Phi_{iI2k}(\bt ,\pi /3)&=&0\quad \mathrm{for\ odd}\ I \label{bc60odd} \\
 \left.\frac{\partial\Phi_{i00}(\bt ,\gm )}{\partial\gm}\right|_{\gm =\pi /3}&=&0 \label{bc600}
\en
\bn
\fl \left.\ba{l} \left(I(I+1)-(2k)^2\right)\Phi_{iI2k}(\bt ,\pi /3) \\
=\frac{1}{2}\left(C_{Ik+1}\Phi_{iI2k+2}(\bt ,\pi /3)+C_{Ik}\Phi_{iI2k-2}(\bt ,\pi /3)\right) \\
\left(I(I+1)-(2k)^2-8\right)\left.(\partial\Phi_{iI2k}(\bt ,\gm )/\partial\gm )\right|_{\gm =\pi /3} \\
=\frac{1}{2}\left(C_{Ik+1}\left.(\partial\Phi_{iI2k+2}(\bt ,\gm )/\partial\gm )\right|_{\gm =\pi /3}\right.\\
\left.+C_{Ik}\left.(\partial\Phi_{iI2k-2}(\bt ,\gm )/\partial\gm )\right|_{\gm =\pi /3}\right) \ea\right\}\quad \mathrm{for\ even}\ I .\nonumber \\
\label{bc60ev}
\en
\endnumparts
\end{itemize}
At $\bt =0$ the boundary conditions are
\bn\label{bcb0}
\fl \left.\frac{\partial\Phi_{iI2k}(\bt ,\gm )}{\partial\bt}\right|_{\bt =0}=0 \quad\mathrm{for}\ I=0 \qquad
\mbox{and} \qquad  \Phi_{iI2k}(\bt =0,\gm )=0 \quad\mathrm{for}\ I\neq 0.
\en
The bound-state wave function  $\Phi_{iI2k}(\bt ,\gm )$ should vanish at
infinity ($\bt\to\infty$)   faster than $1/\bt^2$
to ensure its norm  be finite. 
Equations \eref{setfi} to \eref{bcb0}
are the basic ones 
for direct numerical calculations of $\Phi_{iI2k}$. Note, however, that such
calculations are quite difficult due to the complexity of the system of
equations \eref{setfi} 
(in addition, the number of equations grows as $I/2$) and of the
boundary conditions. 

We end this subsection with the formulae for
the reduced matrix elements of the collective electromagnetic multipole
 operators.
From \eref{Elambda}, \eref{Qin} and \eref{intrwf1} the reduced matrix element of E$\lb$ is
\bn\label{rmeEL}
\fl \langle i'I'\|\hat{M}(\mathrm{E}\lb )\| iI\rangle  =\sqrt{\frac{(2\lb +1)(2I+1)}{16\pi}}\frac{ZeR_0^{\lb}}{2} \nonumber \\
\fl \times  \sum_{k,k',\kappa}\sqrt{(1+\delta_{k0})(1+\delta_{k'0})}(I2k\lb\kappa |I'2k')(\Phi_{i'I'2k'}|Q_{\lb\kappa}^{\mathrm{(charge)(in)}}|\Phi_{iI2k}).
\en
Hence, the E2 reduced matrix element reads
\bn\label{rmeE2}
\fl \langle i'I'\|\hat{M}(\mathrm{E}2)\| iI\rangle = \sqrt{\frac{5(2I+1)}{16\pi}}ZeR_0^{2} 
 \sum_{k\geq 0}\left\{(I2k20 |I'2k)(\Phi_{i'I'2k}|Q_{20}^{\mathrm{(charge)(in)}}|\Phi_{iI2k})\right. \nonumber \\
 +\left.\sqrt{1+\delta_{k0}}\left( (I2k22|I'2k+2)(\Phi_{i'I'2k+2}|Q_{22}^{\mathrm{(charge)(in)}}|\Phi_{iI2k}) \right.\right. \nonumber \\
 +\left.\left.(I2k+22-2|I'2k)(\Phi_{i'I'2k}|Q_{22}^{\mathrm{(charge)(in)}}|\Phi_{iI2k+2})\right)\right\}.
\en
The general formula for the M$\lb$ reduced matrix elements is more complicated.
We give only the formula for the M1 reduced matrix element:
\bn\label{rmeM1}
\fl \langle i'I'\|\hat{M}(\mathrm{M}1)\| iI\rangle =\sqrt{\frac{3(2I+1)}{4\pi}}\frac{Z}{A}\frac{\mu_{\mathrm{N}}}{\hbar} 
\sum_{k\geq 0}\left\{2k(I2k10|I'2k)(\Phi_{i'I'2k}|g_z|\Phi_{iI2k})\right. \nonumber \\
\fl +\left.\left(\sqrt{\frac{(I-2k)(I+2k+1)}{2}}(I2k+11-1|I'2k)\right.\right. \nonumber \\
\fl -\left.\left.\sqrt{\frac{(I+2k)(I-2k+1)}{2}}(I2k-111|I'2k)\right)(\Phi_{i'I'2k}|\frac{g_x+g_y}{2}|\Phi_{iI2k})\right.
\nonumber \\
\fl -\sqrt{1+\delta_{k0}}\left.\left(\sqrt{\frac{(I-2k)(I+2k+1)}{2}}(I2k+111|I'2k+2)(\Phi_{i'I'2k+2}|\frac{g_x-g_y}{2}|\Phi_{iI2k})\right.\right. \nonumber \\
\fl -\left.\left.\sqrt{\frac{(I+2k+2)(I-2k-1)}{2}}(I2k+11-1|I'2k)(\Phi_{i'I'2k}|\frac{g_x-g_y}{2}|\Phi_{iI2k+2})\right)\right\}.
\en

\subsection{Collective sum rules}\label{sumrul}
Quantities measurable experimentally are expressed, generally speaking, through matrix elements of the observables  
$\hat{T}_{\lb\mu}(\al_2,\hat{\pi}_2)$, namely through the quantities
\beq \label{me}
\langle i'I'M'|\hat{T}_{\lb\mu}|iIM\rangle = \int\left(\Psi_{i'I'M'}^{\ast}(\al_2)\hat{T}_{\lb\mu}(\al_2,\hat{\pi}_2)\Psi_{iIM}(\al_2)\right)W\rmd\Omega .
\eeq
By comparing matrix elements calculated in the frame of a theoretical
model with corresponding experimental values one can judge the quality of the model. 
Some restricted tests of the correctness of the theory are provided by the so
called sum rules.  The notion of sum rules is comprehensive and not too
rigorously defined.  The ones which we discuss below are called the
collective sum rules, because they are restricted to the space of collective
states (cf \cite{Sre06}). 
First we derive the so called non-energy weighted sum rules \cite{Dob87}. 
They were proposed by Kumar \cite{Kum72} to determine the intrinsic electric
quadrupole moments of the nuclear states from the experimental data.
Further on,  the energy weighted sum rules
\cite{Pom77} will be discussed as well.

From the closure equation in the collective space:
\beq\label{clo}
\sum_{i,I,M}|iIM\rangle\langle iIM|=1
\eeq
and the Wigner-Eckart theorem \cite{Var88} the mean value of  a tensor 
field $T_{\lb}(\al_2)$ squared can be written as a sum of products of the
reduced matrix elements of $T_{\lb}$
\beq\label{T2}
\langle iIM|(T_{\lb}\cdot T_{\lb})|iIM\rangle =\frac{1}{2I+1}\sum_{i'I'}|\langle iI\|T_{\lb}\| i'I'\rangle |^2.
\eeq
Similarly, for the expectation value of the scalar of the third order 
in $T_{\lb}$ we have
\bn \label{T3}
\fl \frac{(-1)^{\lb}}{\sqrt{2\lb +1}} \langle iIM|([T_{\lb}\times T_{\lb}]_{\lb}\cdot T_{\lb})|iIM\rangle = 
\langle iIM|[[T_{\lb}\times T_{\lb}]_{\lb}\times T_{\lb}]_0|iIM\rangle  \nonumber \\
\fl =\frac{1}{2I+1}\sum_{i_1,I_1,i_2,I_2}
\left\{\ba{ccc}\lb & \lb & \lb \\ I_1 & I_2 & I \ea\right\}
\langle iI\| T_{\lb}\| i_1I_1\rangle\langle i_1I_1\| T_{\lb}\|i_2I_2\rangle\langle i_2I_2\| T_{\lb}\| iI\rangle
\ .
\en
Similar formulae  for higher powers of $T_{\lb}$ also be derived. 
However, for the fourth and 
higher order scalars in $T_{\lb}$ it becomes essential that the field
$T_{\lb}$ should not depend on the momentum $\hat{\pi}_2$ and its components
should commute with each other (cf \cite{Dob87}).  
If it is possible to measure in experiment  sufficiently large number of
the reduced matrix elements of $T_{\lb}$, equations \eref{T2}, \eref{T3} make
it possible to determine the quantities on their left hand sides. Some of
these quantities are otherwise difficult to access experimentally.
When $T_2= \hat{M}(\mathrm{E}2)$  formula
\eref{T2} reads
\beq\label{E2}
\langle iIM|(\hat{M}(\mathrm{E}2)\cdot\hat{M}(\mathrm{E}2))| iIM\rangle =\sum_{i'I'}B(\mathrm{E}2;iI\to i'I')
\eeq
which means that the mean value in the state $|iIM\rangle$ of the intrinsic electric
quadrupole moment squared is equal to the sum of the reduced 
probabilities of the E2 transitions to all accessible states $|i'I'M'\rangle$.  The
sum rules for the E2 operator have been applied to the experimental
determination of the intrinsic quadrupole moments in \cite{Sre06}.

The so called energy weighted collective sum rules are derived from the following double commutation relation:
\beq\label{double}
\left[\al_{2\mu},\left[\hat{H},\al_{2\nu}\right]\right]=\hbar^2A_{2\mu ,2\nu}(\al_2)
\eeq
where $\hat{H}$ is the collective Hamiltonian \eref{pqham}. Taking the mean
values of both sides of \eref{double} in the state  $|iIM\rangle$ and making use of the
Wigner-Eckart theorem one obtains
\bn
\fl \frac{\hbar^2}{2}\frac{\langle iI\|A_{2\lb}(\al_2)\| iI\rangle }{\sqrt{4\lb+1}}=
\sum_{i',I'}\left(E_{i'I'}-E_{iI}\right)(-1)^{I-I'}\left\{\ba{ccc}2 & 2 & 2\lb \\ I & I & I' \ea\right\}\left|\langle iI\|\al_2\|i'I'\rangle\right|^2  \label{enwei}
\en
where
\beq\label{inten}
A_{LM}(\al_2)=\sum_{\mu ,\nu}(2\mu2\nu |LM)A_{2\mu ,2\nu}(\al_2).
\eeq
For $\lb =0$ the energy weighted sum rule \eref{enwei} takes
the form
\bn\label{intrace}
\fl \frac{\hbar^2}{2}\langle iIM|(A_x+A_y+A_z+A_0+A_2)|iIM\rangle =
\frac{1}{2I+1}\sum_{i',I'}(E_{i'I'}-E_{iI})|\langle iI\|\al_2\| i'I'\rangle|^2.
\en
The energy weighted sum rules can therefore be used to estimate the average
values of the inertial functions using the excitation energies and the
reduced matrix elements of $\al_2$ extracted from the experimental data
\cite{Pom77}.

\section{Microscopic theories of the Bohr Hamiltonian\label{sec:micro}}

The present section  focuses on the derivation of the  Bohr collective Hamiltonian
from the microscopic theory  based on an effective nucleon-nucleon
interaction.  There are two main methods which can be used to achieve this
aim.  The first one is the adiabatic approximation of the Time Dependent
Hartree-Fock-Bogolyubov theory (ATDHFB) and the second one is the Generator
Coordinate Method with the Gaussian Overlap Approximation (GCM+GOA).  
Before discussing the details concerning the application to the Bohr 
Hamiltonian case, we briefly recall
basic assumptions underlying these methods  and their general results.

In the following it is assumed that the nuclear many-body Hamiltonian is given
by 
\beq H_{\rm micr}=\sum_{\mu,\nu}K_{\mu\nu}d^+_{\mu}d_{\nu} +
\frac{1}{4}\sum_{\mu,\nu,\alpha,\beta}
{V}_{\mu\nu\alpha\beta}d^+_{\mu}d^+_{\nu}d_{\beta}d_{\alpha} 
\eeq 
with
the two-body interaction term $V$ that can  depend on the nuclear density,
while 
$d^+_\mu$ and $d_\nu$ are the nucleon creation and annihilation operators. 
The methods presented below can be directly applied to most of the effective
non-relativistic nucleon-nucleon interactions.  Some aspects of the Relativistic Mean
Field (RMF) theory need, however, some additional discussion. 

Within the mean field approach the ground state of a nucleus is described by
a generalized product state (in other words a BCS-type state, with the
Slater determinant as a special case).  To describe collective phenomena one
needs a family of such states which depend on properly chosen collective
variables such as quadrupole variables discussed in the previous
sections.  In the next step the collective Hamiltonian is constructed as an
operator in the Hilbert space of functions of the collective variables.

\subsection{The ATDHFB method\label{sec:atdhfb}}
The ATDHFB method is conveniently formulated by using the notion of a generalized
density matrix which 
for a given BCS-type state $|\Psi\rangle$ and a given set of creation and
annihilation operators $d^+_m,d_n$ is defined as follows
\beq
\cR=\left(\begin{array}{cc}
\cR^{11} & \cR^{12}\\
\cR^{21} & \cR^{22}
\end{array}\right)
\eeq
where
\numparts
\bn
\cR^{11}_{mn}=\langle\Psi| d^+_nd_m|\Psi\rangle=\rho_{mn}\\
\cR^{12}_{mn}=\langle\Psi| d_nd_m|\Psi\rangle=\kappa_{mn}\\
\cR^{21}_{mn}=\langle\Psi| d^+_nd^+_m|\Psi\rangle=-\kappa^*_{mn}\\
\cR^{22}_{mn}=\langle\Psi| d_nd^+_m\Psi|\rangle=I-\rho^*_{mn} \ .
\en
\endnumparts
The dimension of the $\cR$ matrix is twice the dimension of the one-particle
space.  A rigorous (but very abstract) definition of a space wherein the
density operator acts can be found e.g.  in \cite{xPR94}. In this section we
will call it the double space.  One should note that there exist some
slightly different sign conventions used to define $\cR$.  The $\cR$ matrix
can be expressed in terms of the coefficients of the Bogolyubov transformation
which connects the $d^+_m,d_n$ operators with the quasiparticle creation and
annihilation operators corresponding to the $|\Psi\rangle$ state, see
\cite{1981DO08,x68ba01}.

The Time Dependent HFB equation of motion reads
\beq\label{tdhfb}
\rmi\hb \dot{\cR}=[\cW(\cR),\cR]
\eeq
with $\cW$ being the self-consistent Hamiltonian (in the double space) 
induced by the density matrix $\cR$
\beq\label{eq:defwu}
\cW=\left(\begin{array}{cc}
K -\lambda I & 0\\
0 & -K^* +\lambda I
\end{array}\right)+
\left(\begin{array}{cc}
\varGamma  & \varDelta\\
-\varDelta^* & -\varGamma^*
\end{array}\right)
\eeq
in which $\lambda$ is the chemical potential and
\beq\label{eq:gamm}
\varGamma_{\mu\nu}=\sum_{\mu', \nu'}
{V}_{\mu\mu'\nu\nu'}\rho_{\nu'\mu'} + \mbox{rearrangement terms}
\eeq
\beq
\varDelta_{\mu\nu}=\frac{1}{2}\sum_{\mu', \nu'}
{V}_{\mu\nu\mu'\nu'}\kappa_{\mu'\nu'} \ .
\eeq
The rearrangement terms in \eref{eq:gamm} appear if  $H_{\rm micr}$ depends on
the density. Typically these terms are considered only for the density dependent
interaction in the particle-hole channel.

In the adiabatic approximation  the $\cR$ matrix is expressed as
$\cR=\exp(\rmi\chi(t))\cR_0(t)\exp(-\rmi\chi(t)) =\cR_0 +\cR_1 +\cR_2 +\dots$,
where $\chi$ is a `small' operator, cf \cite{x78ba01}.  The $\cW$ operator can then be written
as $\cW=\cW_0+\cW_1+\cW_2+ \ldots$, with $\cW_k$ induced by the $\cR_k$
terms.  Note that for $k>0$ only the second term in \eref{eq:defwu} is used. 
By inserting the expansions of $\cR$ and $\cW$ into \eref{tdhfb} one obtains a
sequence of equations from which only the first two are usually considered
\beq \label{eq:basec1} \rmi\hb \dot{\cR_0}=[\cW_0,\cR_1]+[\cW_1,\cR_0] \eeq
\beq \label{eq:basec2}
\rmi\hb\dot{\cR_1}=[\cW_0,\cR_2]+[\cW_1,\cR_1]+[\cW_2,\cR_0]+[\cW_0,\cR_0] \ . 
\eeq 
From the second equation one can infer that $[\cW_0,\cR_0]$ must be of
the second order in the smallness parameter. An important feature of the r.h.s
of the first equation (\ref{eq:basec1}) is its linear dependence on the $\cR_1$
matrix.

In the case when $\cR$ depends on time through several collective variables
${\balpha}=(\alpha_1,...,\alpha_n)$  equation (\ref{eq:basec1}) leads to
\beq\label{eq:maspa}
\rmi\hbar\dot{\alpha_k}\frac{\partial \cR_0}{\partial \alpha_k}=
[\cW_0,\cR^k_1] +
[\cW_1(\cR^k_1),
\cR_0], \ \ k=1,...,n
\eeq
where, since $\cR_1$ depends linearly on $\dot{\cR_0}$,
\beq
\cR_1=\sum_k \cR^k_1 \ .
\eeq
The collective classical Hamiltonian is defined as the expectation value of the $H_{\rm
micr}$ 
Hamiltonian in the $|\Psi(\wec{\alpha})\rangle$ state and can be written as a sum of
the kinetic and potential parts
\beq
H_{\rm cl}=T_{\rm cl}+V_{\rm cl}
\eeq
where
\beq\label{eq:epotcol}
V_{\rm cl}=\langle \Psi_0(\wec{\alpha})| H_{\rm
micr}|\Psi_0(\wec{\alpha})\rangle
\eeq
with $|\Psi_0\rangle$ being the BCS-type state corresponding to the $\cR_0$
matrix
and 
\beq
T_{\rm cl}=\frac{1}{2}\sum_{k,j}B_{kj}(\wec{\alpha})\dot{\alpha_k}\dot{\alpha_j}
\ .
\eeq
The inertial functions  (also called mass parameters) $B_{kj}$ read
\beq\label{eq:massp}
B_{kj}(\wec{\alpha})=\frac{\rmi\hb}{2\dot{\alpha_j}}\Tr_{2d}(\cR^j_1 \ [\frac{\partial \cR_0}{\partial
\alpha_k} ,\cR_0]) 
\eeq
where $\Tr_{2d}$ denotes the trace in the double space. It can be verified
that $B_{kj}=B_{jk}$.  The positive definiteness of the matrix $B$ is not
shown in the most general case, however it can be proved for several
important instances, e.g. for cranking inertial functions, see \cite{1981DO08}. 
The classical expression for $H_{\rm cl}$ needs to be quantized.  
Usually this is done by taking as the
quantum mechanical kinetic energy  the Laplace-Beltrami
operator (times $-\hb^2$) in the space of the  variables $\wec{\alpha}$ with
$B$ as the metric tensor  and as the quantum potential energy
 the operator of multiplication by the function $V_{\mathrm{cl}}$ (this is the so called Podolsky-Pauli method of
quantization, see also section~\ref{bohrham}).

In the following it is assumed that (\ref{eq:basec1}) yields the unique
solution for $\cR_1$ in terms of $\dot{\cR_0}$.  For a discussion of
the existence and uniqueness of such a solution see \cite{1981DO08}.  However,
one should remember that in practice the calculation of $\cR_1$ can be quite
a difficult task.  In the case of vibrational inertial functions only one
collective variable (for axial symmetry systems) was considered, see
\cite{1981DO08} and \cite{1980GI05} (for the case without pairing).  In
\cite{1981DO08}  $\cR_1$ was obtained by matrix inversion while the
method proposed in \cite{1980GI05} (see also \cite{1999YU09}) was based on
the fact that \eref{eq:basec1} can be transformed into (for simplicity we consider here 
 a single variable $\alpha$)
\beq\label{eq:que}
[\cW_0 +\cW_1 - 2\dot{\alpha}\cP, \cR_0+\cR_1]=0
\eeq
where
\beq
\cP=\frac{\rmi\hb}{2}[\partial_\alpha\cR_0,\cR_0] \ .
\eeq
\Eref{eq:que} can be solved by a constrained HFB calculation. In the case
of a rotation (and in the limit of vanishing  angular velocity) it leads to the
Thouless-Valatin expression for the moment of inertia \cite{x62th01}.  For
this reason,  the
term $[\cW_1,\cR_0]$ is sometimes called the Thouless-Valatin (TV) correction.

The trace in \eref{eq:massp} is most easily calculated 
in a quasiparticle basis (in the double space). It can be verified that both 
$\partial_k\cR_0:=\partial \cR_0/\partial
\alpha_k$ and $\cR_1^j$ have the antidiagonal form
\beq 
(\partial_k\cR_0)_{\rm qp}=\bpmax
0 & f_k\\
-f^*_k & 0
\epmax
\eeq

\beq 
(\cR^j_1)_{\rm qp}=\bpmax
0 & z_j\\
-z^*_j & 0
\epmax
\eeq
so that 
\beq
\Tr_{2d}(\cR^j_1 \ [\partial_k \cR_0
 ,\cR_0])
=\Tr
( z_j f^*_k -z^*_j f_k) \ .
\eeq

Neglecting the TV term  brings in a substantial simplification, because then
\beq
\rmi\hbar \dot{\alpha_k}f_{k,\mu\nu}=(E_\mu+ E_\nu)z_{k,\mu\nu}
\eeq
where $E_\mu$ are quasiparticle energies,  and to 
calculate the inertial functions  in this approximation one needs only 
$\partial_k\cR_0$ and $E_\mu$:
\beq\label{eq:bkj}
B_{kj}=
\frac{\hbar^2}{2}\sum_{\mu,\nu}\frac{f_{j,\mu\nu}f^*_{k,\mu\nu}
+f^*_{j,\mu\nu}f_{k,\mu\nu}}{(E_\mu+ E_\nu)} \ .
\eeq
Inertial functions calculated
without considering the Thouless-Valatin term are sometimes called the
cranking inertial functions.

Matrix elements $f_{k,\mu\nu}$ can be expressed in terms of
the derivatives of the density matrix $\rho$ and the pairing tensor $\kappa$ in the
canonical basis
\beq\label{eq:efkamunu}
f_{k,\mu\nu}= 
s_{\nu}(\partial_k\rho)_{\mu\bar{\nu}}(u_{\mu}v_{\nu}+ v_{\mu}u_{\nu})
+
(\partial_k\kappa)_{\mu\nu}(u_{\mu}u_{\nu}-v_{\mu}v_{\nu}) .
\eeq
In this formula $s_\nu$ is the phase factor with the properties $s_\mu s^*_\mu=1$, 
$s_\mu=-s_{\bar\mu}$, while 
$\mu$ and $\bar{\mu}$ denote canonically conjugated states.
Another, perhaps more familiar, expression with the explicit dependence on
the derivative of the BCS state is
\beq
f_{k,\mu\nu}=\langle {\Psi_0}| a_\nu a_\mu |\partial_k\Psi_0\rangle
\eeq
where $a_\mu$ are the quasiparticle annihilation operators.
If the condition $[\cW_0,\cR_0]\approx 0$ is used one  gets yet another,
sometimes quite useful, form
\beq\label{eq:rokahade}
\fl f_{k,\mu\nu}= 
-\frac{1}{E_{\mu}+E_{\nu}}[s_\nu(\partial_k
h_0)_{\mu\bar{\nu}}(u_{\mu}v_{\nu}+ v_{\mu}u_{\nu}) +
(\partial_k\varDelta)_{\mu\nu}(u_{\mu}u_{\nu}-v_{\mu}v_{\nu})]
\eeq 
where $h_0=K-\lambda I+\varGamma$, see (\ref{eq:defwu}).  The references
cited in \sref{sec:bhspec} contain many examples of application of 
formulae (\ref{eq:bkj}) to (\ref{eq:rokahade}).  

\subsection{The GCM+GOA method\label{subs:gcmgoa}}

This method is based on the variational principle.  Its simplest version
consists of taking as test functions the integrals
 $\int d\wec{\alpha} f(\wec\alpha)\Phi(\wec{\alpha})$   with the BCS-type
states $|\Phi(\wec{\alpha})\rangle$ and the weights $f(\wec{\alpha})$ which
are to be determined.  The assumption of the Gaussian overlaps (GOA)
\beq\label{eq:overlap}
\fl\langle\Phi(\wec{\alpha''})|\Phi(\wec{\alpha'})\rangle
=\exp(-\sum_{k,j}g_{kj}(\wec{\alpha})(\alpha''_k-\alpha'_k)(\alpha''_j-\alpha'_j)/2),
\ \ \ \wec{\alpha}=(\wec{\alpha''}+\wec{\alpha'})/2
\eeq
enables one to transform the Hill-Wheeler-Griffin equation into the
Schrodinger-like, second order
differential equation for the weight functions $f(\wec{\alpha})$
\beq
\hat{H}_{\rm GCM}f(\wec{\alpha})=Ef(\wec{\alpha}) \ .
\eeq
Below, only some of the  final results for the GCM collective Hamiltonian
$\hat{H}_{\rm
GCM}$ are presented. Details can be found in e.g. \cite{Oni75} (see
also \cite{RiSch80,Goz85}). A specific
 application to the Bohr Hamiltonian problem was studied in \cite{Une76}.
The GCM Hamiltonian reads 
\beq\label{eq:hgcm}
\hat{H}_{\rm GCM}=\hat{T}_{\rm GCM}+V_{\rm GCM}
\eeq
where
\beq\label{eq:hkingcm}
\hat{T}_{\rm GCM}=-\frac{\hb^2}{2}\frac{1}{\sqrt{\det g}}
\sum_{k,j}\frac{\partial}{\partial \alpha_k}
\sqrt{\det g}\,
(B^{-1}_{\rm GCM})^{kj}\frac{\partial}{\partial \alpha_j} \ .
\eeq
Note that in the case of the quadrupole collective variables 
the Hamiltonian \eref{eq:hgcm} has the same structure as the Hamiltonian \eref{pqham} discussed
earlier, with $\sqrt{\det g}=W$ and $B^{-1}_{\rm GCM}=A$.
The inverse of the inertial tensor is given by
\beq\label{eq:bgcm}
(B^{-1}_{\rm GCM})^{kj}=\frac{1}{2\hb^2}(\Rel\, h_{12} - \Rel\, h_{11})^{kj}
\eeq
where
\numparts
\begin{eqnarray}
&h_{11,mn}=D_{\alpha''_m}D_{\alpha''_n}h(\wec{\alpha''},\wec{\alpha'})\wara\\
&h_{12,mn}=D_{\alpha''_m}D_{\alpha'_n}h(\wec{\alpha''},\wec{\alpha'})\wara
\end{eqnarray}
\endnumparts
with
\beq
h(\wec{\alpha''},\wec{\alpha'})=\langle\Phi(\wec{\alpha''})|H_{\rm
micr}|\Phi(\wec{\alpha'})\rangle/\langle\Phi(\wec{\alpha''})|\Phi(\wec{\alpha'})\rangle
\ .
\eeq
$D$ denotes here a covariant derivative and the indices on the right hand side
of \eref{eq:bgcm} are raised using
the inverse of the metric tensor $g_{kj}$. The tensor $g_{kj}$ can be written as
\beq
g_{kj}(\wec{\alpha})=
\langle\partial_{\alpha_k}\Phi(\wec{\alpha})|\partial_{\alpha_j}\Phi(\wec{\alpha})\rangle
\ .
\eeq 

The potential energy can be transformed into the form
\beq\label{eq:gcmpot}
V_{\rm GCM}= V_{\rm cl}(\wec{\alpha})
+ V_{\rm ZPE}(\wec{\alpha})
\eeq
where
\bn
V_{\rm cl}(\wec{\alpha})=\langle\Phi(\wec{\alpha})|H_{\rm micr}|\Phi(\wec{\alpha})\rangle\\
V_{\rm ZPE}(\wec{\alpha})=-
\frac{\hb^2}{2}\sum_{k,j}g^{kj}(B^{-1}_{\rm GCM})_{kj} -
\frac{1}{8}\sum_{k,j}g^{kj}D_{\alpha_k} D_{\alpha_j} V_{\mathrm{cl}}(\wec{\alpha}) \ .
\en
The first term $V_{\rm cl}(\wec{\alpha})$ in (\ref{eq:gcmpot}) is
essentially the same as in the ATDHFB method.  The second one, $V_{\rm
ZPE}(\wec{\alpha})$,
specific for the GCM method, is called the zero point energy correction. 
One of the advantages of the GCM over the ATDHFB method is that it avoids 
ambiguities of the quantization procedure.

Let us now add two comments.  Firstly, the derivation of (\ref{eq:hgcm}) is
based on the assumption that the space with the metric tensor $g$ is
essentially flat, which means that one can introduce such variables that $g$
is constant.  In some references, e.g.  \cite{1987RE04}, it is assumed
from the very beginning that collective variables do already have such a
property.  Secondly, the overlaps \eref{eq:overlap} can in principle contain
imaginary terms linear in $\wec{\alpha''}-\wec{\alpha'}$, which would lead
to  extra first order derivative terms in \eref{eq:hgcm}, see
\cite{Oni75,1985GO15}.

It seems that the GCM formulae for mass parameters, which in the case of the
Bohr Hamiltonian were discussed for the first time in \cite{Une76}, have never been
used in practical calculations.  There are two possible reasons for this.
The  first one is their
technical complexity as compared to the cranking inertial functions of
the ATDHFB method.  The second one is the known deficiency of the simplest version
of the GCM, which can be easily seen in the case of the translational motion,
for which the GCM fails to give the correct mass.  This deficiency can be
amended by considering a richer set of test functions with the so called
conjugate parameters.  The even more sophisticated version of the GCM
presented in \cite{x80GO01,1987RE04} leads to the inertial functions which
are the same as those in ATDHFB and to the potential energy containing the ZPE
terms, see also \cite{x79gi01}.

\subsection{Application to the Bohr Hamiltonian case \label{sec:bhspec}}

Studies  on the microscopic foundation of the Bohr Hamiltonian have a long
history.  In fact they were an important stimulus for the development of the
ATDHFB method  (see \cite{Bel65} and \cite{x68ba01}, where the general
microscopic formulae for the inertial functions and potential energy
entering the Bohr Hamiltonian were obtained).  The first attempt to apply
these results to the real nuclei (in the W-Pt region) and to compare them
with experimental data was performed by Kumar and Baranger in
\cite{x68ku01}, see also \cite{Kum74}.  They used a spherical mean field
potential plus a separable quadrupole-quadrupole interaction and the
seniority (constant $G$) pairing.  In the following years their
quadrupole+pairing model has been applied to several regions of nuclei, 
to cite only the most recent papers \cite{2001KU18,2002GU18}.

The second attempt, based on a picture of nucleons moving in a deformed
single-particle Nilsson potential was presented in \cite{1976KA30,1977RO30}. 
The collective potential energy was calculated by means of 
macroscopic-microscopic method using the Strutinsky shell correction method. 
More recently this model has been extended by an approximate inclusion of
the effects of pairing vibrations \cite{Pro99,1999ZA10,Sre06}.

There are several technical difficulties, including a substantial
computational
effort, if one wishes to apply the theories presented in previous
subsections to the case of a nuclear mean field obtained self-consistently
from realistic effective interactions, e.g.  of the Gogny or Skyrme type. 
The first results with the Gogny force (GCM+GOA) were announced in
\cite{1978GI15}, but more extensive calculations were done only in the 90's
\cite{1994DE25,1999LI38}, see also more recent \cite{2006DE23,2009GIxx}. 
The Skyrme interaction was used in the framework of the ATDHFB in \cite{Del89},
and next, in a more extended way in \cite{2004PR01,2008PR05,2009PRxx}. Let us also
mention the papers \cite{2004FL04,2008KL04}, where the Skyrme force (with GCM+GOA) was
used with an interesting, although somewhat simplified treatment of the Bohr
Hamiltonian. In addition, recent years brought applications of the RMF to
describe quadrupole collective states \cite{2004PR03,2009NI04}.  The range
of nuclei studied in the cited papers is quite wide: from the krypton
isotopes to transactinides.

An important advantage of the microscopic approaches is a clear physical
interpretation of the quadrupole variables, which  in most cases  are simply
proportional to the components of the quadrupole tensor of the mass distribution,
that
is in the laboratory frame
\beq \alpha_{2\mu}\sim \langle \Phi |{Q}^{\rm (mass)}_{2\mu}|\Phi\rangle 
\eeq 
with
\beq 
{Q}^{\rm (mass)}_{2\mu}=\sum_{i=1}^A r_i^2 Y_{2\mu}(\theta_i,\phi_i) 
\eeq
where the sum runs over all nucleons and $A$ is the mass number.
A slightly
different  model is the one studied in \cite{1976KA30,1977RO30}, where the collective
variables describe an elliptic deformation of the single particle potential.
In the intrinsic frame,
instead of $Q^{\rm (mass)}_{2\mu}$,
one often uses the $Q_0,Q_2$ operators:
\numparts
\bn
Q_0=\sum^A_{i=1} (3z^2_i-r_i^2)=\sqrt{16\pi/5}\,{Q}^{\rm (mass)}_{20} \\
Q_2=\sum^A_{i=1} \sqrt{3}(x^2_i-y^2_i) \ .      
\en
\endnumparts
The intrinsic coordinates $(a_0,a_2)$ and $(\bt,\gm)$ (see \eref{a0},
\eref{a2}, \eref{bg}) are
given by the mean values of the $Q_0,Q_2$ operators
\numparts
\bn
a_0=\beta\cos\gamma=c q_0=c \langle\Phi| Q_0 |\Phi\rangle\\
a_2=\beta\sin\gamma=c q_2=c \langle\Phi| Q_2 |\Phi\rangle
\en
\endnumparts
with  the coefficient $c$ conventionally chosen as
\beq
c=\sqrt{\pi/5}/A\overline{r^2}
\eeq
to get some correspondence with phenomenological models (see e.g.
\cite{Eis87}). For $\overline{r^2}$  the liquid drop model estimate 
$\frac{3}{5} (r_0 A^{1/3})^2$ with $r_0=1.2\, {\rm fm}$,
is usually adopted.

All microscopic theories start with  BCS-type states constructed in the
intrinsic
 frame which are subsequently rotated to a desired orientation.  In
the self-consistent theory a set of states parametrized by the quadrupole
variables $q_0,q_2$ (or equivalently $\bt$, $\gm$) is obtained from the
HFB calculations with constraints on the mean values of $Q_0$ and $Q_2$
operators 
\bn 
\delta \langle \Phi|H_{\rm micr} - \lambda_0 Q_0 - \lambda_2
Q_2|\Phi\rangle=0 \label{eq:conshfb}\\
 \langle\Phi| Q_0|\Phi \rangle=q_0, 
\quad \langle\Phi| Q_2|\Phi\rangle =q_2 \ .
\en
For the fixed values of $q_0$,
$q_2$
 the states $|\Phi\rangle$ in the intrinsic frame  should be invariant against the discrete symmetry
group ${\rm D}_2 \subset \mbox{O}$,  which is generated by rotations by  $\pi$  around three
orthogonal axes.  This property is obvious if one recalls that such rotations do
not change the values of $q_0$ and $q_2$.  An extensive analysis of the
consequences of this and other symmetries for various quantities of the mean
field theory can be found in \cite{2000DO14,2000DO15}.  \Eref{eq:conshfb} is
solved numerically by expanding the BCS-type states in a properly
constructed basis.  The basic ingredient of such a construction are the
eigenstates of the three-dimensional harmonic oscillator in  Cartesian
\cite{1983GI06,2009NI04} or cylindrical coordinates \cite{1999SA19}.

Several properties of the kinetic part of the Bohr Hamiltonian in the intrinsic
frame,
e.g.  the block structure of the inertia tensor, can be inferred from the
general theory ---  see  formulae \eref{inclhrv}, \eref{eq:hclvib}.  
However, in view of the approximations  adopted, it is instructive
to prove these properties directly using the definitions of inertial functions
given in \sref{sec:atdhfb}.  We sketch the proof for the ATDHFB inertial functions in the
Appendix~B, for the GCM+GOA case cf \cite{Une76}.

In the intrinsic frame the rotational part of the collective space is
identical with the SO(3) group.  For the tangent space (in the differential
geometry sense) of the SO(3) group it is more convenient to use a basis
determined by the group structure itself, consisting of elements of the Lie
algebra (in other words the angular momentum operators), rather than the
usual coordinate basis (e.g.  derivatives with respect to the Euler angles). 
The rotational part of the Hamiltonian (see (3.4)) is then expressed in
terms of the collective angular momentum operators and the moments of
inertia.  The dependence of the BCS-type state
$|\Phi(\bt,\gm,\varphi,\vartheta,\psi)\rangle$ on the Euler angles
is explicitly known (see Appendix B),
and if  the TV term is neglected, one arrives at
the Inglis-Belyaev formula for the moment of inertia
\beq\label{eq:ingbel}
I_k=\hbar^2 \sum_{\mu,\nu}\frac{|\langle\nu|J_k|\bar{\mu}\rangle|^2(u_{\mu}v_{{\nu}}
-u_{\nu}v_{{\mu}})^2}{(E_\mu+ E_\nu)}
\eeq
where $J_k$ is the angular momentum operator in the Fock space. Note that
the sum in \eref{eq:ingbel} is over the whole one-particle space basis.

The vibrational parameters are a bit more complicated, even if the TV
correction term is neglected.  To find them one can use (\ref{eq:efkamunu}) but
the derivatives of $\rho$ and $\kappa$ have to be in this case calculated numerically, see
papers \cite{1999YU09,2004PR01}, where parameters obtained in this way were
called the $M^P$ masses.  More frequently, another approximation is 
adopted, which consists in using  (\ref{eq:rokahade}) but neglecting the
derivative of $\varDelta$.  The derivatives of $\cR_0$ with respect to the
Lagrange multipliers $\lambda_{i}$ can be then calculated as
\beq 
f_{\lambda_{i},\mu\nu}= \frac{s_\nu}{E_{\mu}+E_{\nu}}\langle \mu| Q_i|
\bar{\nu}\rangle(u_{\mu}v_{\nu}+ v_{\mu}u_{\nu}) 
\eeq 
and next the 
derivatives of $q_i$ with respect to $\lambda_j$
\beq\label{eq:difla} \frac{\partial q_i}{\partial
{\lambda_j}}=\sum_{\mu,\nu}\frac{\langle \mu |Q_i| \bar{\nu}\rangle
\langle \bar{\nu}|Q_{j}| \mu\rangle} {(E_\mu+ E_\nu)}(u_{\mu}v_{\nu}
+u_{\nu}v_{\mu})^2 \ .
\eeq 
As a result, one  gets the  widely used formula for the
vibrational inertial functions
\beq\label{eq:bqq}
B_{q_iq_j}=\hbar^{2}(S^{-1}_{(1)}S_{(3)}S^{-1}_{(1)})_{ij} \eeq where  $S_{(n)}$ 
denotes  the $2\times 2$ matrix 
\beq\label{eq:ssn}
(S_{(n)})_{ij}=\sum_{\mu,\nu}\frac{\langle \mu |Q_i| \bar{\nu}\rangle
\langle \bar{\nu}|Q_{j}| \mu\rangle} {(E_\mu+ E_\nu)^{n}}(u_{\mu}v_{\nu}
+u_{\nu}v_{\mu})^2 \ .
\eeq

The indices $\mu,\nu$ in \eref{eq:ingbel}, \eref{eq:difla}, \eref{eq:ssn} run over
all one-particle basis states, but in many cases, due to symmetry
properties, their range can be limited to  half of the basis, so one should
take care when comparing formulae in various papers.
The transition to the inertial functions for $\bt$ and $\gm$ variables, entering
e.g. \eref{hvibbg} is straightforward. Note also that the symmetry properties of 
these functions, which can be inferred from the 
discussion of the symmetric quadrupole bitensors in \sref{funcoor},
provide useful tests for correctness of
numerical calculations.

Let us make one remark about the two kinds of nucleons. As long as one does not consider the
proton-neutron pairing, the state $|\Phi\rangle$ (see \eref{eq:conshfb}) is
a product of the BCS-type states for protons and neutrons and consequently
the ATDHFB mass parameters are simply the sum of the corresponding
parameters for protons and
neutrons.

Once the mass parameters and the potential energy are determined, the
eigenvalue problem of the Bohr Hamiltonian is solved numerically yielding
the collective wave functions and
energies, which can be compared directly with experimental data.  The wave functions are then used to calculate matrix elements
of collective electromagnetic transition operators, primarily of the E2
operator, which in the microscopic approach is equal to 
\beq
q^{\rm (charge)}_{2\mu}=
\langle \Phi|e\textstyle \sum_{i=1}^{Z}r_i^2 Y_{2\mu}(\theta_i,\phi_i)|\Phi\rangle
\eeq
with the sum over protons only,  and
has the structure discussed in \sref{colmom}.  On the experimental side, modern
techniques  provide a large amount of data about the electromagnetic
transitions.  In several cases not only  transition probabilities are
measured but also relative signs of the matrix elements are determined,
see e.g.  \cite{Sre06}.  The correct interpretation of rich experimental
data is  challenging and provides a stringent test of theoretical models.

A very general conclusion stemming from numerous papers on microscopic
calculations of the Bohr Hamiltonian is that it leads to a good qualitative
agreement with experiment, correctly reproducing the main features of
collective spectra.  These results are remarkable, especially if one
remembers that they are obtained without any free parameters fitted to
collective quantities.  At the quantitative level one finds, however, several
substantial discrepancies between the theory and measurement.  For example,
quite often the theoretical spectrum is `stretched' compared to the
experimental one, although the two are similar in their general patterns.

There are several possible reasons for such discrepancies.  A natural
question is what is the optimal nucleon-nucleon effective interaction in
both particle-hole and particle-particle channels.  As it is easily seen in
the case of the Skyrme forces, various parametrizations can lead to rather
different potential energies.  On the other hand the inertial functions are
quite sensitive to the strength of the pairing interaction.  There are also
some open problems in the collective part of the theory.  Let us mention two
of them --- the question of the Thouless-Valatin corrections and of the extension
of the collective space by including variables describing pairing degrees of
freedom.  The TV correction in the case of the moments of inertia can be
estimated by performing cranked HFB calculations as described in
\cite{1999LI38}.  A study of some test cases led to the conclusion that on
average the TV term increases the moment of inertia by 30\%.  To account for
this effect the constant factor 1.3 was used in \cite{1999LI38} to scale the
moments obtained with the Inglis-Belyaev formula.  Much less is known about
the TV correction in the case of vibrational parameters.  There exist only
old papers discussing the axial case \cite{1980GI05,1981DO08}, where the
reported effects, depending on the nucleon-nucleon interaction, were of a
size comparable to those for moments of inertia.  In our opinion, in order
to respect the symmetry conditions, when simulating the TV corrections in the
simplified approach, all inertial functions, not only the moments of
inertia, should be rescaled by a constant multiplicative factor.  Such a
common factor, equal to 1.2, was used e.g.  in \cite{2004PR01}.  The
question of extending the collective space is discussed in section
\ref{subs:ext}.

\subsection{Example of microscopic calculations}

The theory presented in the previous sections is now applied for twelve
heavy even-even nuclei \nucs{}.  The considered chain of nuclei provides a
good example of a transition from  deformed axial (\nuc{178}{Hf}) to
almost spherical (\nuc{200}{Hg}) nuclei.  Theoretical results were obtained
using the SLy4 version of the Skyrme interaction \cite{1997CH49} and the
constant $G$ (seniority) pairing.  The inertial functions and the potential
energy entering the Bohr Hamiltonian were calculated using the ATDHFB theory
(the formulae
(\ref{eq:epotcol}), (\ref{eq:ingbel}), (\ref{eq:bqq})).  The general outline of
calculations is similar to that of \cite{2004PR01}, where one can also find
more about the basis we used and other numerical details. Here, however, we do
not use the scaling
factor discussed in the end of the previous subsection.
The pairing
strength was determined by comparing the `experimental' pairing gaps from the five
point formula \cite{1988MA04} for Hf and W isotopes with the minimal
quasiparticle energies calculated for the deformation corresponding to the minimum of
the potential.  To get the inertial functions and collective potential
energy, the HF+BCS calculations with linear constraints on $q_0$ and $q_2$
were performed for 155 points forming a regular grid in the sextant
$0<\bt\le 0.7$, $0 \le \gm \le 60^{\circ}$, with the distance between points
equal to $0.05$ and $6^\circ$ in the $\bt$ and $\gm$ direction, respectively.

\begin{figure}[htb]
\includegraphics{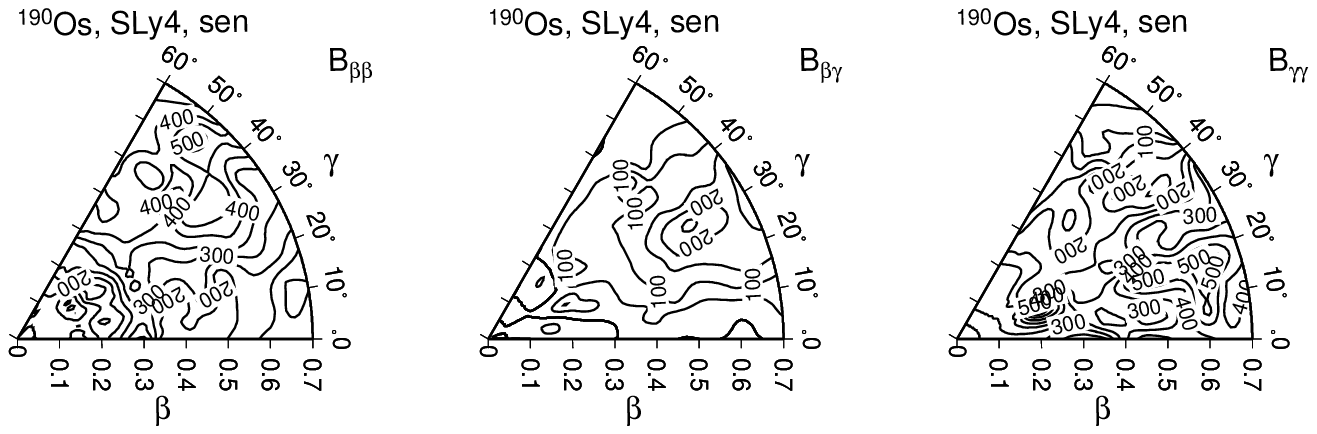}
\caption{Vibrational inertial functions $B_{\bt\bt}$, $B_{\bt\gm}$,
$B_{\gm\gm}$   (in
$\hbar^2$/MeV)
for the \nuc{190}{Os} nucleus.\label{fig:vibpar} }
\end{figure}

\begin{figure}[htb]
\includegraphics{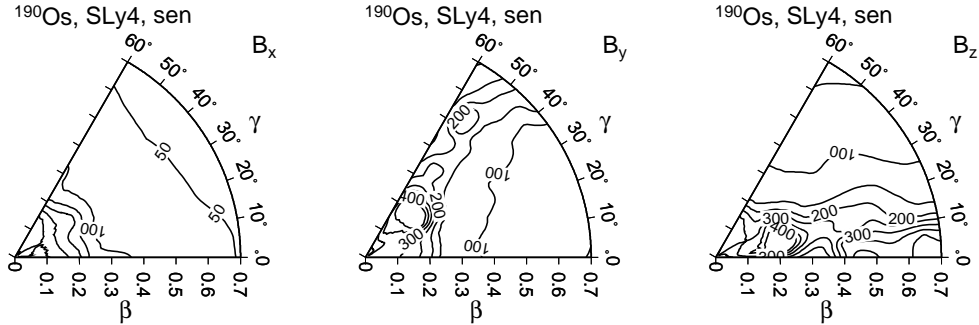}
\caption{Rotational  inertial functions $B_u$, $u=x,y,z$ (in
$\hbar^2$/MeV) for the \nuc{190}{Os} nucleus.
\label{fig:rotpar} }
\end{figure}

\begin{figure}
\includegraphics{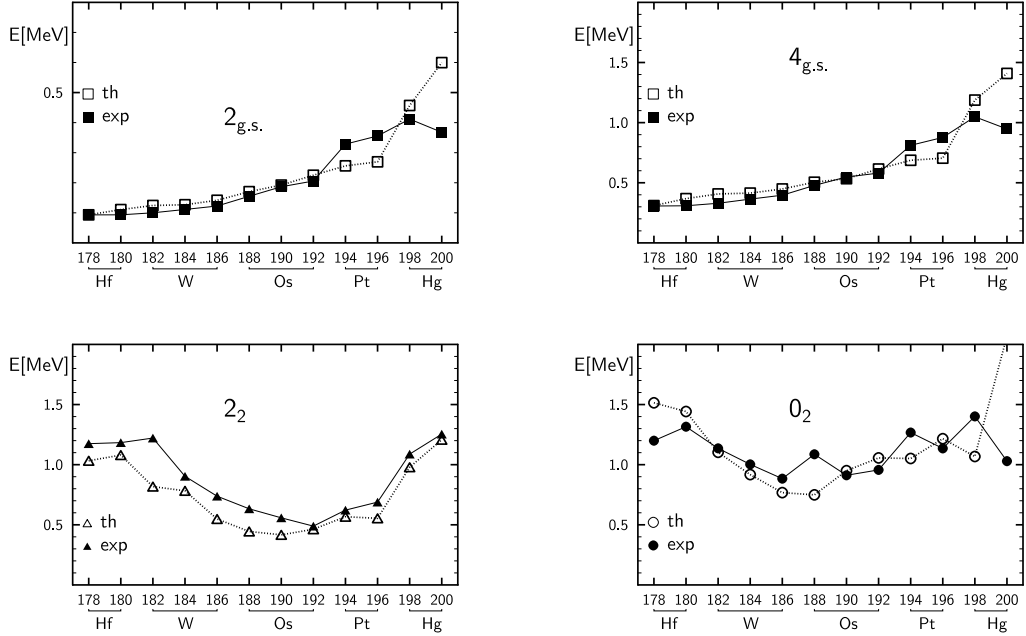}
\caption{Lowest collective energy levels for \nucs{} nuclei. 
Upper panels: $2_1$ and $4_1$
levels in the g.s. band. Lower panels: $2_2$ and $0_2$ levels. Experimental
data taken from \cite{nndc0209}.\label{fig:lev}}
\end{figure}

\begin{figure}[htb]
\includegraphics{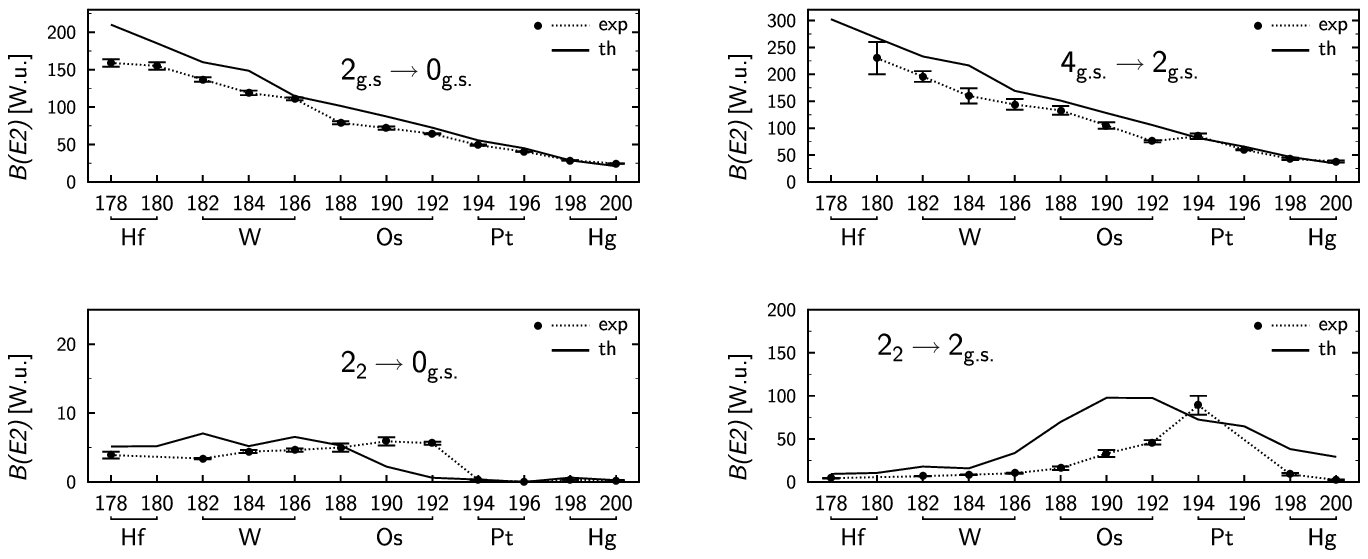}
\caption{Selected $B(E2)$  transition probabilities (in W.u.) in the \nucs{} nuclei.
Upper panels: transitions within the ground state band, lower panels:
$2_2\rightarrow 0_{\rm g.s.}$ and $2_2\rightarrow 2_{\rm g.s.}$. Note the
different scales in various panels. Experimental
data taken from \cite{nndc0209}.\label{fig:e2} }
\end{figure}

A sample of the results is presented below. 
Figures~\ref{fig:pot1}--\ref{fig:pot3} show the collective potential energy
relative to that of the spherical shape for three nuclei: \nuc{178}{Hf},
\nuc{190}{Os} and \nuc{200}{Hg}.  The inertial functions $B_{\bt\bt}$,
$B_{\bt\gm}$, $B_{\gm\gm}$, $B_{x,y,z}$ are plotted in
figures~\ref{fig:vibpar} and \ref{fig:rotpar} for the \nuc{190}{Os} nucleus
only.  They exhibit quite strong dependence on the $\bt$, $\gm$ variables
and differ significantly from the functions adopted in most of the
phenomenological models, in which $B_{\bt\bt}{=}B_{\gm\gm}{=}B_{x,y,z}{=}B$
and $B_{\bt\gm}=0$ (i.e.  in which there is only one constant mass parameter
$B$).  Next, the selected theoretical energy levels and the E2 transition
probabilities are compared with experimental data taken from
\cite{nndc0209}.  The levels are labelled with their spin $I$ and the number
$k$ for a given spin as $I_k$.  \Fref{fig:lev} shows the levels $2_1$ and
$4_1$ belonging to the ground state band and the levels $2_2$ and $0_2$,
which are often treated as bandheads of the quasi $\gm$ and $\bt$ bands. 
The reduced probabilities of two transitions within the g.s.  band, namely
$2_1\rightarrow 0_1$, $4_1\rightarrow 2_1$ as well as of two inter-band
transitions $2_2\rightarrow 0_1$ and $2_2\rightarrow 2_1$ are plotted in
\fref{fig:e2}.  Of course, the results shown here and those not shown
deserve a more detailed discussion which we postpone to a subsequent
publication.  However, one should admit that an overall agreement between the
theory and the experiment seen in figures \ref{fig:lev} and \ref{fig:e2} is
quite remarkable.  It must be stressed once more that the sole input for the
calculations is the effective nucleon-nucleon interaction without any
additional parameters, effective charges, etc.

\begin{figure}
\includegraphics{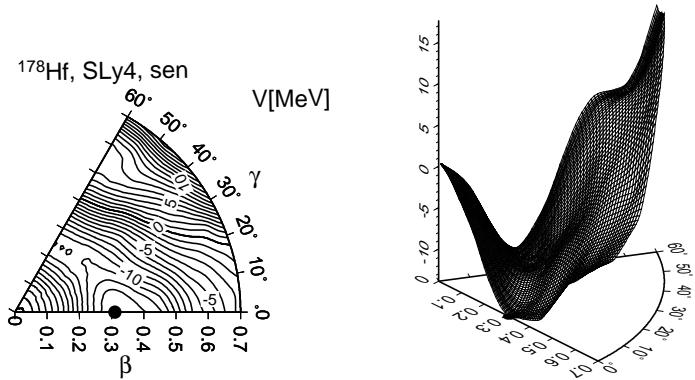}
\caption{Potential energy (relative to that of the spherical shape) 
for the \nuc{178}{Hf} nucleus calculated using the SLy4 Skyrme
interaction plus the constant $G$ pairing.\label{fig:pot1}}
\end{figure}

\begin{figure}
\includegraphics{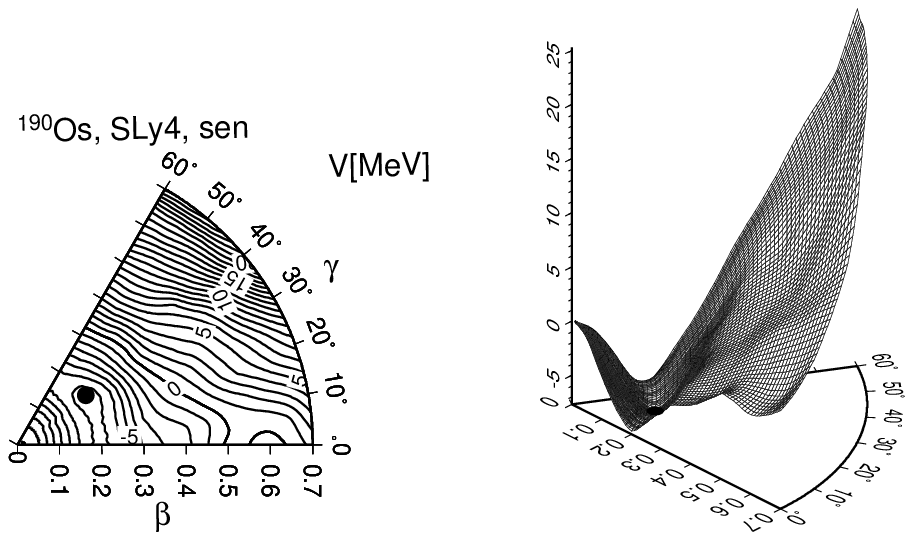}
\caption{Potential energy for the \nuc{190}{Os} nucleus, see also 
caption to \fref{fig:pot1}.\label{fig:pot2}}
\end{figure}

\begin{figure}
\includegraphics{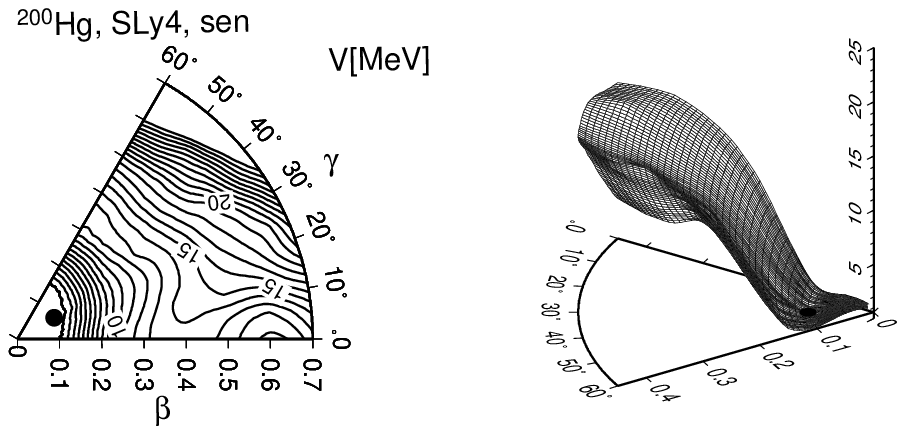}
\caption{Potential energy for the \nuc{200}{Hg} nucleus, see also 
caption to \fref{fig:pot1}.\label{fig:pot3}}
\end{figure}

\subsection{Extension of the collective space\label{subs:ext}}

We end this section with a brief discussion of ideas concerning the
extension of the collective space to variables connected with the pairing
degrees of freedom.  Some vague suggestions about such possibility can be
found already in \cite{Bel65,x68ba01}.  In the simplest version of the
collective approach to pairing, appropriate for the constant $G$
interaction, the pairing gap $\Delta$, which enters the formulae for the
$v_\mu$, $u_\mu$ coefficients, is treated as a variable and not a solution
of the BCS equations.  In this way a family of states is parametrized by
$\Delta$ and the ATDHFB \cite{Bes70} or the GCM+GOA \cite{Une76,1985GO15}
methods allow to obtain the collective pairing Hamiltonian.  The case of the
state dependent pairing needs a slightly different definition of a
collective variable instead of $\Delta$, see e.g.  \cite{2007PR07}.  Usually
one also considers a second variable, the gauge angle, which is used to
implement the projection on a fixed number of particles.  An important
consequence of the collective treatment of the pairing is that one gets
probability distributions in the $\Delta$ variable (coming from
eigenfunctions of the collective pairing Hamiltonian) instead of a single
value of the pairing gap.  It is well known that values of 
the inertial functions entering the Bohr Hamiltonian strongly depend
on the value of the pairing gap.  Hence 
it is important to estimate the influence of the
extension of the collective space on the spectrum and other properties 
of the  Hamiltonian.

If  both quadrupole and pairing variables are combined, the resulting
extended collective space is nine dimensional (four pairing variables for
two kinds of nucleons).  In principle one can use the methods from sections
\ref{sec:atdhfb} or \ref{subs:gcmgoa} to construct the collective
Hamiltonian also in this case \cite{2007PR07}.  However, a full analysis of
such Hamiltonian would be very difficult so that one must resort to some
approximations \cite{Pro99,2007PR07}.  Firstly, the terms which explicitly
couple quadrupole and pairing variables are neglected.  Secondly, the ground
state of the pairing part of the Hamiltonian is determined to get the
probability distribution in the $\Delta$ variable (for both protons and
neutrons).  In \cite{Pro99,1999ZA10} the inertial functions of the Bohr
Hamiltonian were calculated using the value of the pairing gap corresponding
to the maximum of the distribution while in \cite{2007PR07} the inertial
functions obtained for various values of $\Delta$ were averaged with the
weight given by this distribution. Results of the cited papers show that
the collective treatment of pairing can bring significant changes in the energy
spectrum, but more extensive studies (also on more consistent methods of
determining the pairing strength) are still needed.  


\section{Summary}\label{summary}

The Bohr collective model was originally invented to describe  quadrupole
oscillations of nuclear surface in the spherical nuclei.  It has been
subsequently developed and successfully applied to the description of the
quadrupole collective excitations in various deformed nuclei.  The Bohr
Hamiltonian can be treated either at the phenomenological level or can be
derived from a microscopic many-body theory.  In the present review we have
leaned towards the latter treatment of the Bohr model.  Microscopic theories
give, as a rule, quite complicated collective Hamiltonians which are
determined by a number of functions of the dynamical variables rather than
by a number of parameters as it used to be in the phenomenological
approaches.  Therefore, it is important to study the most general under some 
 natural conditions, form of the Bohr Hamiltonian.  The basic assumption
of the Bohr collective model is that the dynamical variables form a real
electric  quadrupole tensor.  The detailed
analysis of the isotropic tensor fields, presented in this review, allows us
to study the tensor structure of various ingredients of the general Bohr
collective model.  Its specific version is determined by several scalar
functions.  The collective Hamiltonian is defined by the six scalar inertial
functions, the scalar weight and the (obviously scalar) potential --- eight
scalar functions of two scalar variables altogether.  To define other
collective observables we still need a number of additional scalar
functions.  The knowledge of the universal tensor structure of the
collective wave functions allows us to express them also through scalar
functions specific for a definite eigenstate of a given Hamiltonian.  

The scalar functions defining the model can be calculated in the framework
of a microscopic theory.  In practice it is the components of the inertial
bitensor rather than the scalar inertial functions themselves that are
calculated.  There are the two fundamental methods to obtain the Bohr
Hamiltonian from the many-body theory.  One is the Adiabatic Time Dependent
Hartree-Fock-Bogolyubov approach which, since it treats the collective
motion as the classical motion of a wave packet, gives the classical and not
the quantum Hamiltonian.  Quantization of the classical Hamiltonian
obtained in this method can lead to an ambiguity in the collective
potential.  The other method is provided by the Gaussian Overlap
Approximation of the Generator Coordinate Method.  It directly gives the
quantum second order differential collective Hamiltonian (without the
Gaussian Overlap Approximation the Generator Coordinate Methods leads to 
integral equations).  Both methods mentioned above can be applied to 
different microscopic models.  A whole range of the nuclear many-body
theories are available: the phenomenological mean field models with different
single-particle potentials such as the Nilsson or the Woods-Saxon ones, the
self-consistent approaches with the Skyrme or the Gogny effective
nucleon-nucleon forces and the Relativistic Mean Field theory.  The
presented example
of application of the reviewed approach to the \nucs{} nuclei shows that the
Bohr Hamiltonian derived from the effective interaction is a very useful
tool for explaining nuclear collective properties.  In addition, having the
Bohr Hamiltonian obtained from the microscopic theory, one can try to examine
another question which recently attracted attention, that is  of
the possible foundations of phenomenological models inspired partly by the
Interacting Boson Model \cite{Ia00,For04}.  This would require, however, a
more extensive and detailed discussion which goes beyond the scope of the
present review.

\ack
The authors are  grateful to Piotr
Chankowski for a careful reading of the manuscript.
This work was supported in part by the Polish Ministry of Science
under Contract No.~N~N202~328234. 

\appendix
\section{Spherical tensors and tensor fields}\label{tensor}
In the nuclear collective models the operands are most often tensors under
the O(3) group of orthogonal transformations in the three-dimensional
physical space.  The models employ tensor fields such as potentials,
inertial functions, etc., which are usually isotropic functions of the dynamical
variables.  Here, in order to fix the notions and notation, we collect the definitions and properties of quantities and operations
appearing in the theory of spherical tensors. 
We also briefly present the structure of the isotropic tensor fields,
which is not so widely known.

\subsection{Spherical tensors}  \label{sphten}

A set of $2\lambda +1$ covariant  components $\alpha_{\lambda\mu}$ with projections
(magnetic numbers) $\mu =-\lambda ,\, -\lambda +1,\,\dots\, ,\lambda -1,\,
\lambda$, representing an object (a c-number, an operator, etc.)
$\alpha_{\lambda}$ in a three-dimensional coordinate system U, forms the
SO(3) irreducible spherical tensor of rank $\lambda$ if
\begin{equation}
\label{deften}
\alpha^{\prime}_{\lambda\mu^{\prime}}=\sum_{\mu=-\lambda}^{\lambda}D^{\lambda\ast}_{\mu\mu^{\prime}}(\varphi ,\vartheta ,\psi )\alpha_{\lambda\mu}
\end{equation} 
where $\alpha^{\prime}_{\lambda\mu^{\prime}}$ are the components
representing the same object $\alpha_{\lambda}$ in the coordinate system
U$'$ rotated with respect to U by the Euler angles $\varphi$, $\vartheta$, $\psi$, and
$D^{\lambda}_{\mu^{\prime}\mu}(\varphi ,\vartheta ,\psi )$ are the Wigner
functions (rotation matrices) \cite{Var88,BM69}.  The Wigner functions
defined in \cite{Var88} are complex conjugate of those of Bohr and Mottelson
\cite{BM69}. In nuclear physics
the convention of Bohr and Mottelson is most widely used and we adopt it in
this work.  We do not quote here various useful 
properties of the Wigner functions 
as they can be found easily for instance in
\cite{Var88}.  Only integer ranks $\lb$ are considered here.
 
In the special case of a rotation by the Euler angle $\varphi$ with $\vartheta
=0$, $\psi =0$ (i.e.  of a rotation about $z$-axis) formula \eref{deften}
reduces to
\beq\label{rotz}
 \alpha^{\prime}_{\lambda\mu^{\prime}}=\sum_{\mu}\rme^{-\rmi\mu\varphi}\delta_{\mu\mu '}\al_{\lb\mu}
\eeq
A tensor whose only nonvanishing components is $\al_{\lb 0}$ (i.e. 
$\al_{\lb\mu} =0$ for $\mu\neq 0$) is invariant under such rotations, and
can be called the axially symmetric tensor ($z$ is the symmetry axis).  In
general, a SO(3) tensor transforming under rotations as in \eref{deften},
does not have a definite transformation rule under the inversion of the
coordinate system, which supplements the rotations to the O(3) group. 
However, every SO(3) tensor can be decomposed into the two terms
\beq
\al_{\lb\mu}=\al_{\lb\mu}^{(\mathrm{E})}+\al_{\lb\mu}^{(\mathrm{M})}
\eeq
called the proper (polar or electric) and the pseudo- (axial or magnetic)
tensor, respectively, with the following transformation properties:
\beq\label{E}
{\al_{\lb\mu}^{(\mathrm{E})}}^{\prime\prime}=(-1)^{\lb}\al_{\lb\mu}^{(\mathrm{E})}
\eeq
and
\beq\label{M}
{\al_{\lb\mu}^{(\mathrm{M})}}^{\prime\prime}=(-1)^{\lb +1}\al_{\lb\mu}^{(\mathrm{M})}
\eeq
where the double prime denotes the components of the corresponding tensors
in the coordinate system U$''$ whose axes are reflected with respect to
those of U.  The sign factors on the right-hand sides of \eref{E} and
\eref{M} are the parities.  Further below, the superscripts (E) or (M) will be
omitted when the parity is not relevant.  In general, the components
$\al_{\lb\mu}$ are complex numbers.  However, in physical application most
frequently real spherical tensors appear whose components are determined by
$2\lb +1$ real numbers up to  a single phase factor. More precisely, the complex components of such
tensors satisfy the following relations
\begin{equation}\label{phase} \alpha^{\ast}_{\lambda\mu}=\rme^{2\rmi\delta
(\lb )} (-1)^{-\mu}\alpha_{\lambda -\mu}.  
\end{equation} 
Relations \eref{phase} are
preserved by the transformations \eref{deften}.  In the following the common phase is
set to zero  ($\delta
(\lb )=0$)   as in the case of the standard spherical harmonics
$Y_{\lambda\mu}$.  \Eref{phase} suggests that the tensor
$\alpha_{\lambda}$ can be presented in the following form:
\bn\label{reim}
\fl \alpha_{\lambda 0}= a_{\lambda 0} \qquad
\alpha_{\lambda\mu} = \frac{1}{\sqrt{2}}(a_{\lambda\mu}+ib_{\lambda\mu}) \qquad
\alpha_{\lambda -\mu} = \frac{(-1)^{\mu}}{\sqrt{2}}(a_{\lambda\mu}-ib_{\lambda\mu}) 
\en
for $\mu >0$.  Thus, the tensor $\al_{\lb}$ is indeed determined by the $2\lambda +1$
real numbers: the real and imaginary parts,
$a_{\lambda\mu}\; (\mu =0,\, 1,\,\dots ,\lambda )$ and $b_{\lambda\mu}\;
(\mu =1,\,\dots ,\lambda )$, respectively, of the complex components
$\alpha_{\lambda\mu}$ for $\mu\geq 0$.

Coupling two real tensors, $\alpha_{\lambda}$ and $\chi_{\kappa}$, of ranks
$\lambda$ and $\kappa$ and the zero phases in the following way
\begin{equation}\label{vc}
[\alpha_{\lambda}\times\chi_{\kappa}]_{LM}=\sum_{\mu\nu}(\lambda\mu\kappa\nu|LM)\alpha_{\lambda\mu}\chi_{\kappa\nu} ,
\end{equation}
where $(\lambda\mu\kappa\nu|LM)$ is the Clebsch-Gordan coefficient, one
obtains a real tensor of rank $L$, where $|\lambda -\kappa |\leq L\leq \lambda
+\kappa$, called an irreducible tensor product.  Its phase is determined
 by the relation
\beq\label{totphase}
[\alpha_{\lambda}\times\chi_{\kappa}]_{LM}^{\ast}=(-1)^{L-M-\lb -\kappa}[\alpha_{\lambda}\times\chi_{\kappa}]_{L-M} . 
\eeq
The scalar product of the two real tensors of the same rank, $\alpha_{\lambda}$ and
$\chi_{\lambda}$, is defined as
\begin{equation}\label{scalartensor}
(\alpha_{\lambda}\cdot\chi_{\lambda})=(-1)^{-\lambda}\sqrt{2\lambda +1}[\alpha_{\lambda}\times\chi_{\lambda}]_{00}.
\end{equation}

It is also convenient to formally introduce the contravariant components
$\alpha_{\lambda}^{\phantom{\lb}\mu}$ of a tensor $\alpha_{\lambda}$ by relation
\begin{equation}\label{contra}
\alpha_{\lambda}^{\phantom{\lb}\mu}= \alpha_{\lambda\mu}^{\ast} .
\end{equation}
The scalar product can then be written as
\begin{equation}\label{scalar}
(\alpha_{\lambda}\cdot\chi_{\lambda})=\sum_{\mu}(-1)^{\mu}\alpha_{\lambda\mu}\chi_{\lambda -\mu}
=\sum_{\mu}\alpha_{\lambda\mu}\chi_{\lambda \mu}^{\ast}=\sum_{\mu}\alpha_{\lambda\mu}\chi_{\lambda}^{\phantom{\lb}\mu}.
\end{equation}

The square of a tensor $\alpha_{\lambda}$ expressed in terms of the 
variables $a_{\lambda\mu}$ and $b_{\lambda\mu}$ reads
\begin{equation}\label{square}
(\alpha_{\lambda}\cdot\alpha_{\lambda})=\sum_{\mu =0}^{\lambda}a^2_{\lambda\mu}
+\sum_{\mu =1}^{\lambda}b^2_{\lambda\mu}.
\end{equation}
This shows that $a_{\lambda}$'s and $b_{\lambda}$'s can play the role of the Cartesian coordinates in 
the $(2\lambda +1)$-dimensional Euclidean
space, whose  volume element is
\begin{equation}\label{vol}
\mathrm{d}\Omega = \prod_{\mu =0}^{\lambda}\mathrm{d}a_{\lambda\mu}
\prod_{\mu =1}^{\lambda}\mathrm{d}b_{\lambda\mu}.
\end{equation}
Let this Euclidean space be infinite, i.e. $-\infty
<a_{\lb\mu},b_{\lb\mu}<\infty$ for all $\mu$'s.  The scalar product in the
Hilbert space H$_{\Omega}$ of functions 
of the Cartesian coordinates
$a_{\lb\mu},\ b_{\lb\mu}$ is defined in the usual way by
\beq\label{Hilbert}
\langle\Psi |\Phi\rangle =\int\Psi^{\ast}(a_{\lb},b_{\lb})\Phi (a_{\lb},b_{\lb})\rmd\Omega .
\eeq
The integral is over the whole space.

Apart from the spherical tensors, the bitensors or matrices with  tensor
indices also appear in the collective models.  The spherical bitensor is defined
in a similar way as the tensor, namely
\begin{equation}\label{biten}
M^{\prime}_{\lambda\mu^{\prime},\kappa\nu^{\prime}}
=\sum_{\mu\nu}D^{\lambda\ast}_{\mu\mu^{\prime}}(\varphi ,\vartheta ,\psi )
D^{\kappa\ast}_{\nu\nu^{\prime}}(\varphi ,\vartheta ,\psi )
M_{\lambda\mu ,\kappa\nu}
\end{equation}
where $M_{\lambda\mu,\kappa\nu}$ and
$M^{\prime}_{\lambda\mu^{\prime} ,\kappa\nu^{\prime}}$ are the components of the  bitensor $M_{\lambda ,\kappa}$
in the coordinate systems U and U$'$, respectively.  The bitensor $M_{\lambda ,\kappa}$ can always be
decomposed into a number of tensors $T^{(\lambda\kappa)}_{LM}$ of ranks $L$,
where 
$|\lambda -\kappa |\leq L\leq\lambda + \kappa$, 
as follows:
\begin{equation}\label{bidecom}
M_{\lambda\mu ,\kappa\nu}=\sum_{LM}(\lambda\mu \kappa\nu |LM)T^{(\lambda\kappa)}_{LM}.
\end{equation}
If a matrix of the bitensor $M_{\lb ,\lb}$ is symmetric the summation in
\eref{bidecom} is restricted to even $L$'s.  The scalar product of a
bitensor $M_{\lb ,\kappa}$ and a tensor $\alpha_\kappa$ is defined similarly to \eref{scalar}
\bn
 (M_{\lambda\mu ,\kappa}\cdot\al_{\kappa})&=&\sum_{\nu}(-1)^{\nu}M_{\lambda\mu ,\kappa -\nu}\al_{\kappa\nu}
=\sum_{\nu}M_{\lambda\mu ,\kappa}^{\phantom{\lb\mu\kappa}\nu}\al_{\kappa\nu} \label{scalright} \\
 (\al_{\lb}\cdot M_{\lambda ,\kappa\nu})&=&\sum_{\mu}(-1)^{\mu}\al_{\lb\mu}M_{\lambda -\mu ,\kappa\nu}
=\sum_{\mu}\al_{\lb\mu}M_{\lambda\phantom{\mu},\kappa\nu}^{\phantom{\lb}\mu}. \label{scalleft} 
\en

The differential operators\footnote{Differential operators acting in the Hilbert space of functions of variables $a_{\lb\mu},\ b_{\lb\mu}$ are denoted with the hat throughout the paper.}
\beq\label{cartmom}
\hat{p}_{\lb\mu}=-\rmi\hbar\frac{\partial}{\partial a_{\lambda\mu}} \qquad \hat{q}_{\lb\mu}=-\rmi\hbar\frac{\partial}{\partial b_{\lambda\mu}}
\eeq
are  Hermitian in H$_{\Omega}$ and play the role of the momenta  canonically
conjugate to the coordinates since they fulfill the following commutation relations
\bn\label{cartcomm}
\fl [a_{\lb\mu},\hat{p}_{\lb\nu}]=\rmi\hbar\delta_{\mu\nu} \qquad [b_{\lb\mu},\hat{p}_{\lb\nu}]=0 \qquad
[a_{\lb\mu},\hat{q}_{\lb\nu}]=0 \qquad [b_{\lb\mu},\hat{q}_{\lb\nu}]=\rmi\hbar\delta_{\mu\nu}.
\en
The covariant tensor operator of momentum is defined as
\begin{equation}\label{momcov}
\hat{\pi}_{\lambda\mu}=\frac{1}{\sqrt{2}}(\hat{p}_{\lb\mu}+\rmi \hat{q}_{\lb\mu})= -\rmi\hbar\frac{\partial}{\partial\alpha^{\ast}_{\lambda\mu}}=
 -\rmi\hbar\frac{\partial}{\partial\alpha_{\lambda}^{\phantom{\lb}\mu}}.
\end{equation}
The relations inverse to \eref{reim} have been used here to express the
derivatives with respect to  $\al_{\lb\mu}$ through the derivatives 
with respect to $a_{\lb\mu}$
and $b_{\lb\mu}$.  The momentum tensor Hermitian adjoint to
$\hat{\pi}_{\lambda\mu}$ is
\begin{equation}\label{momcontra}
\hat{\pi}_{\lambda\mu}^{\dag}=\frac{1}{\sqrt{2}}(\hat{p}_{\lb\mu}-\rmi \hat{q}_{\lb\mu}) =-\rmi\hbar\frac{\partial}{\partial\alpha_{\lambda\mu}}.
\end{equation}
It plays the role of the momentum tensor canonically conjugate to the 
coordinate tensor $\alpha_{\lambda}$ since it fulfills the
following commutation relation:
\begin{equation}\label{comm}
[\alpha_{\lb\mu},\hat{\pi}_{\lambda\nu}^{\dag}]=\rmi\hbar\delta_{\mu\nu}.
\end{equation}
It follows from  (\ref{deften}) that the Hermitian adjoint momentum
operator (\ref{momcontra}) has the transformation properties of the
contravariant and not the covariant tensor in spite of the convention of
notation of (\ref{contra}).  According to the convention (\ref{contra}) 
\begin{equation}\label{astdag}
\hat{\pi}_{\lambda}^{\phantom{\lb}\mu}=\rmi\hbar\frac{\partial}{\partial\alpha_{\lambda\mu}}=-\hat{\pi}_{\lambda\mu}^{\dag}.
\end{equation}
Obviously, the momentum tensor is of the same parity as the conjugate
coordinate tensor.  It is easy to check that the differential operator
\begin{equation}\label{angmom}
\hat{L}_{1\mu}^{(\lambda )}=\rmi (-1)^{\lambda}\sqrt{\frac{\lambda (\lambda +1)(2\lambda +1)}{3}}[\alpha_{\lambda}\times\hat{\pi}_{\lambda}]_{1\mu}
\end{equation}
obeys the following commutation relations:
\begin{equation}\label{commla}
[\hat{L}_{1\nu}^{(\lambda )},\alpha_{\lambda\mu}]= \sqrt{\lambda (\lambda +1)}(\lambda\mu 1\nu |\lambda\mu +\nu )\alpha_{\lambda\mu +\nu}   
\end{equation}
and
\begin{equation}\label{commlp}
[\hat{L}_{1\nu}^{(\lambda )},\hat{\pi}_{\lambda\mu}]= \sqrt{\lambda (\lambda +1)}(\lambda\mu 1\nu |\lambda\mu +\nu )\hat{\pi}_{\lambda\mu +\nu}
\ .
\end{equation}
This means that $\hat{L}_{1\mu}^{(\lambda )}$ is the angular momentum
operator acting in the space of functions of $\alpha_{\lambda\mu}$ (cf
\cite{Var88}).  The angular momentum (\ref{angmom}) is always an axial
vector irrespective of the parity of $\al_{\lb}$ and $\hat{\pi}_{\lb}$. 
Equations \eref{angmom}, \eref{commla} and \eref{commlp} lead to the obvious
commutation relations for the components of the angular momentum
\beq\label{commll}
[\hat{L}_{1\mu}^{(\lambda )},\hat{L}_{1\nu}^{(\lambda )}]=-\hbar\sqrt{2}(1\mu 1\nu |1\kappa )\hat{L}_{1\kappa}^{(\lambda )}.
\eeq

\subsection{Isotropic tensor fields}\label{field}

In continuum mechanics there appears the notion of an isotropic function of
variables which are the components of a tensor (cf \cite{Trues52}).
Roughly speaking, the function is isotropic if no parameters with tensor
properties (e.g.  material tensors)  enter its definition.  The isotropic
functions which are tensors themselves, are called the isotropic
tensor fields.  The isotropic tensor field $T_{ LM}(\al_{\lambda})$ of rank
$L$
obeys  the following transformation rule:
\begin{equation}\label{isofield}
T'_{LM'}(\al_{\lb\mu})=T_{LM'}(\al_{\lb\mu}^{\prime})
= \sum_{M=-L}^{L} D^{L\ast}_{MM'}(\varphi ,\vartheta ,\psi) T_{LM}(\al_{\lb\mu}) 
\end{equation}
where $T_{LM}$ and $\al_{\lb\mu}$ are the components of tensors 
$T_L$ and $\al_{\lb}$ in 
the original system U  whereas $T'_{LM'}$ and $\al^{\prime}_{\lb\mu}$ 
are the respective components in the rotated system U$'$. 
\Eref{isofield} states that to transform the field $T_{ LM}(\al_{\lambda})$ it is sufficient to transform its tensor argument $\al_{\lb\mu}$. It would not 
be so for a tensor field which is not isotropic.


The question arises to what extent the form of functional dependence of the
isotropic tensor field $T_{LM}(\al_{\lb})$ on the tensor $\al_{\lb}$ is
determined solely by their tensor properties.
To answer this question we present below a scheme of constructing an
arbitrary spherical tensor as a function of another spherical tensor.  
It turns out that
for a given rank $\lambda$ there exists a finite complete system of
$i_{\lambda}$ irreducible  
(i.e.  none of them can be expressed rationally
and integrally in terms of the others)
elementary tensors
$\varepsilon_{ilm}(\al_{\lb})$, $i=1,...,i_{\lb}$, of different ranks
$l$. 
 The highest components
$(m=l)$ of the elementary tensors are called the elementary factors
\cite{Chac77}.  The elementary tensors are constructed by 
coupling  successively $n$ tensors $\al_{\lambda}$
to  different $l$'s, $l<n\lambda$:
\begin{equation}
\label{elem-fact}
\varepsilon_{ilm}(\al_{\lb})\equiv\varepsilon_{lm}^{(n)}(\al_{\lb}[\mathrm{c}])=[\underbrace{\al_{\lambda}\times\dots\times
\al_{\lambda}}_{n}]_{lm}^{[\mathrm{c}]},
\end{equation}
where the symbol [c], redundant in most cases,
specifies the coupling scheme.  Tensor
$\al_{\lambda}=\varepsilon_{\lambda}^{(1)}$ itself is an elementary tensor. 
The scalar product of two tensors $\alpha_{\lb}$, 
$(\al_{\lb}\cdot\al_{\lb})=\varepsilon_{0}^{(2)}$ as well as all other independent
scalars are the elementary factors.  Completeness of the system of
elementary factors does not mean that relations between them must not exist. 
On the contrary, some relations between them, called syzygies, do exist. 
They take the form:
\begin{equation}
\label{syz}
S_j(\varepsilon_{1ll},\dots ,\varepsilon_{i_{\lambda}l'l'})=0
\end{equation}
with $j=1,\dots ,j_{\lambda}$, where $S_j$ are some  rational integral functions. 
The number of independent syzygies is finite and depends on the
rank $\lambda$.
It turns out that  to obtain a tensor of a given rank $L$, it is sufficient
to align (i.e.  to couple to the maximal angular momentum) the elementary
tensors.  The alignment of the elementary tensors means the multiplication
of the elementary factors (the relevant Clebsch-Gordan coefficients are all
equal to 1).  Because of the existence of the syzygies \eref{syz}, some of the
aligned tensors
\begin{displaymath} 
[\varepsilon_{l}^{(n)}(\al_{\lb}[\mathrm{c}])\times\varepsilon_{l'}^{(n')}(\al_{\lb}[\mathrm{c'}])\times\dots]_{L=l +l'+\dots}
\end{displaymath}
can be expressed rationally and integrally in terms of the others.  Using
the syzygies the finite number $k_{\lambda L}$ of independent aligned
tensors $\tau_{kLM}(\varepsilon )$ with $k= 1,\dots ,k_{\lambda L}$ can be
found.  These independent aligned tensors are called fundamental tensors. 
This is because an arbitrary tensor field of rank $L$, being the isotropic
function of tensor $\al_{\lambda}$, can  always be written in the form:
\begin{eqnarray}
\label{L-field}
T_{LM}(\al_{\lambda})&=&\sum_{k=1}^{k_{\lambda L}}f_k(\varepsilon_0)\tau_{kLM}(\varepsilon)
\end{eqnarray}
where argument $\varepsilon_0$ of $f_k$ stands  for all independent
scalars.  Scalar functions $f_k$ can be, in general, arbitrary.  However,
the form of some of $f_k$ can be restricted.  For instance, some scalar
$\varepsilon_0$ cannot appear in it in too high  power.  This is because
a syzygy can make the expression
$(\varepsilon_{i00})^n(\varepsilon_{i'll})^{n'}$ for some $i$, $i'$, $n$ and
$n'$ dependent on other elementary factors.  When $\al_{\lb}$ has the
positive parity, an arbitrary field of the form (\ref{L-field}) has obviously
the positive parity as well.  However, when $\al_{\lb}$ has the negative parity,
the field $T_{L}$ has, in general, no definite parity.  To decompose it into
the parts of the positive and the negative parity, the parities of the
scalars and the fundamental tensors should be taken into account.  This can 
easily be
done by noticing that the parity of an elementary tensor
$\varepsilon_{l}^{(n)}([\mathrm{c}])$ is $(-1)^n$.

A justification of the procedure presented above can be traced back to the
theory of covariants of algebraic forms given by Dickson \cite{Dic14},
which, however, uses  different notions than these used in the present
paper.  Let us sketch the main points of this theory (cf \cite{Tur46,Olv99}). 
Let the binary (in variables $x_1,\, x_2$) $2\lambda$-ic (of order
$2\lambda$) algebraic form be
\bn
\label{form}
&& F_{2\lambda}(x_1,x_2;\al_{\lambda -\lambda},\dots ,\al_{\lambda\lambda}) 
=\sum_{\mu=-\lambda}^{\lambda}\left(\ba{c}2\lambda \\ \lambda - \mu \ea\right)^{1/2}
\al_{\lambda\mu}x_1^{\lambda +\mu}x_2^{\lambda -\mu}
\en  
with a set of coefficients $(\al_{\lambda\mu}$,\ $\mu=-\lambda ,\dots
,\lambda)$.  Changing in \eref{form} the variables by a linear non-singular
transformation
\begin{equation}
\label{lintrans}
x_i=\sum_{k=1}^2A_{ik}y_k
\end{equation}
where $i=1,2$,   one obtains
\bn
\label{trform}
\fl F_{2\lambda}(x_1,x_2;\al_{\lambda} )=G_{2\lambda}(y_1,y_2;\al_{\lambda}) 
=\sum_{\mu=-\lambda}^{\lambda}\left(\ba{c}2\lambda \\ \lambda - \mu \ea\right)^{1/2}
\chi_{\lambda\mu}(\al_{\lambda})y_1^{\lambda +\mu}y_2^{\lambda -\mu} 
\en  
with a new set of coefficients $(\chi_{\lambda\mu},\, \mu=-\lambda ,\dots
,\lambda)$.  The $2l$-ic form
\bn
\label{cov}
H_{2l}^{(n)}(x_1,x_2;\al_{\lambda})&=&\sum_{m=-l}^{l}\left(\ba{c}2l \\ l - m \ea\right)^{1/2}
h_{lm}^{(n)}(\al_{\lambda})x_1^{l +m}x_2^{l -m}, \nonumber \\
\en  
such that
\begin{equation}
\label{covdef}
H_{2l}^{(n)}(y_1,y_2;\chi_{\lambda})=(\mathrm{det}(A_{ik}))^wH_{2l}^{(n)}(x_1,x_2;\al_{\lambda})
\end{equation}
is called  a (homogeneous) covariant with the weight $w$ of $F_{2\lb}$. Coefficients
$h_{lm}^{(n)}$ are homogeneous polynomials of order $n$ (called the degree
of the covariant) such that $\lambda n=l+w$.  For $l=0$ the form
$H_0^{(n)}=h_{00}^{(n)}$ is called an invariant of $F_{2\lb}$.  The polynomial in front of
the highest power of $x_1$ in \eref{cov}, $h_{ll}^{(n)}$, is called a
semi-invariant of $F_{2\lb}$.  It can be shown in the theory of covariants of
algebraic forms that any covariant can be expressed rationally and
integrally in terms of a finite, irreducibly complete set of covariants,
which can be related rationally and integrally to each other by a finite
system of independent syzygies (Gordan-Hilbert Finiteness Theorem, see
\cite{Tur46}).  The number of the basis covariants for a few lowest $\lb$'s
is listed in \cite{Olv99}.

It remains to answer the question of what the covariants of algebraic forms
have in common with the tensor fields being functions of tensors.  It turns
out that the semi-invariant $h_{ll}^{(n)}(\al_{\lambda})$ forms the
component of the highest projection of a tensor field of rank $l$ depending
on the tensor of rank $\lambda$ (the highest weight of an irreducible
representation of SO(3) embedded in an irreducible representation of
SU($2\lambda +1$)).  The other polynomials $h_{lm}^{(n)}(\al_{\lambda})$ in
(\ref{cov})
 are the remaining components of the same tensor.  A proof of this statement
is demonstrated briefly in reference \cite{Roh78}.  Constructive proofs of
the form (\ref{L-field}) can be carried through by explicitly constructing a
basis in the space of functions of $\al_{\lambda}$ and showing that it
has the structure as in (\ref{L-field}).  In the case of $\lambda =1$ it is
well known from  textbooks of quantum mechanics that an arbitrary tensor
field of rank $L$ which is a function of the position vector $\bi{r}$ takes
the form:
\begin{equation}
\label{field-1}
T_{LM}(\bi{r})=f_L(r)Y_{LM}(\frac{\bi{r}}{r}) 
\end{equation}
where $r^2=\bi{r}{\cdot}\bi{r}$ and $Y_{LM}$ is the corresponding
spherical harmonic.  From \eref{field-1} it is seen immediately that in the
case of $\lambda =1$ there are the two elementary tensors $(i_{1}=2)$,
namely the vector $\bi{r}$ and the scalar $r^2$, and no syzygy $(j_1=0)$.  
Moreover, for every
given $L$ there is only one fundamental tensor $(k_{1L}=1)$ of the form
\cite{Var88}:
\beq
\label{sphar}
\tau_{LM}(\bi{r})=[\underbrace{\bi{r}\times\bi{r}\times\dots\times\bi{r}}_{L}]_{LM}  
\propto  r^LY_{LM}(\frac{\bi{r}}{r}).
\eeq
Constructions of the oscillator bases in the cases of $\lambda =2$ (see
\cite{Eis87} and references quoted therein) and $\lambda =3$ \cite{Roh78}
have been also performed confirming the form \eref{L-field}.

\section{Properties of the ATDHFB inertial functions in the intrinsic
frame\label{app:iner}}

\subsection{(In)dependence on the Euler angles.}

In this Appendix the three Euler angles $(\varphi,\vartheta,\psi)$ will be
compactly denoted as $\wec{\varphi}$ or $\varphi_j$, $j{=}1,2,3$.  The
BCS-type states depending on the deformation and the Euler angles
$\wec{\varphi}$ are constructed as $U(\wec{\varphi})|\Phi(\wec{q})\rangle$
where $|\Phi(\wec{q})\rangle$ is the BCS state in the intrinsic frame and
$\wec{q}$ stands for any of the pairs of the deformation variables:
$(a_0,a_2)$, $(q_0,q_2)$ or $(\bt,\gm)$.  $U(\wec{\varphi})$ is the rotation
operator in the Fock space.  The argument $\wec{\varphi}$ will sometimes be
dropped for the sake of brevity.  The density matrix
$\cR_0(\wec{q},\wec{\varphi})$ can be written as
\beq\label{eq:room}
\cR_0(\wec{q},\wec{\varphi})=\cU\cR_0(\wec{q})\cU^+
\eeq
 where
\beq
\cU=\left(\begin{array}{cc}
u &0\\
0 & u^*
\end{array}\right) \ .
\eeq
$u$ is a matrix of $U(\wec{\varphi})$ in a fixed basis of the one-particle
space. It
 fulfills the relation
\beq\label{eq:doom}
U^+d^+_\mu U=\sum_\nu u^*_{\mu\nu}d^+_\nu \ .
\eeq
\Eref{eq:room}  can be derived as follows
(the $(11)$ block of the $\cR_0$ is  taken as an example):
\bn
\nonumber\fl(\cR^{11}_0(q,\wec{\varphi}))_{mn}=
\langle U\Phi(\wec{q})|d^+_nd_m\, |U\Phi(\wec{q})\rangle =
\langle \Phi(\wec{q})|U^+d^+_nUU^{+}d_m\, U|\Phi(\wec{q})\rangle =\\
=\sum_{jk}u^*_{nj}u_{mk}\langle \Phi(\wec{q})|d^+_jd_k\, |\Phi(\wec{q})\rangle =
(u\cR^{11}_0(\wec{q})u^+)_{mn} \ .
\en

Now we show that the ATDHFB inertial functions do not depend on the Euler
angles. To calculate the vibrational functions $B_{q_iq_k}$ one needs the 
derivatives of  $\cR_0(\wec{q},\wec{\varphi})$ with respect to
the deformation variables.
It is easy to show that the derivatives
 can  be written analogously to \eref{eq:room},
i.e. 
$\partial_{q_j}\cR_0(\wec{q},\wec{\varphi})=\cU\partial_{q_j}\cR_0(\wec{q})\cU^+$. 
The same is true for $\cR^{k}_1(\wec{q},\wec{\varphi})$ in view of
\eref{eq:basec1} and the properties of the induced self-consistent
Hamiltonian $\cW$.
Consequently, all
vibrational functions
 (which are proportional to the trace
$\Tr_{2d} (R^{k}_1[\partial_{q_j}{\cal  R}_0,\cR_0]$))
do not depend on the Euler angles.

The rotational functions deserve  a  detailed treatment. 
Derivatives of $\cR(\wec{q},\wec{\varphi})$ with respect to the Euler angles
$\varphi_j$ are equal
\bn
\fl\partial_{\varphi_j}\cR^{11}_{mn}(\wec{q},\wec{\varphi})=
\langle \partial_{\varphi_j} U\Phi(\wec{q})|d^+_nd_m |U\Phi(\wec{q})\rangle 
+
\langle U\Phi(\wec{q})|d^+_nd_m |\partial_{\varphi_j}U\Phi(\wec{q})\rangle =\\
\fl=\langle U^{-1}\partial_{\varphi_j} U\Phi(\wec{q})|U^{+}d^+_nd_m U|\Phi(\wec{q})\rangle 
+
\langle \Phi(\wec{q})|U^{+}d^+_nd_mU|U^{-1} \partial_{\varphi_j}U\Phi(\wec{q})\rangle =\\
=(u\tilde{\cR}^{11}_{0,j}(\wec{q},\wec{\varphi})u^+)_{mn}
\en
where
\beq
(\tilde{\cR}^{11}_{0,j})_{mn}(\wec{q},\wec{\varphi})=
\langle \Phi(\wec{q})|d^+_nd_m|U^{-1} \partial_{\varphi_j}U\Phi(\wec{q})\rangle + {\rm
h.c.}
\eeq
The operator
$U^{-1}\partial_{\varphi_j}U$ can be expressed in terms of the generators of the rotation
group
\beq
U^{-1}\partial_{\varphi_j}U=\sum_k P_{kj}(\wec{\varphi})Y_k
\eeq
where $P(\wec{\varphi})$ is the universal (i.e. the same for all representations of
the rotation group) matrix
\beq
\label{wz:macp}
P=\left(\begin{array}{ccc}
-\sin\vartheta\cos\psi , &\sin\psi, & 0 \\
\sin\vartheta\sin\psi , & \cos\psi, & 0 \\
\cos\vartheta, &  0, & 1
\end{array}\right)
\eeq
and $Y_k=-iJ_k/\hb$, where $J_k$ are  components of the angular momentum
operator in the Fock space. Finally
\beq
\tilde{\cR}^{11}_{0,j}(\wec{q},\wec{\varphi})=\sum_k P_{kj}(\wec{\varphi})
u{\cF}^{11}_k(\wec{q})u^+
\eeq 
where the matrices
\beq
(\cF_k^{11}(\wec{q}))_{mn}=
-\rmi\hb^{-1}\langle \Phi(\wec{q})|d^+_nd_m J_k|\Phi(\wec{q})\rangle + {\rm
h.c.}
\eeq
do not depend on the Euler angles.
Of course, all the blocks of the $\tilde{\cR}_{0,j}(\wec{q}, \wec{\varphi})$ matrix exhibit
the same dependence on the Euler angles. Moreover, the matrices $\cR_1$
behave in an analogous way, that is
\beq
\cR_{1}^l=\dot{\varphi_l}\sum_rP_{rl}(\wec{\varphi}){\cU} {\cZ}_r(\wec{q}){\cU}^+
\eeq
where $\cZ_r(\wec{q})$ do not depend on the Euler angles.

Taking into account that  the angular velocity in
the body fixed frame is given by
\begin{equation}
\omega_j=\sum_k P_{jk}\dot{\varphi_k}
\end{equation}
the rotational classical Hamiltonian can be written as
\begin{equation}\label{eq:erot}
H_{\rm cl,rot}=\frac{1}{2}\sum_{kj} I_{kj}\omega_k\omega_j 
\end{equation}
with
\beq\label{eq:ikj}
I_{kj}=\frac{\rmi\hbar}{2}\Tr_{2d} (\cZ_k[\cF_j,\cR_0]) 
\eeq
which obviously  do not depend on the Euler angles.

\subsection{Block structure of the inertia tensor in the intrinsic frame}

Taking into account the ${\rm D}_2$ symmetry of the $|\Phi(\wec{q})\rangle$
state, one can prove the ATDHFB inertial functions give the classical
kinetic energy which has the form identical with that of \eref{inclham}
\beq H_{\rm cl,kin}=H_{\rm cl,kin,vib}+H_{\rm cl,rot}
\eeq 
or, more precisely, that the vibration-rotational terms are absent and, moreover, that
\beq\label{eq:ikjdelta} I_{kj}=\delta_{kj}I_k \ .  
\eeq 
The sketch of the proof is as follows.  The operators $S_k=\exp(-\rmi\pi
J_k/\hb)$ are the generators of the group ${\rm D}_2$ and invariance of the
$|\Phi(\wec{q})\rangle$ with respect to the ${\rm D}_2$ group means that
$S_k|\Phi(\wec{q})\rangle=|\Phi(\wec{q})\rangle$, which in the density
matrix language can be rewritten as $\cS_k\cR_0 \cS^+_k=\cR_0$.  The
connection between $S_k$ and $\cS_k$ is analogous to that between $U$ and
$\cU$ (cf \eref{eq:room} to \eref{eq:doom}).  For the angular momentum in the
Fock space one has $S^+_kJ_jS_k= J_j$ for $k=j$ and $S^+_kJ_jS_k=-J_j$
otherwise.  In consequence, the matrices $\cF_j$ and $\cZ_l$ have the same
property, i.e.  $\cS_k \cF_j \cS^+_k=\pm \cF_j$.  The formula
\eref{eq:ikjdelta} follows from the definition \eref{eq:ikj} of $I_{kj}$ and
the properties of the trace.  In a similar way one can verify that the mixed
vibration-rotational parameters $B_{q_i\varphi_j}$ vanish.

If  $H_{\rm cl,rot}$ \eref{eq:erot} is treated as an expression for the
metric tensor in the space of rotation matrices and \eref{eq:ikjdelta} is
valid, the corresponding Laplace-Beltrami operator has the form
\begin{equation}
\sum_k\frac{1}{I_k}\tilde{L}^2_k
\end{equation}
where $\tilde{L}_k$ are the generators of the right regular representation
of the SO(3) group.  They are proportional to the collective angular
momentum operators in the intrinsic frame, cf \eref{inangmom}.

\end{document}